\documentclass[a4]{aa}
\usepackage{longtable,lscape}
\usepackage{graphicx,amssymb,latexsym,amsmath}
\usepackage[authoryear]{natbib}
\bibpunct{(}{)}{;}{a}{}{,} 
\begin{document}

\title{REM near-IR and optical photometric monitoring of Pre-Main Sequence 
Stars in Orion.
\thanks{Based on observations collected at the ESO REM telescope 
(La Silla, Chile) and at the MMT Observatory, a joint facility of the Smithsonian Institution and the University of Arizona.}
\fnmsep\thanks{Tables~\ref{tab:VRIJHK} and \ref{tab:references} and the light curves of the variable stars are only available 
		in electronic form at the CDS via anonymous
		ftp to {\tt cdsarc.u-strasbg.fr (130.79.128.5)} or via 
		   {\tt http://cdsweb.u-strasbg.fr/cgi-bin/qcat?J/A+A/}}
\fnmsep\thanks{Figures \ref{fig:flare_11}--\ref{fig:spe_mg_all} are only available 
			in electronic form at {\tt http://www.aanda.org}}}
\subtitle{Rotation periods and starspot parameters}

\author{A. Frasca\inst{1} \and E. Covino \inst{2} \and L. Spezzi\inst{1,3} 
\and J.M. Alcal\'a\inst{2} \and E. Marilli\inst{1} \and G. F\H{u}r\'{e}sz\inst{5} \and D. Gandolfi\inst{1,4} } 
\offprints{A. Frasca}
\mail{antonio.frasca@oact.inaf.it}

  \institute{INAF, Osservatorio Astrofisico di Catania, via S. Sofia, 78, 95123 Catania, Italy
\and  INAF, Osservatorio Astronomico di Capodimonte, via Moiariello 16, 80131 Napoli, Italy
\and European Space Agency (ESTEC), PO Box 299, 2200 AG Noordwijk, The Netherlands 
\and Th\"uringer Landessternwarte Tautenburg, Sternwarte 5, 07778 Tautenburg, Germany
\and Center for Astrophysics, 60 Garden Street, Cambridge, MA 02138, USA} 

   \date{Received 21 September 2009 / Accepted 24 October 2009 }

\abstract
{}
{We aim at determining the rotational periods and the starspot properties in very young 
low-mass stars belonging to the Ori OB1c star forming region, contributing to the study 
of the angular momentum and magnetic activity evolution in these objects. }  
{We performed an intensive photometric monitoring of the PMS stars falling in a field 
 of about 10$\arcmin\times 10\arcmin$ in the vicinity of the Orion Nebula Cluster (ONC), 
also containing the BD eclipsing system 2MASS\,J05352184-0546085.
Photometric data were collected between November 2006 and January 2007 with the REM telescope 
in the $VRIJHK'$ bands.
The largest number of observations is in the $I$ band (about 2700 images)
and in $J$ and $H$ bands (about 500 images in each filter).  
From the observed rotational modulation, induced by the presence of surface inhomogeneities, 
we derived the rotation periods. 
The long time-baseline (nearly three months) allowed us 
to detect rotation periods, also for the slowest rotators, with sufficient accuracy ($\Delta P/P<2\%$).
The analysis of the spectral energy distributions and, for some stars, of high-resolution
spectra provided us with the main stellar parameters (mass, luminosity, effective temperature,
mass, age, and $v\sin i$) which are essential for the discussion of our results.
Moreover, the simultaneous observations in six bands, spanning from optical to 
near-infrared wavelengths, enabled us to derive the starspot properties for these very
young low-mass stars.}
{In total, we were able to determine the rotation periods for 29 stars,
spanning from about 0.6 to 20 days. Thanks to the  relatively long
time-baseline of our photometry, we derived periods for 16 stars and
improved previous determinations for the other 13. We also report the
serendipitous detection of two strong flares in two of these objects.
In most cases, the light-curve amplitudes decrease progressively from the $R$ to $H$ band 
as expected for cool starspots, while in a few cases, they can only be modelled by the presence 
of hot spots, presumably ascribable to magnetospheric accretion. 
The application of our own spot model to the simultaneous light curves in different bands allowed us
to deduce the spot parameters and particularly to disentangle the spot temperature and size
effects on the observed light curves. 
}
{}


\titlerunning{REM observations of PMS stars in Orion}
\authorrunning{A. Frasca et al.}
\maketitle

\section{Introduction}
\label{sec:Intro}

The study of the evolution of angular momentum of low-mass stars in the pre-main sequence (PMS) 
phase has gained great advantage from high-precision photometry of star forming regions (SFR). 
While during the main sequence (MS) evolutionary phase the total angular momentum of a solar-mass star 
is lost efficiently via the magnetic-wind breaking process \citep[e.g.,][]{Chabo95}, in the PMS phase 
the presence of a circumstellar disc coupled to the star through magnetic fields permits a more efficient 
angular momentum transfer, thus preventing the spin-up driven by the rapid mass-accretion and contraction 
of the star \citep{Shu94}. 

An accurate knowledge of the rotation period distribution of the solar and low-mass stars in different SFR 
environments is of paramount importance for understanding how the disc-locking mechanism is effective in 
modifying the angular momentum in the PMS stage, and for testing the reliability of the theoretical models 
of PMS star interiors and their evolutionary tracks.

Measurements of periodic light-variations with typical amplitudes in the range 0.05--0.40 mag in the 
$V$ and $R$ bands, due to uneven 
starspot distributions on the stellar photospheres, are a direct way to obtain this physical parameter. 	 
There has been a rapid increase in the number of PMS objects with measured rotation periods over 
the last decade \citep[see, e.g.,][and references therein]{Mat04,Herbst07,Irwin09}.
A broad, bimodal distribution 
exists at the earliest observable phases (about 1~Myr) for stars more massive than 0.4~M$_{\sun}$. 
The fast rotators (50--60\% of the sample) evolve to the ZAMS with little or no angular momentum loss. 
The slow rotators continue to lose substantial amounts of angular momentum for up to 5 Myr, creating the even 
broader bimodal distribution characteristic of 30--120 Myr old clusters. Accretion disc signatures, such as 
H$\alpha$ emission and infrared (IR) excess, are observed more frequently among slowly rotating PMS stars, 
suggesting a connection between accretion and rotation \citep{rebu06}. According to the present picture, discs 
appear to influence rotation only for a small percentage of the solar-type stars and only during the first 5 Myr 
of their life. This time interval is comparable to 
the maximum life-time of accretion discs derived from NIR studies and may be a useful upper limit to the 	 
time available for forming giant planets. For the sparse population of young stars (mainly weak-T Tauri
stars, wTTS, with ages between 5 and 30 Myr) in the Orion complex, \citet{Mari07} found no evidence of 
bimodality for masses higher than 0.7~M$_{\sun}$, and a median rotation period of $\sim$\,1.5 days, 
suggesting that the spin-up process at this stage has become dominant over the disc-locking effect. 
It is then mandatory to explore the rotational properties of as many clusters/associations as possible 
with ages less than 5 Myr. 
Indeed, in this age range, the different scenarios for angular momentum evolution proposed by \citet{lamm}  
and related to different time-scales of disc-locking and stellar contraction \citep{Hart02} are expected 
to produce their stronger effects.
 	 
There is growing evidence that the same mechanisms operating in solar-mass stars also regulate the rotation 
of very-low mass (VLM) stars and brown dwarfs (BDs). Indeed, very recently many BDs showing accretion discs 	 
with physical properties similar to their analogs around higher-mass stars have been discovered 
\citep[e.g.,][and reference therein]{Luhman05,Alc06,Muz06,Mer07}. 
However, it seems that both disc interactions and stellar winds are less efficient in braking these objects, 
although the statistics available for this VLM regime is rather poor. 

In some cases, both VLM and BD objects in star forming regions exhibit photometric light curves 
with high amplitudes and irregular modulations \citep{Scholz04,Scholz05} that are usually explained 
by hot spots caused by mass-accretion flow from the disc \citep{Herbst94}.
The same authors remark that wTTS and some classical T Tau stars (cTTS) display instead more regular 
light curves with smaller amplitudes ascribable to cool photospheric spots that can survive for several 
stellar rotations. 

In this paper, we report the results of an intensive photometric monitoring, 
in near-infrared (NIR) and optical bands, of a sample of young low-mass stellar and sub-stellar
objects (YSOs) in a small area flanking the Orion Nebula Cluster (ONC).
The main goal of our observations was the study of the
NIR light curve of \object{2MASS\,J05352184-0546085}, an eclipsing system
composed of two BD components recently discovered by \citet{stass06}. 
The long-lasting and intensive simultaneous optical and NIR
monitoring also allowed us to study the photometric variability of many
YSOs in the field of view around the eclipsing BD. Therefore, in this study
we report the characterization of these objects and determine their
rotational periods, as well as the properties of their starspots.
The present work thus contributes to the study of the early phases of angular momentum evolution and of 
the behaviour of magnetic activity in young, VLM objects which still needs observational 
and interpretation efforts.

We derived the rotation periods of our targets by means of time-series analysis of our large 
photometric data-set gathered in the $VRIJHK'$ bands. 
 The physical characterization (radius, luminosity, effective temperature, mass, and age) of the objects, 
which is essential for any discussion on angular momentum evolution and stellar activity, is based on 
our photometric data as well as on literature mid-IR data from the Spitzer space telescope 
\citep{rebu06} through an analysis of their spectral energy distribution (SED).
For some objects, accurate physical parameters and the projected rotational velocity,
$v\sin i$, were determined by means of high-resolution spectra.

The use of simultaneous multi-band light curves allows an accurate analysis of the 
spot properties because the different amplitude of the light curves constrains the spot temperatures and 
filling factors.

\section{Observations and data reduction} \label{sec:Obs}

The observations were performed with REM (Rapid Eye Mount), a 60-cm robotic telescope located at 
the ESO-La Silla Observatory (Chile), on 66 nights from November 2nd, 2006 to January 20th, 2007.
By means of a dichroic, REM feeds simultaneously two cameras at the two Nasmyth focal stations, 
one for the NIR (REMIR) and one for the optical (ROSS). 
The cameras have nearly the same field of view of about 10$\arcmin\times 10\arcmin$ and use wide-band filters 
($z'$, $J$, $H$, and $K'$ for REMIR and $V$, $R$, and $I$ for ROSS).
The main scientific aim of REM is the study of the early phases of after-glow of gamma-ray bursts 
detected by space-borne high-energy alert systems. 
Even if the schedule of REM is optimized to collect as many triggers as possible from active space-borne 
observatories, a considerable amount of time can be spent on additional science.

We selected a 10$\arcmin$ field flanking the ONC that included the eclipsing BD binary 
\object{2MASS\,J05352184-0546085} \citep{stass06}.
To this purpose, we observed mainly in the $I$-band with ROSS and in $JHK'$ with REMIR. 
In total, we collected 184, 175, 2769, 514, 461, and  250 usable images in $V$, $R$, $I$, $J$, $H$, and 
$K'$ bands, respectively. 
Exposure times were typically of 60--120 seconds in the optical bands and of 30--60 seconds in $JHK'$. 
Longer exposure times were not adopted in order to avoid image blurring by the guiding/de-rotating system.

\begin{figure}[ht]
   \centering{\hspace{1cm}\includegraphics[width=7.5cm,height=9cm]{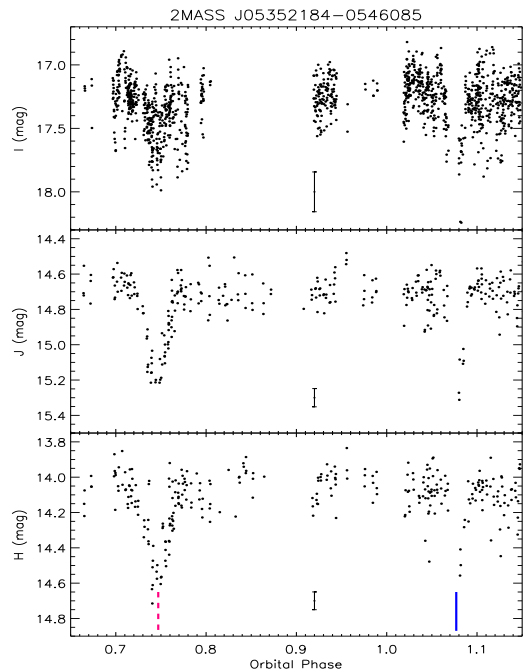} }                                                                                               
    \caption{REM light curves of the eclipsing binary 2MASS\,J05352184-0546085 composed of two BDs, 
    in the $I$, $J$, and $H$ bands.
    Photometric points are phased according to the ephemeris of \citet{stass06}.
    The average error bar for each band is also shown. The phases of the primary and secondary minimum
    are marked, in the lower box, by vertical continuous and dashed lines, respectively.}
   \label{Fig:BD}
    \end{figure}

Full phase coverage of the primary eclipse was not achieved during the observing season due to bad weather and 
technical problems. Also, the telescope size and relatively short exposure times were not sufficient to obtain 
high quality light curves of 2MASS\,J05352184-0546085 worthy of being analyzed (see Fig.~\ref{Fig:BD}).
During the final redaction of the present manuscript, we saw the very recent paper of \citet{Gomez09} where complete 
and more accurate $JHK$ light curves of 2MASS\,J05352184-0546085, are presented and analyzed.

However, the long-lasting and intensive simultaneous optical and NIR monitoring allowed us to detect and
study the photometric variability of several objects in this 10$\arcmin$ field of view.
In particular, we measured the magnitudes for 97 stars brighter than, or comparable to, 2MASS\,J05352184-0546085 
in $JHK'$. For 58 of these stars, we were also able to measure $RI$ (and for 47 of them also $V$) magnitudes. 
Some stars were too faint to be detected in the optical bands, or situated near the edges 
of the REMIR field of view and outside of the optical frames.  Therefore, due to slight random shifts in the 
field centering from one pointing to another, the closer to the border, the poorer was the data sampling for these stars.
Very close pairs (separation $\leq 7\arcsec$) are not considered in the following analysis.	

\begin{figure} 
     \includegraphics[width=9cm]{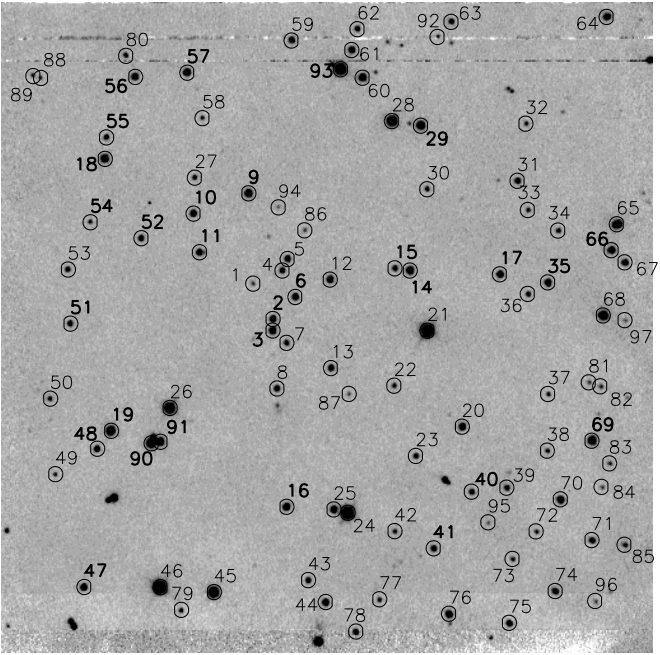}	
  \caption{Finding chart of the objects investigated in the present paper. The identification code (Id in 
	   Table~\ref{table:periods}) is written in boldface for the stars which display periodic variations.
           The image is taken with the REMIR camera in the $H$ band. The central 
	   coordinates at J2000 are RA=$05^{\rm h}35^{\rm m}16\fs7$, 
	   DEC=$-05\degr46\arcmin43\arcsec$. Orientation: North is up, East to the left. }
  \label{fig:field}
  \end{figure}

The field of view, as observed in the $H$ filter, together with the identification code given by us 
to the above mentioned stars, is displayed in Fig.~\ref{fig:field}. 

The pre-reduction of the REMIR images is automatically done by the AQuA pipeline \citep{Testa04} and the co-added and 
sky-subtracted frames, resulting from five individual ditherings, are made available to the observer. 

For the ROSS camera, the data reduction was more complex. Indeed, the field of view is vignetted and 
the non-uniform illumination of the CCD changes with the de-rotator position.
Since twilight flat-fields were not available, we built up master-flats at the four different de-rotator orientations
by using scientific images of our field and several standard-star fields separately for the $V$, $R$, and $I$ filters.
Each scientific image, after subtraction of the dark-frame, was divided by the proper master-flat, depending on the
filter and de-rotator orientation.
We verified that the scatter in the data is strongly reduced this way.

\subsection{Differential photometry}

Aperture photometry for all the selected stars was performed with DAOPHOT by using the IDL\footnote{IDL (Interactive Data 
Language) is a trademark of Research Systems Incorporated (RSI).} routine  \textsc{Aper}.
The photometric errors due to the photon statistics in the NIR bands are of about 0.04--0.06 mag for the faintest stars
($H\,\simeq 14$ mag), while they are as low as 0.001 mag for the brightest stars ($H\,\simeq 9$ mag) in the field. 
In the $I$ band, a faint star ($I \,\simeq 17$ mag) has a photometric error of about 0.10 mag, while for a bright one 
($I \,\simeq 12$ mag) the photon statistics gives rise to a 0.001-mag error.

We calculated the differential magnitude of each target with respect to an artificial comparison star 
built-up using all the non-variable stars present in the field.
We used the approach of \citet{broeg}, which consists in comparing each star with an artificial comparison 
composed by all the other stars in the field. An iterative process allowed us to identify the variable objects 
and separate them from those that are best suited to construct the artificial comparison, by weighting 
them down according to their variability. 
This method has also the advantage of providing reliable error bars.
The final errors derived with this algorithm turn out to be of about 0.10 mag for a star with 
$I \,\simeq 17$ mag, while for the brightest stars in $I$-band we obtain an error of $\simeq 0.02$ mag.
In the NIR bands, we have a typical error of $\sim 0.09$ mag for a star with $H=14$ mag, while an 
error of about 0.02 mag is found for the brightest stars. 
 The photometric accuracy achieved is mainly limited by the quality of the flat-field correction, 
 due to the complexity of the instrument.

\subsection{Standard photometry}

In order to transform the $V, R$ and $I$ instrumental magnitudes to the standard Johnson-Cousins system, 
some stars in the Landolt standard fields SA 98, SA 94, and SA 96 (Landolt 1992), observed by REM during 
the season of our observations, were used. 
The standard $V$, $R$, and $I$ magnitudes were determined using the transformation equations:

\begin{equation}V =v - \kappa_V \cdot AM + c_V \cdot (V-R)+ZP_V \end{equation} 

\begin{equation}R =r - \kappa_R \cdot AM + c_R \cdot (R-I)+ZP_R \end{equation} 

\begin{equation}I =i - \kappa_I \cdot AM + c_I \cdot (R-I)+ZP_I \end{equation} 

{\noindent where $v$, $r$, and $i$ are the instrumental magnitudes, $\kappa$ the coefficients of atmospheric extinction, $AM$ the airmass, 
and $c_V$ and $ZP_V$, $c_R$ and $ZP_R$, and $c_I$ and $ZP_I$ the color terms and zero-points for the $V$, $R$, and $I$ bands, respectively.} 

 We determined the atmospheric extinction coefficients using the non-variable stars in our Orion field observed, 
during some nights, across a suitable airmass range. 
Their values, listed in Table~\ref{tab:stand}, are however not very different from the mean extinction 
coefficients for La Silla.

The standard magnitudes from \citet{stet00} were used to determine the zero-points and color terms for the SA~98 area (Table~\ref{tab:stand}). 
Forty-five sufficiently bright stars with standard magnitudes in the \citet{stet00} catalogue fall within the REM $10\arcmin\times10\arcmin$ field 
centered around the star \object{SA 98-978}. 
Unfortunately, this standard field was observed six hours later than the Orion field on December 23, 2006.
The zero-points determined on January 17th and 18th, 2007 by means of the stars \object{SA~94-242} and \object{SA~96-36} are also listed in 
Table~\ref{tab:stand}.
We adopted these zero-points for the standardization of the magnitudes on January 17 and 18, 2007. 
Indeed, these standard stars were observed within 30 minutes from the Orion field. 
The standard magnitudes evaluated in these two dates allowed us to transform the differential 
light curves into standard ones.

\begin{table}
\caption{Atmospheric extinction and photometric calibration coefficients.} 
\label{tab:stand} 
\begin{tabular}{l l l l l l} 
\hline\hline 
\noalign{\medskip}
 Filter   & $\kappa$ &  $c^{\rm a}$  &   $ZP^{\rm a}$  &   $ZP^{\rm b}$  &   $ZP^{\rm c}$ \\
\noalign{\medskip}
\hline 
\noalign{\medskip}
$V$  &  0.187  & $-0.108$  & 20.899$\pm$0.024  & 20.874  &  20.954 \\ 
$R$  &  0.135  & $-0.110$  & 21.208$\pm$0.030  & 21.202  &  21.220 \\
$I$  &  0.072  & $-0.147$  & 20.629$\pm$0.020  & 20.577  &  20.608 \\
\noalign{\medskip}
\hline 
\end{tabular}
\begin{list}{}{}		         	                	                 
\item[$^{\mathrm{a}}$]  From SA~98 standard area observed on December 23, 2006. 
\item[$^{\mathrm{b}}$]  From SA~94-242 observed on January 17, 2007. 
\item[$^{\mathrm{c}}$]  From SA~96-36 observed on January 18, 2007. 
\end{list}
\end{table}

The conversion of the instrumental magnitudes $j$, $h$, and $k'$ to the standard $JHK'$ was performed 
by defining, for each frame, a zero-point by means of all the non-variable stars in the Orion field, 
whose magnitudes were taken from the 2MASS catalogue \citep{cutri03}. 
Since these stars were observed simultaneously with the variable ones, no correction for the 
airmass was needed. 

The standard $VRIJHK'$ average magnitudes derived by us are listed in Table~\ref{tab:VRIJHK} together with the IRAC 
Spitzer magnitudes in mid-IR bands (3.6, 4.5, 5.8, 8 $\mu$m) reported for a few of them by \citet{rebu06}.

\begin{figure}[tbh]
    \centering{\hspace{1cm} \includegraphics[width=7.5cm,height=9cm]{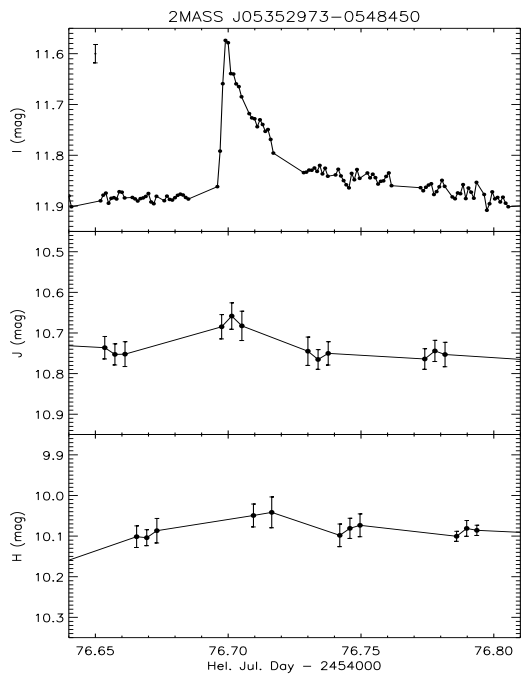} }
    \caption{The flare event observed on Star \#19 (2MASS~J05352973-0548450) on December 7th, 2006 at 4:41 UT in the $I$ 
    band (\textit{upper panel}).
    The simultaneous $J$ and $H$ light curves are also displayed in the two bottom panels.}
    \label{fig:flare}
    \end{figure}

A very interesting outcome of the intensive monitoring in the $I$ band was the detection of a strong flare 
of the star \#19 (\object{2MASS~J05352973-0548450} = \object{V498~Ori}) occurred on December 7th 2006, reaching the peak intensity 
($\Delta I \approx 0.3$ mag) at about 04:41 UT, with a total duration of nearly 2 hours (Fig.~\ref{fig:flare}).
Unfortunately, we lack simultaneous $V$ and $R$ data. 
Notwithstanding the smaller time resolution, a small flux intensification ($\Delta J \approx 0.1$ mag) seems to be 
present in the $J$-band as well, while no clear enhancement emerges in the $H$-band.
The data points corresponding to this flare event were excluded from the 
following time-series analysis. 
Another strong flare ($\Delta I \approx 0.8$ mag) was observed on star \#11 (\object{2MASS~J05352550-0545448} = \object{NR~Ori})
on December 10th 2006 at $\approx$\,01:30 UT and lasted for about 2 hours (Fig.~\ref{fig:flare_11})\footnote{Available 
in electronic form only.}. In this case, no 
significant enhancement emerges in the simultaneous $J$ light curve. 

\subsection{Complementary spectroscopic data} \label{sec:Spec}

Spectroscopic observations of some of the stars observed with REM 
were carried out using the Hectochelle multi-object spectrograph \citep{Szent98}
at the 6.5-m MMT telescope in Arizona. These data were used by \citet{Furesz08} and 
\citet{Tobin09} to study the kinematics of the Orion nebula clusters.

The resolution was $R\simeq 34\,000$ and the spectra observed in 2004 and 2005 \citep{Furesz08} 
are centered at H$\alpha$ and span 190\,\AA, while those acquired in 2006 and 2007 \citep{Tobin09}
cover the spectral range $\lambda$5150--5300\,\AA, which includes the \ion{Mg}{i}\,b triplet.
The reader is referred to the aforementioned papers for further details 
about observations and data reduction.

\section{Time-series analysis and rotation periods}

We derived the periods of the light variations by applying to the best light curves (typically $R$, $I$, $J$, and $H$ 
bands) a periodogram analysis \citep{sca} and the CLEAN deconvolution algorithm \citep{rob}, which allowed us to 
reject aliases generated by the spectral window of the data. 
We also evaluated the period uncertainty following the prescriptions of \citet{hoba}. 
The frequency uncertainty can be written as
\begin{equation}
 	\Delta\nu = \frac{3\sigma_{\rm N}}{4\sqrt{N_0}\,T\,A},
\label{Eq:Delta_nu}
\end{equation}	

\noindent{where $\sigma_{\rm N}^2$ is the data variance, $T$ the total time-span of the data, $A$ the amplitude of the signal,
and $N_0$ the number of independent points. We considered two data points independent (uncorrelated) if those were spaced in 
time by more than 0.05 in period units. Another estimate of the frequency uncertainty is provided by the full width at
half maximum (FWHM) of the periodogram peak. In this case, the values are typically 10--20 times larger than
those calculated through Eq.~\ref{Eq:Delta_nu} and reported in Table~\ref{table:periods}. However, even adopting uncertainties 
based on the FWHM of the peak, our results do not change.}

The period of variation was taken as the reciprocal of the frequency of the highest peak in the {\it cleaned} power spectrum.
Usually, the most accurate periods were found for the $I$ light curves, mainly because of the very large number of data-points and 
relatively large amplitudes.
For some of the reddest objects we found instead better period determinations (smaller errors) in the $J$ or $H$ filters.
The variation periods are listed in Table~\ref{table:periods} together with the bands in which they have been detected 
(ordered according to increasing period errors). The light-curve amplitudes in the $VRIJHK'$ bands and previous period
determinations from the literature \citep[][]{stass99,Carpenter01,rebu01,rebu06} are also reported in this table. 

\begin{table*}
\caption{Rotation periods ($P$), false-alarm probabilities (FAP), and light-curve amplitudes ($\Delta V$, $\Delta R$, $\Delta I$, $\Delta J$, 
$\Delta H$, and $\Delta K'$) of our targets.  A colon marks periods with FAP$>$\,5\%.} 
\label{table:periods} 
\centering 
\begin{tabular}{l l l l l l l l l l l l l l} 
\hline\hline 
  Id   &   $P$    &   Err    & FAP &  Band & $\Delta V$ & $\Delta R$ & $\Delta I$ & $\Delta J$ & $\Delta H$ & $\Delta K'$ & $P_{\rm Stas}^{\mathrm{a}}$ & $P_{\rm Reb}^{\mathrm{b}}$ & $P_{\rm Carp}^{\mathrm{c}}$ \\ 
       &  (days)  &  (days)  &	   &       &  (mag)     &  (mag)     &  (mag)	 &  (mag)     &    (mag)   &	(mag)	&  (days)     &   (days)  &   (days)     \\ 
\hline	     				    									 

   1   & 9.78$^{\mathrm{c}}$ &   ...    &        ...	 &   ...    &	...    &   ...    &    ...   &    ...	 &    ...    &   ...	&   9.78$^{\mathrm{d}}$  &   ...  &   ...   \\ 
   2   &   7.751  &   0.108  & 3$\cdot 10^{-5}$  &  $RIJH$  &	...    &  0.085   &  0.065   &   0.040   &   0.035   &  0.025	&   ...   &   ...  &   ...   \\ 
   3   &   1.311  &   0.003  & 1$\cdot 10^{-10}$ &  $IHJ $  &	0.050  &   ...    &  0.040   &   0.030   &   0.030   &  0.020	&   ...   &   ...  &   ...   \\ 
   5   &   ...    &   ...    &        ...	 &   ...    &	...    &   ...    &    ...   &    ...	 &    ...    &   ...	&   ...   &   0.60 &   ...  \\ 
   6   &   5.993  &   0.070  & 6$\cdot 10^{-3}$  &  $IRJH$  &	0.090  &  0.065   &  0.040   &   0.020   &   0.020   &  0.020	&   ...   &   ...  &   ...   \\ 
   9   &   0.5627 &   0.0002 & 1$\cdot 10^{-11}$ &  $IJHR$  &	0.090  &  0.100   &  0.065   &   0.040   &   0.040   &  0.045	&   0.57  &   ...  &   ...   \\ 
  10   &   5.816  &   0.010  & 1$\cdot 10^{-6}$  &  $RIHJ$  &	0.365  &  0.345   &  0.280   &   0.140   &   0.150   &  0.125	&   ...   &   5.90 &   5.88  \\  
  11   &   7.383  &   0.003  & 1$\cdot 10^{-6}$  &  $HJIR$  &	0.850  &  0.970   &  1.150   &   1.250   &   1.180   &  0.900	&   ...   &   7.50 &   7.46  \\  
  14   &   4.047  &   0.003  & 5$\cdot 10^{-10}$ &  $IHJR$  &	0.390  &  0.345   &  0.265   &   0.135   &   0.145   &  0.125	&   4.02  &   4.04 &   3.98  \\ 
  15   &  11.86:  &   0.08   & 9$\cdot 10^{-2}$  &  $IJH $  &	...    &   ...    &  0.160   &   0.030   &   0.035   &  0.030	&   ...   &   ...  &   ...   \\ 
  16   &   6.351  &   0.072  & 2$\cdot 10^{-2}$  &  $IRJH$  &	0.145  &  0.130   &  0.055   &   0.035   &   0.035   &  0.030	&   ...   &   ...  &   ...   \\ 
  17   &   8.435  &   0.020  & 1$\cdot 10^{-4}$  &  $IHJR$  &	0.395  &  0.290   &  0.245   &   0.125   &   0.130   &  0.115	&   7.70  &   4.18 &   ...  \\ 
  18   &  12.461  &   0.069  & 1$\cdot 10^{-7}$  &  $RIHJ$  &	0.285  &  0.225   &  0.110   &   0.055   &   0.055   &  0.035	&   ...   &   ...  &   ...   \\ 
  19   &   0.605  &   0.035  & 4$\cdot 10^{-4}$  &  $JI  $  &	...    &   ...    &  0.060   &   0.030   &    ...    &   ...	&   ...   &   ...  &   ...   \\ 
  29   &   8.468  &   0.046  & 1$\cdot 10^{-2}$  &  $IJR$   &	0.175  &  0.150   &  0.145   &   0.040   &   0.050   &   ...	&   ...   &   ...  &   ...   \\ 
  35   &  12.521  &   0.218  & 3$\cdot 10^{-4}$  &  $HJ  $  &	...    &   ...    &    ...   &   0.070   &   0.070   &   ...	&   ...   &  12.41 &   ...  \\ 
  40   &   5.297  &   0.007  & 5$\cdot 10^{-7}$  &  $HJ  $  &	...    &   ...    &    ...   &   0.275   &   0.270   &  0.250	&   ...   &   ...  &   ...   \\  
  41   &   6.783  &   0.040  & 2$\cdot 10^{-2}$  &  $JHIR$  &	0.260: &  0.205   &  0.125   &   0.075   &   0.090   &  0.110	&   ...   &   ...  &   ...   \\ 
  47   &  10.02   &   0.21   & 5$\cdot 10^{-2}$  &  $HJ  $  &	...    &   ...    &    ...   &   0.055   &   0.055   &   ...	&   9.60  &  10.08 &   ...   \\ 
  48   &   7.721  &   0.019  & 2$\cdot 10^{-4}$  &  $IHJ $  &	...    &   ...    &  0.280   &   0.270   &   0.230   &   ...	&   ...   &   ...  &   ...   \\ 
  51   &   2.763  &   0.010  & 2$\cdot 10^{-2}$  &  $IJH $  &	...    &   ...    &  0.040   &   0.025   &   0.025   &  0.025	&   ...   &   2.81 &   ...   \\ 
  52   &   1.696  &   0.004  & 1$\cdot 10^{-2}$  &  $IJHR$  &	0.170  &  0.090   &  0.070   &   0.030   &   0.035   &  0.050	&   1.69  &   1.70 &   ...   \\ 
  54   &   2.514  &   0.002  & 5$\cdot 10^{-12}$ &  $JHI $  &	...    &   ...    &  0.315   &   0.265   &   0.235   &  0.190	&   ...   &   ...  &   ...   \\ 
  55   &   8.432  &   0.035  & 5$\cdot 10^{-6}$  &  $IRJH$  &	0.300: &  0.230   &  0.135   &   0.070   &   0.120   &   ...	&   8.06  &   ...  &   ...   \\ 
  56   &   3.80   &   0.06   & 3$\cdot 10^{-2}$  &  $R$     &	...    &  0.065   &    ...   &    ...	 &    ...    &   ...	&   3.86  &   3.84 &   ...   \\ 
  57   &   4.647  &   0.038  & 1$\cdot 10^{-2 }$ &  $JIR $  &	0.185  &  0.120   &  0.050   &   0.040   &    ...    &   ...	&   ...   &   ...  &   ...   \\ 
  59   &   ...    &   ...    &        ...	 &   ...    &	...    &   0.5    &    0.4   &    0.3	 &    ...    &   ...	&   7.33  &   7.33 &   ...   \\  
  60   &   ...    &   ...    &        ...	 &   ...    &	...    &   ...    &    ...   &    ...	 &    ...    &   ...	&   4.04  &   4.11 &   4.09  \\ 
  61   &   ...    &   ...    &        ...	 &   ...    &	...    &   1.5    &    0.8   &    0.6	 &    ...    &   ...	&   ...   &  18.68 &   ...   \\ 
  62   &   ...    &   ...    &        ...	 &   ...    &	...    &   ...    &    ...   &    ...	 &    ...    &   ...	&   2.05  &   2.05 &   ...   \\ 
  63   &   ...    &   ...    &        ...	 &   ...    &	...    &   ...    &    ...   &    ...	 &    ...    &   ...	&   1.10  &   1.09 &   ...   \\ 
  66   &  21.68:  &   0.30   & 1$\cdot 10^{-1}$  &  $HJ  $  &	...    &   ...    &    ...   &   0.40	 &   0.40    &   ...	&   ...   &  59.87 &  14.62  \\  
  69   &   6.550  &   0.068  & 5$\cdot 10^{-3}$  &  $JH  $  &	...    &   ...    &    ...   &   0.060   &   0.050   &   ...	&   ...   &   ...  &   ...   \\  
  75   &   ...    &   ...    &        ...        &   ...    &	...    &   ...    &    ...   &    ...    &    ...    &   ...	&   ...   &  10.71 &   ...   \\  
  90   &   4.376  &   0.013  & 1$\cdot 10^{-5}$  &  $IJRH$  &	0.205  &  0.135   &  0.120   &   0.055   &   0.040   &   ...	&   ...   &   ...  &   ...   \\ 
  91   &  10.176  &   0.055  & 2$\cdot 10^{-5}$  &  $JHIR$  &	...    &  0.070   &  0.080   &   0.090   &   0.095   &  0.070	&   ...   &   ...  &   ...   \\ 
  93   &   5.70:  &   0.12   & 2$\cdot 10^{-1}$  &  $IJ  $  &	...    &   ...    &  0.150   &   0.200   &    ...    &   ...	&   ...   &   5.53 &   5.53  \\  
				          	                	                    	         
\hline				         	                	                 
\end{tabular}			         	                	                 
\begin{list}{}{}		         	                	                 
\item[$^{\mathrm{a}}$]  Rotational periods derived by \citet{stass99}. 
\item[$^{\mathrm{b}}$]  Rotational periods derived by \citet{rebu01} or \citet{rebu06}. 
\item[$^{\mathrm{c}}$]  Rotational periods derived by \citet{Carpenter01}. 
\item[$^{\mathrm{d}}$]  Orbital period from \citet{stass06}, and confirmed by us.
\end{list}
\end{table*}

In total, we determined the rotation periods for 29 stars.  Thirteen of these had periods previously detected, 
while for all the others the period is derived here for the first time.
 In fact, observations in different observing seasons may be necessary to detect as many rotation periods 
as possible, as shown, e.g., by \citet{Padma09} for some ONC stars. Indeed, depending on the spottedness level and 
the spot distribution over the photosphere, the amplitude of the light variation may sometimes be too small for 
detecting a periodic modulation.

On the other hand, we could not detect any reasonable periodicity in our data of 7 stars (also  included  in  
Table~\ref{table:periods}) for which the  aforementioned  authors  did. However, we note that, except for star\,\#5, 
all of them lie near the edges or out of the field, and/or close to bad columns of the ROSS
and REMIR detectors (cf. Fig.~\ref{fig:field}) and, consequently, we have for them a lower number of useful data points. Two of 
them (\#59 and \#61) display large non-periodic variations, as discussed later in this section, while, for the remaining five,  
no significant variation has been detected.

False-alarm probabilities (FAP) for the peaks in the periodograms were also computed according to the definitions 
given by \citet{sca} and by \citet{hoba}. As stressed by \citet{hoba}, the FAP depends on the data sampling and 
can be lowered by a large number of data points taken so close in time that they do not really represent
independent measurements and tend to inflate the power artificially. 
Therefore, we adopted the number of independent points $N_0$, evaluated as described above, for the FAP computation.

With the exception of only three stars (i.e., \#15, \#66 and \#93), the periods are detected 
with a FAP$\leq$\,5\%, i.e. with a high confidence level (1-FAP$\geq$\,95\%).
The periods of the three stars with a high FAP are considered only as possible detections and are
marked with a colon in Table~\ref{table:periods}. 
In order to check the reliability of these FAPs, we run 1000 Monte Carlo simulations of photometric sequences with 
only noise and with the same time sampling of $I$ data and measured the power of the highest peak in the periodogram of each 
sequence. Comparing periodogram peak heights of the real data with the cumulative distribution of the peak heights in
the simulated data, we found that all stars with very low FAPs in Table~\ref{table:periods} have indeed a FAP$<$0.001.

\begin{figure*}[tbh]
\hspace{-0.5cm}
     \includegraphics[width=9.4cm]{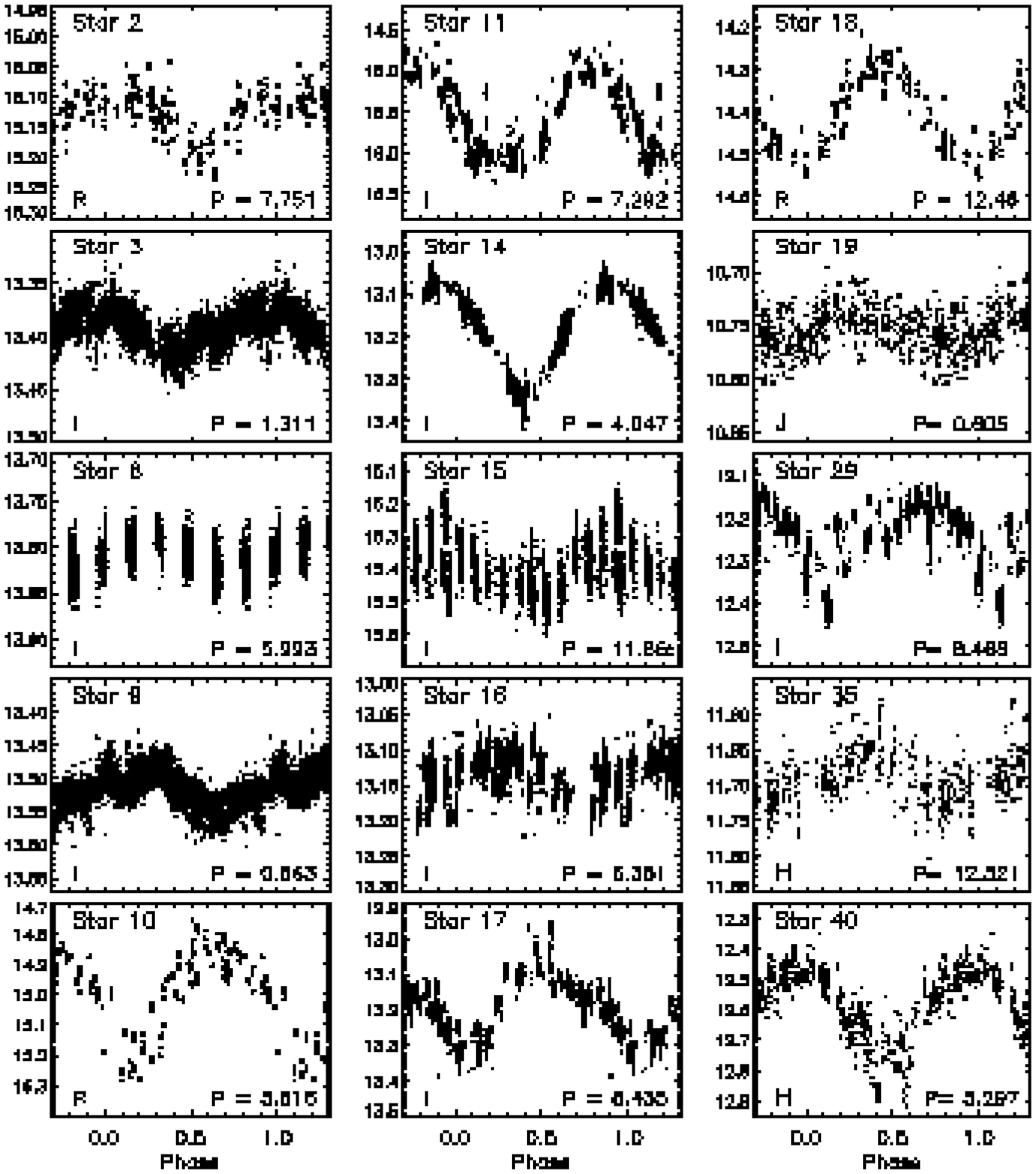} 
\hspace{-0.2cm}
     \includegraphics[width=9.4cm]{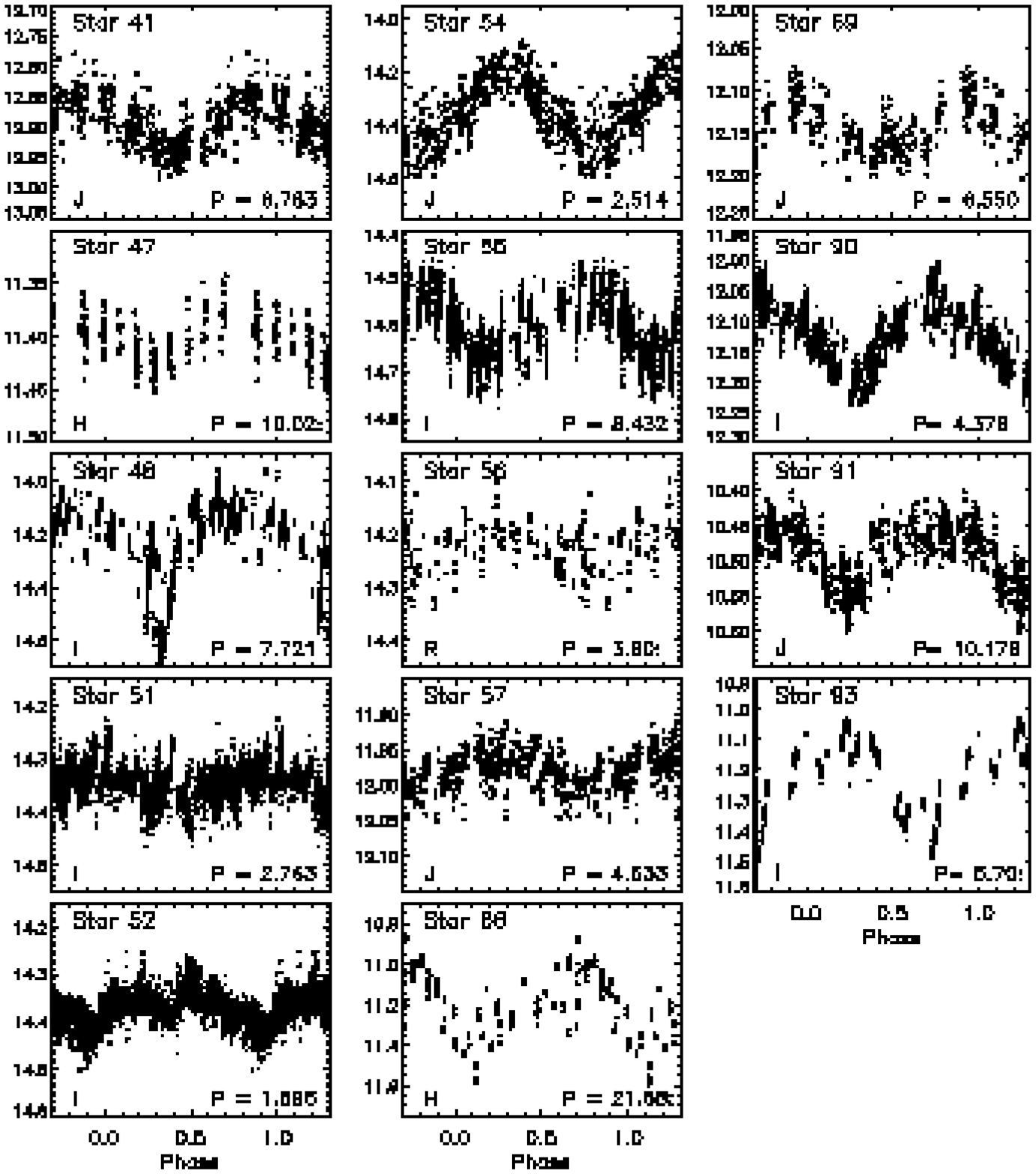}
   \caption{Phase-folded light curves of the objects with detected period in the 2MASS\,J05352184-0546085 field. 
   Source Id and observed band are also indicated in each box. }
    \label{fig:curves}
    \end{figure*}

The LC collection is displayed in Fig.~\ref{fig:curves}, where the source identifiers and the photometric 
periods are indicated\footnote{Light curves are available at the CDS.}.  

A very peculiar light curve, which shows two diagonal strips resembling eclipse egresses around phase 0.2
and 0.4, is displayed by star \#48 (\object{2MASS\,J05353047-0549037}). At the beginning we thought that this star could
be an eclipsing binary.
However, we could not find any period suitable to fold these steep variations into a phased eclipsing binary light curve. 
We have no clear explanation for such a behaviour and we can only speculate that these features might be produced by 
distinct objects/dust-clumps (protoplanetary condensations?) transiting  over the apparent stellar disc. More repeated 
observations are needed to settle this point.

\begin{figure*}[ht]
  \begin{center}
    \includegraphics[width=5.5cm]{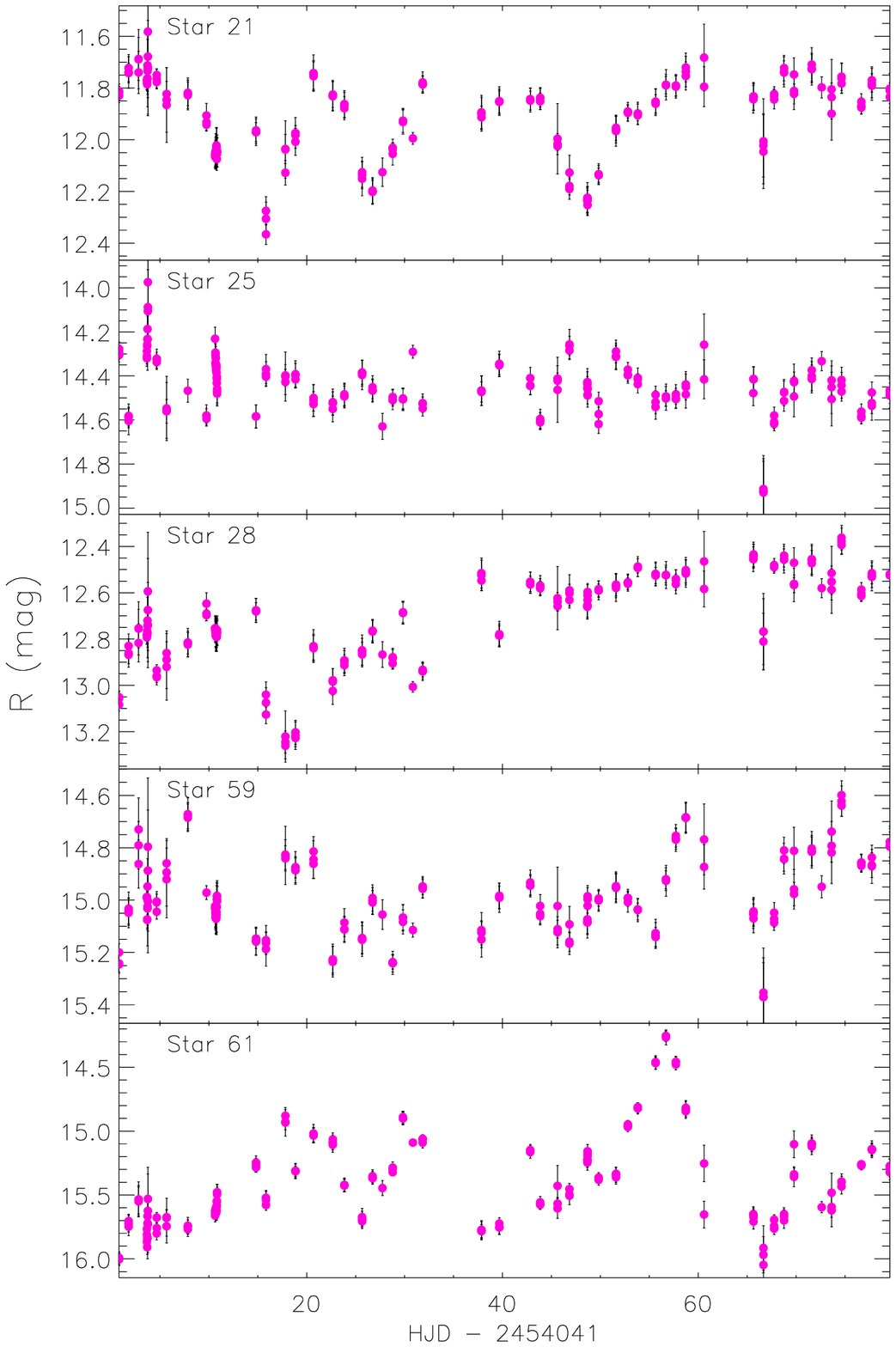}
    \includegraphics[width=5.5cm]{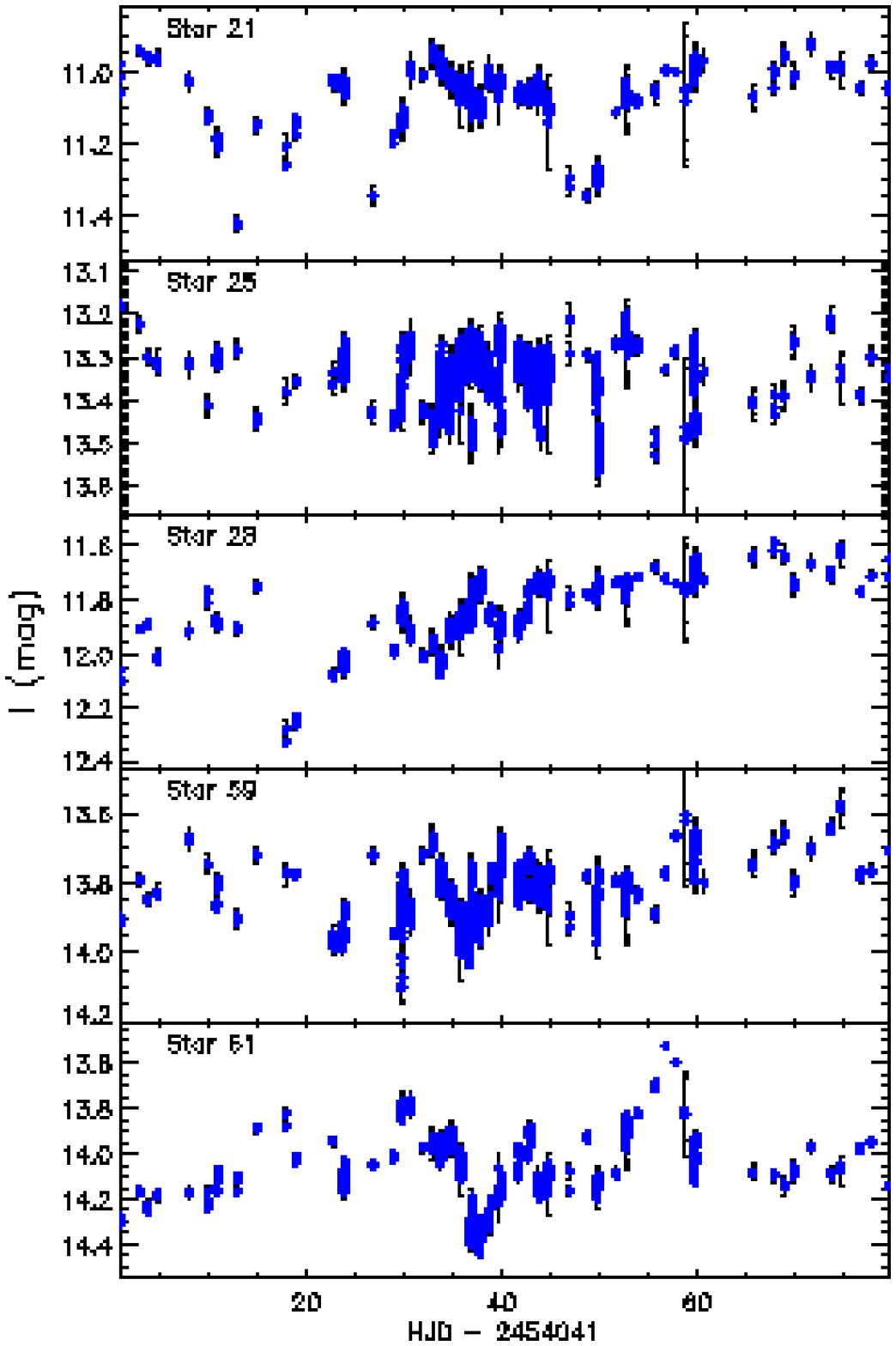}
    \includegraphics[width=5.5cm]{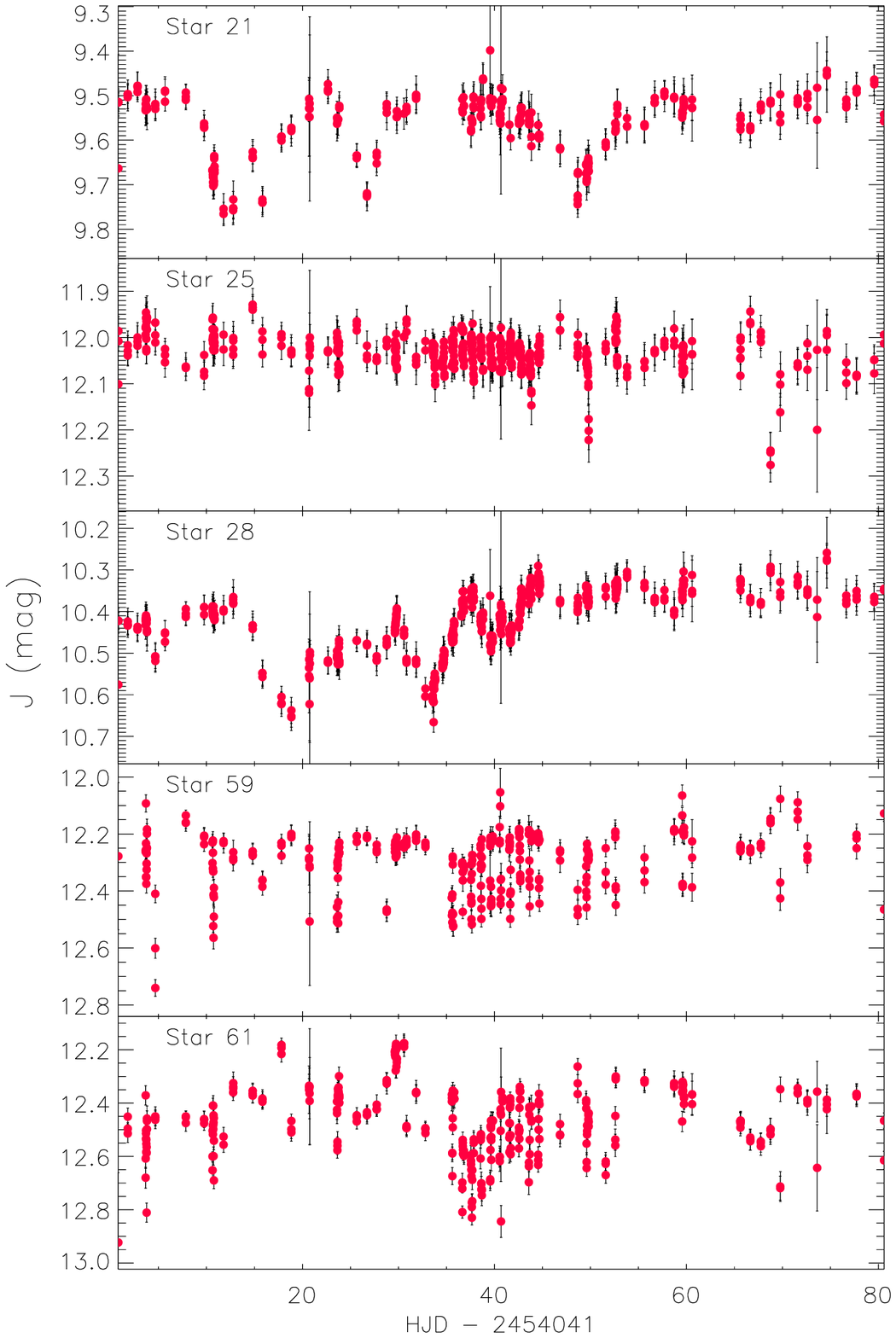}
  \end{center}
\caption{Light curves in the $R$, $I$, and $J$ band of the five variable stars for which we could not find
any reliable period.
}
\label{fig:erratic}
\end{figure*}

In order to pick up all the variable stars in our field, we applied a $\chi^2$ test
on the $I$-band photometric sequences. We found only five bona-fide variables (with a reduced chi-square 
$\chi_{\nu}^2>3$) showing non-periodic variations as large as 0.6 magnitudes (\#21, \#25, \#28, \#59, and \#61). 
Their $R$, $I$, and $J$ light curves are displayed as a function of the heliocentric Julian day in 
Fig.~\ref{fig:erratic}\footnote{Available in electronic form only.}. 
We note that, with the exception of star\,\#25, all the others were already classified as variable stars.

\section{Spectral Energy Distributions and stellar parameters}	
\label{sec:param}

\begin{figure*}
 \centering{\hspace{0.1cm}\includegraphics[width=17.5cm]{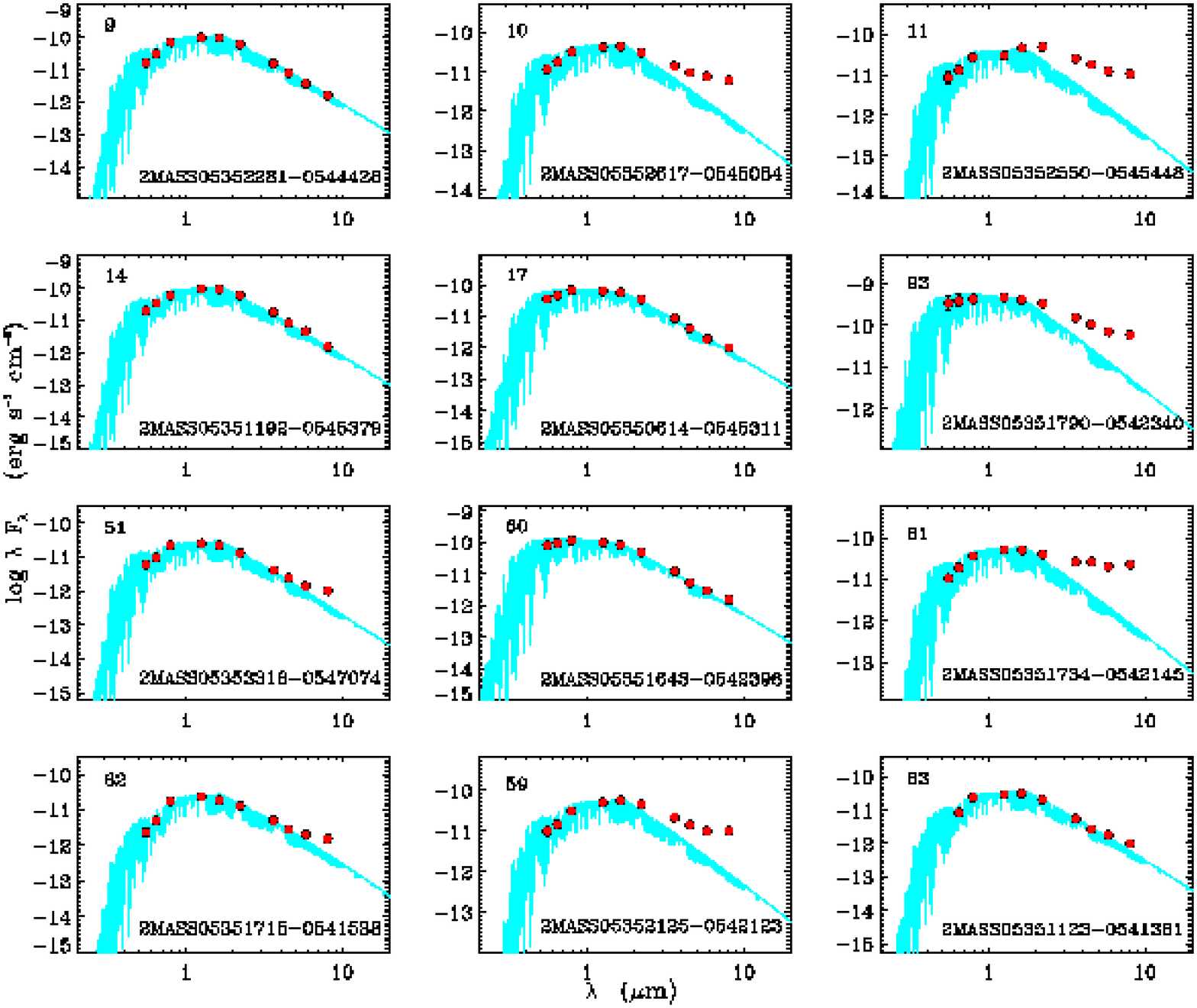}} 	
\caption{Spectral energy distributions for the 12 stars with available IRAC mid-IR fluxes (dots). 
The best-fit NextGen spectrum is over-plotted with continuous lines in each box. The entry 
number in Table~\ref{table:periods} and the 2MASS identifier are also indicated in each box.} 
  \label{Fig:SED}
    \end{figure*}

The standard $VRIJHK'$ photometry (Table~\ref{tab:VRIJHK}) allowed us to derive
the spectral energy distribution (SED), from the optical to NIR, for a 
sub-sample of 66 sources. For 58 of these we used our own ROSS and REMIR 
photometry, while for 8 objects we appended $V$ and/or $I$ magnitudes retrieved 
from the literature \citep{rebu00,rebu01,DENIS05,lask08} to our own $JHK'$ data. 
For 14 objects we made also use of the Spitzer IRAC magnitudes from \citet{rebu06}, 
extending the SED to the mid-IR.
We then adopted the grid of NextGen low-resolution synthetic spectra, with 
$\log g = 3.5$ and solar metallicity by \citet{hau99}, to perform a three-parameter 
fit to the SEDs. The stellar radius ($R_*$), effective temperature ($T_{\rm eff}$), 
and interstellar extinction ($A_V$) are free parameters of the fit. The \citet{carde89} 
extinction law with $R_V=3.1$ was used. To convert the apparent magnitudes 
into absolute ones, we assumed the ONC distance $d= 414\pm7$\,pc \citep{Menten07}.
The best solution was found by minimizing the $\chi^2$ of the fit, which was 
performed only when $VRIJHK'$ data were available. Figure~\ref{Fig:SED} displays 
the results of the fitting procedure for 12 of the stars with IRAC mid-IR flux measurements
available. The stellar luminosity was then obtained by integrating the best-fit model 
spectrum.

\begin{figure}
\centering{\includegraphics[width=9cm]{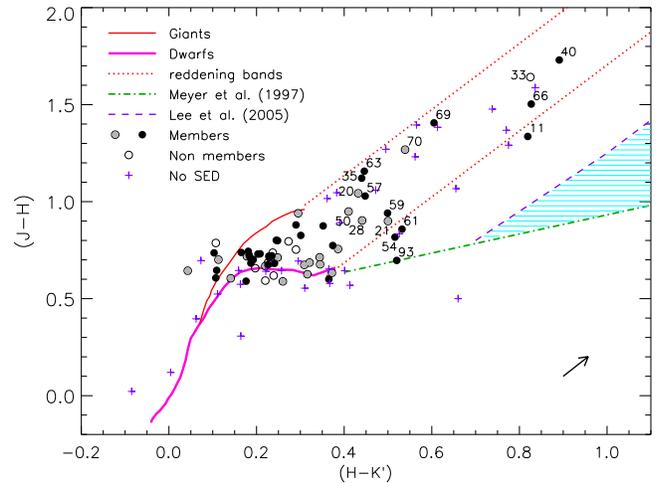}}	
 \caption{NIR color-color diagram for the sources in our sample. The continuous
 lines represent the colors for dwarfs and giants as labelled. The reddening
 direction is indicated by the dotted lines. The $A_{\rm V}=1$ mag reddening vector is also shown.
 The locus of strong NIR excess (hatched area) is defined by the criteria of \citet{meyer97} and \citet{Lee05} as labelled.
 Members and non-members are represented by filled and open circles, respectively.
 Black-filled circles indicate objects with a period determination. Crosses represent sources 
 with only $JHK'$ data.} 
  \label{IR-cc-diag}
  \end{figure}

\begin{figure}
 \centering{\includegraphics[width=8.8cm]{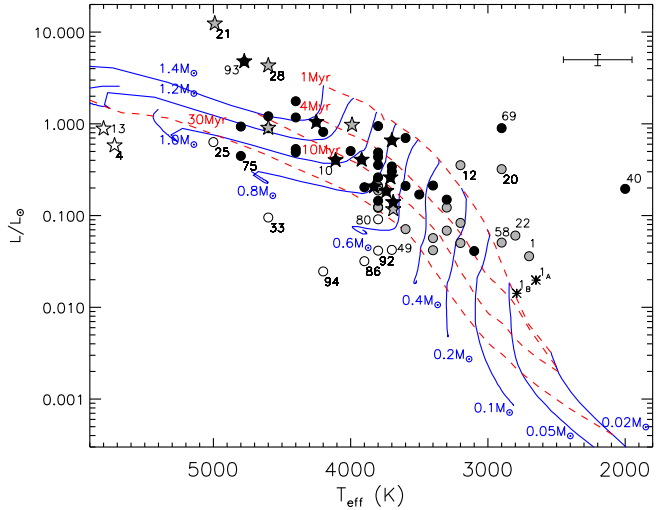}}	
\caption{HR diagram of the objects with standard photometry listed in Table~\ref{tab:VRIJHK}.
Star symbols have been used for the objects with $T_{\rm eff}$ derived from spectra,
while the other objects are represented with circles.
Black symbols refer to stars with period determination. Objects falling in a position 
not consistent with the Orion cloud (open symbols) are also labelled with their Id number. 
The two asterisks mark the positions of the individual components of the BD eclipsing 
binary according to \citet{stass06}.
The evolutionary PMS tracks and isochrones by \citet{baraf98} and \citet{chab00} are also
displayed in the figure by continuous and dotted lines, respectively.} 
        \label{Fig:HR}
    \end{figure}

For the 15 stars with Hectochelle spectra and for star \#93 \citep{wolf}, we performed the SED fit by 
fixing the effective temperature to the value derived from the corresponding 
high-resolution spectrum (see Sect.~\ref{sec:vsini}). These values are reported 
in Table~\ref{tab:spec}, while in Table~\ref{tab:param} the Id and $T_{\rm eff}$ 
values for these stars are listed in italic characters. By comparing these 
values with those coming from the SED fits, i.e. assuming $T_{\rm eff}$ 
as a free parameter, they result to be in agreement within 200\,K, which is the
typical uncertainty on $T_{\rm eff}$ when derived from SED fits \citep{gan08}.
However, the fits may underestimate $T_{\rm eff}$ 
by several hundred degrees in the case of high extinction and/or strong 
NIR excess. To investigate this, we use the NIR color-color diagram
shown in Fig.~\ref{IR-cc-diag}, where the stars with the parameters derived from the SED analysis 
are represented by circles, while those ones with only $JHK'$ photometry are indicated by crosses. 
 Although none of the sources fall in the region typical of very strong NIR excess, 
most objects appear moderately reddened ($A_V <$\,1 mag); however, 17 objects among those 
 with SED available (labelled with their Id in Fig.~\ref{IR-cc-diag}) have probably a higher extinction. In particular, 
two of the brightest variable stars (\#21 and \#28) as well as star \#70 seem to have a higher reddening. 
In fact, the SED analysis, without the constraint on $T_{\rm eff}$ provided by the spectra, 
led to underestimate the temperature of these objects by about 600\,K. For the other stars lying in the locus 
of high extinction
we do not have Hectochelle spectra, thus the parameters derived from the SED analysis
should be regarded as crude estimates, particularly for objects with incomplete SEDs
(e.g, \#33, \#40, \#66, and \#69). 
Remaining undetected in the ROSS bands, \#40 is definitely the reddest 
object in the diagram. Unfortunately, only its $I$ magnitude is found in 
the GSC 2.3 catalogue \citep{lask08}, preventing an accurate SED fit. A tentative fit 
yields a temperature of the order of 2000\,K, which is highly unreliable.

The results of the SED fits are in general agreement with our findings from the color-color diagram. 
In particular, ten of the seventeen objects falling in the region of higher extinction of the 
$(J-H)$--$(H-K')$ diagram (Fig.~\ref{IR-cc-diag}) turned out to have $A_V>1$~mag, while other four  
(\#11, \#59, \#61, and \#93) mainly owe their position in the color-color diagram to a significant 
infrared excess (see Fig.~\ref{Fig:SED}).
Notably, all of these stars are variable, either periodic (\#11 and \#93) or erratic (\#59 and \#61).

With the exception of seventeen objects, the $A_V$ values derived from the SEDs 
(Table~\ref{tab:param}) are less than 1 mag, as expected for 
objects at the ONC distance that are not embedded in the nebula. 
Extinction values of $\approx 1$\,mag and ranging from about 0.5 to 2\,mag 
were found by \citet{greve} from measurements of emission line ratios in the 
Orion Nebula. Thus, the 17 stars with higher values of $A_V$ are presumably mainly
objects embedded into the nebula or background stars, as \#4 and \#13.
The latter are both classified spectroscopically as non-members. 
The high extinction values found by us ($A_V\simeq 2.6$ mag)
are in agreement with background objects observed behind the Orion cloud. The SED analysis without a temperature 
constraint tends to underestimate $A_V$ yielding a best-fit with a cooler ($\sim$ 1000\,K) template.

All in all, we estimate that the typical errors on effective temperature and 
luminosity are $\approx$250\,K and 15\%, respectively. The accuracy of the radii 
determinations mainly depends on the uncertainties on the distance and $T_{\rm eff}$
that give rise to errors on the radii of the order of 10\%. 
The $A_V$ error is always smaller than 0.5\,mag except for the four values 
marked with a colon in Table~\ref{tab:param} for which it is as high as 
$\sim$\,1\,mag.

We placed all the objects on the Hertzsprung-Russel (HR) diagram (Fig.~\ref{Fig:HR}) 
and estimated their masses and ages by comparison with the set of theoretical PMS 
tracks and isochrones by \citet{baraf98} and \citet{chab00}. 
An age spread is apparently observed which is significantly larger than that ascribable 
to the errors on luminosity and $T_{\rm eff}$. We find that our sample contains both 
very young objects (age $\sim 0.5$--5\,Myr), whose age is quite consistent with the ONC,
and older stars with ages up to 30\,Myr. 

We assigned to each star a ``photometric membership" to the Orion cloud according to the position 
on the HR diagram. We considered as \textit{likely members} of the Orion cloud 
(denoted with a ``Y'' in Table~\ref{tab:param}) all the objects with an age 
$\leq$\,10 Myr, \textit{possible members} (``P'' in Table~\ref{tab:param}) those 
with 10 Myr $<$\,age $\leq$\,30 Myr or lying above the 1\,Myr isochrone 
(possible foreground stars), and \textit{non-members} (``N'' in Table~\ref{tab:param}) 
the ones with age $>$\,30 Myr or falling below the zero-age main-sequence 
(likely background stars). The latter are plotted with open circles in Fig.~\ref{Fig:HR}, 
while the members and possible members are represented by grey or black-filled circles. 
The black-filled circles indicate members and possible members with known period, either 
from this work or from the literature. The non-members and some particular objects, 
like those falling above the 1\,Myr isochrone, are also labelled with their Id 
in Fig.~\ref{Fig:HR}. We note that all the stars with determined period appear 
consistent with the distance of the Orion SFR, implying that variability is a 
powerful tool to pick up young stars in associations, as stressed, e.g., 
by \citet{briceno05}. 

In Table~\ref{tab:param}, we also report the \textit{spectroscopic} membership for 
the objects studied by \citet{Furesz08}, that rests upon the radial velocity and/or 
H$\alpha$ emission, and the \textit{astrometric} membership based on proper 
motions \citep{Dias06}.

The two BD components of the eclipsing binary 2MASS\,J05352184-0546085 (star \#1) are 
also displayed in the HR diagram with asterisks. We adopted the effective temperatures 
and radii reported by \citet{stass06}. Their position is fairly well consistent with 
that derived by us from the SED analysis, accounting for binarity. We note that another 
two objects, namely star \#22 and \#58, could be young BDs according to their location 
in the HR diagram. For these objects no periodicity was detected, while star \#40, 
the coolest one, exhibits a nice modulation with a period of 5.3 days in both the $J$ and $H$ 
bands.

\begin{table*}
\vspace{-0.3cm}
\caption{Physical parameters derived from the SED analysis and from the HR diagram. 
Effective temperatures constrained from spectroscopy are indicated in italics.
} 
\label{tab:param}
\vspace{-0.2cm}
\begin{tabular}{ l l l l l l l l r c c c} 
\hline\hline 
  Id  &   2MASS		    & Other  & $T_{\rm eff}$ &  $L$           & $A_V$	  & $R$      & $M$	      &  Age  & \multicolumn{3}{c}{Membership} \\
      &                     & Id &  (K)          &  ($L_{\sun}$)  & (mag)	  & ($R_{\sun}$) & ($M_{\sun}$)   &  (Myr)  & Ph$^{\mathrm{a}}$ 
      & Sp$^{\mathrm{b}}$ & As$^{\mathrm{c}}$ \\ 
\hline 
\noalign{\smallskip}
  ~1  &  {\scriptsize 05352184-0546085}   &                   &  2700	   &   0.036 &	1.10  &   0.87 &     0.04 & $\leq 1$ &  Y &  ... &  ...	\\ 
{\it ~2}&  {\scriptsize 05352029-0546399} &                   & {\it 3710} &   0.260 &	0.20  &   1.24 &     0.70 &	7.1  &  Y &  Y   &  ...	\\ 
   ~3  &  {\scriptsize 05352021-0546510}  &                   &  3800	   &   0.436 &	0.13  &   1.52 &     0.85 &	5.1  &  Y &  ... &  Y	\\ 
{\it ~4}&  {\scriptsize 05352007-0545526} &\object{Parenago 1980}& {\it 5720}& 0.581 &	2.55  &   ...  &     ...  &	...  &  N &  N   &  ...	\\ 
   ~5  &  {\scriptsize 05351981-0545409}  & \object{V1732 Ori}&  3400	   &   0.213 &	0.07  &   1.33 &     0.40 &	2.8  &  Y &  ... &  Y	\\ 
   ~6  &  {\scriptsize 05351903-0546163}  &        	      &  3700	   &   0.303 &  0.12  &   1.34 &     0.62 &	4.0  &  Y &  ... &  ... \\ 
   ~7  &  {\scriptsize 05351924-0547012}  &        	      &  3300	   &   0.122 &  0.88  &   1.07 &     0.25 &	2.5  &  Y &  Y   &  ... \\ 
   ~8  &  {\scriptsize 05351950-0547457}  &        	      &  3800	   &   0.188 &  0.19  &   1.00 &     0.80 &    20.1  &  P &  ... &  ... \\ 
{\it ~9}&  {\scriptsize 05352281-0544428} & \object{V1529 Ori}& {\it 3480} &   0.658 &	0.89  &   1.98 &     0.70 &	1.6  &  Y &  Y   &  ...	\\ 
{\it 10}&  {\scriptsize 05352617-0545084} &                   & {\it 4110} &   0.403 &	1.06  &   1.25 &     1.00 &    10.0  &  Y &  Y   &  ...	\\	   
  11  &  {\scriptsize 05352550-0545448}   & \object{NR Ori}   &  3800	   &   0.261 &	0.56  &   1.18 &     0.80 &    10.0  &  Y &  ... &  ...	\\ 
  12  &  {\scriptsize 05351695-0545558}   &         	      &  3200	   &   0.354 &  0.70  &   1.94 &     0.30 &	1.0  &  Y &  Y   &  ... \\ 
{\it 13}&  {\scriptsize 05351621-0547201} &         	      & {\it 5800} &   0.887 &  2.72  &   ...  &     ...  &    ...   &  N &  N   &  ... \\ 
{\it 14}&  {\scriptsize 05351192-0545379} & \object{V1490 Ori}& {\it 4250} &   1.046 &	1.21  &   1.89 &     1.30 &	4.0  &  Y &  Y   &  Y	\\ 
  15  &  {\scriptsize 05351290-0545376}   &                   &  3500	   &   0.170 &  1.42  &   1.12 &     0.40 &	4.0  &  Y &  Y   &  ... \\ 
  16  &  {\scriptsize 05351797-0549375}   & \object{V793 Ori} &  4000	   &   0.506 &  0.17  &   1.48 &     1.00 &	6.4  &  Y &  ... &  ... \\ 
  17  &  {\scriptsize 05350614-0545311}   & \object{V482 Ori} &  4400	   &   0.531 &  0.28  &   1.26 &     1.10 &    16.0  &  P &  ... &  Y	\\ 
  18  &  {\scriptsize 05353222-0544265}   &\object{Parenago 2103} &  3600  &   0.700 &  0.11  &   2.16 &     0.62 &	1.0  &  Y &  ... &  Y	\\ 
  19  &  {\scriptsize 05352973-0548450}   & \object{V498 Ori} &  4400	   &   1.763 &  0.37  &   2.29 &     1.40 &	5.0  &  Y &  ... &  ... \\ 
  20  &  {\scriptsize 05350738-0548010}   &                   &  2900	   &   0.321 &  0.67  &   2.25 &      ... &     ...  &  P &  Y   &  ... \\ 
{\it 21}&  {\scriptsize 05351033-0546335} & \object{AA Ori}   & {\it 4830} &  12.39  &	2.50  &   4.72 &     2.7  & $\leq 1$ &  Y &  Y   &  Y	\\ 
  22  &  {\scriptsize 05351205-0547296}   &         	      &  2800	   &   0.061 &  0.35  &   1.05 &     0.06 & $\leq 1$ &  Y &  ... &  ... \\ 
{\it 23}&  {\scriptsize 05351011-0548342} &         	      & {\it 3690} &   0.117 &  0.10  &   0.84 &     0.62 &    18.0  &  P &  Y   &  ... \\ 
  25  &  {\scriptsize 05351493-0549348}   &         	      &  5000	   &   0.633 &  0.17  &   1.06 &     0.95 &    40.0  &  N &  ... &  ... \\ 
  27  &  {\scriptsize 05352637-0544340}   &         	      &  3600	   &   0.071 &  0.07  &   0.69 &     0.60 &    28.0  &  P &  Y   &  ... \\ 
{\it 28}&  {\scriptsize 05351423-0543175} & \object{AB Ori}   & {\it 4260} &   4.812 &	1.88  &   3.28 &     1.7  &	9.0  &  Y &  Y   &  Y	\\ 
  29  &  {\scriptsize 05351236-0543184}   & \object{V486 Ori} &  4400	   &   1.174 &	0.21  &   1.87 &     1.40 &	5.0  &  Y &  ... &  Y	\\ 
  30  &  {\scriptsize 05351146-0544183}   &         	      &  3200	   &   0.083 &  0.52  &   0.94 &     0.18 &	2.3  &  Y &  Y   &  ... \\ 
  31  &  {\scriptsize 05350575-0544001}   &         	      &  3800	   &   0.210 &  0.22  &   1.06 &     0.70 &    10.1  &  P &  ... &  ... \\ 
  32  &  {\scriptsize 05350567-0543046}   &         	      &  3200	   &   0.050 &  0.01  &   0.73 &     0.18 &	4.5  &  Y &  ... &  ... \\ 
33$^*$ & {\scriptsize 05350486-0544267}   &         	      &  4600	   &   0.095 &  3.58: &   0.49 &     ...  &	...  &  N &  ... &  ... \\ 
  35  & {\scriptsize 05350303-0545333}    &         	      &  4600	   &   0.896 &  4.75  &   1.49 &     1.20 &    16.0  &  P &  Y   &  ... \\
  39  &  {\scriptsize 05350406-0548540}   &         	      &  3800	   &   0.121 &  0.06  &   0.80 &     0.70 &    28.2  &  P &  Y   &  ... \\ 
40$^*$ &  {\scriptsize 05350627-0549021}  &         	      &  2000	   &   0.196 &  4.18: &   3.69 &     ...  &	...  &  P &  ... &  ... \\ 
  41  &  {\scriptsize 05350825-0550003}   &         	      &  3900	   &   0.204 &  0.02  &   0.99 &     0.80 &    15.9  &  P &  ... &  ... \\ 
  42  &  {\scriptsize 05351088-0549485}   &         	      &  3400	   &   0.057 &  0.01  &   0.69 &     0.40 &    20.1  &  P &  ... &  ... \\ 
  47  &  {\scriptsize 05353023-0551169}   &         	      &  4400	   &   0.489 &  0.14  &   1.21 &     1.05 &    20.0  &  P &  ... &  ... \\  
{\it 48}&  {\scriptsize 05353047-0549037} &         	      & {\it 3830} &   0.208 &  0.10  &   1.04 &     0.75 &    12.5  &  P &  Y   &  Y	\\ 
  49  &  {\scriptsize 05353294-0549326}   &         	      &  3700	   &   0.042 &  0.05  &   0.50 &     0.50 &    71.3  &  N &  ... &  ... \\ 
  50  &  {\scriptsize 05353385-0548211}   &         	      &  3300	   &   0.069 &  0.91  &   0.80 &     0.25 &	5.6  &  Y &  ... &  ... \\ 
{\it 51}&  {\scriptsize 05353316-0547074} & \object{V1744 Ori}& {\it 3690} &   0.140 &	0.10  &   0.92 &     0.57 &    11.3  &  P &  Y   &  ...	\\	    
{\it 52}&  {\scriptsize 05352930-0545381} & \object{V1551 Ori}& {\it 3740} &   0.185 &	0.03  &   1.03 &     0.62 &	8.9  &  Y &  Y   &  ...	\\ 
  53  &  {\scriptsize 05353372-0546162}   &          	      &  3800	   &   0.124 &  0.06  &   0.81 &     0.70 &    25.4  &  P &  ... &  ... \\ 
  54  &  {\scriptsize 05353266-0545284}   &          	      &  3100	   &   0.041 &  0.05: &   0.70 &     0.10 &	2.5  &  Y &  Y   &  ... \\ 
  55  &  {\scriptsize 05353229-0544060}   & \object{V1564 Ori}&  3800	   &   0.145 &	0.22  &   0.88 &     0.75 &    28.2  &  P &  ... &  ...	\\ 
{\it 56}&  {\scriptsize 05353092-0543053} & \object{V1556 Ori}& {\it 3920} &   0.874 &	0.05  &   1.38 &     0.90 &	6.4  &  Y &  Y   &  Y	\\ 
  57  &  {\scriptsize 05352765-0542551}   &\object{Parenago 2059}& 4200	   &   0.816 &	1.69  &   1.71 &     1.20 &	5.0  &  Y &  ... &  Y   \\ 
  58  &  {\scriptsize 05352634-0543364}   &                   &  2900	   &   0.051 &  0.23  &   0.89 &     0.08 &	1.0  &  Y &  ... &  ... \\ 
  59  &  {\scriptsize 05352125-0542123}   & \object{V416 Ori} &  3700	   &   0.343 &  0.01  &   1.43 &     0.70 &	4.5  &  Y &  Y   &  Y	\\ 
  60  &  {\scriptsize 05351643-0542396}   & \object{V1506 Ori}&  4800	   &   0.934 &  0.03  &   1.40 &     1.15 &    20.0  &  P &  ... &  Y	\\ 
  61  &  {\scriptsize 05351734-0542145}   & \object{V411 Ori} &  3800	   &   0.356 &  0.73  &   1.38 &     0.75 &	5.0  &  Y &  Y   &  ... \\ 
  62  &  {\scriptsize 05351715-0541538}   & \object{V1507 Ori}&  3300	   &   0.149 &  0.41  &   1.18 &     0.35 &	3.6  &  Y &  ... &  ... \\ 
  63  &  {\scriptsize 05351123-0541361}   & \object{V1489 Ori}&  3600	   &   0.210 &  2.65  &   1.18 &     0.50 &	4.5  &  Y &  ... &  ... \\ 
{\it 65}&  {\scriptsize 05345908-0544303} & \object{KT Ori}   & {\it 3990} &   0.975 &	0.48  &   2.07 &     1.10 &	2.5  &  Y &  Y   &  ...	\\  
66$^*$  &  {\scriptsize 05345923-0544553} &	    	      &  4600	   &   1.210 &  4.96: &   1.74 &     1.40 &	9.0  &  Y &  ... &  ... \\  
69$^*$  &  {\scriptsize 05345898-0547596} &	    	      &  2900	   &   0.897 &  3.47  &   3.76 &     ...  &    ...   &  P &  ... &  ... \\  
{\it 70}&  {\scriptsize 05350054-0548591} &         	      & {\it 4600} &   0.904 &  4.07  &   1.50 &     1.20 &    16.0  &  P &  Y   &  ... \\  
\hline
\end{tabular}			         	                	                 
\begin{list}{}{}		         	                	                 
\item[$^{\mathrm{a}}$] Photometric membership (present work). 
\item[$^{\mathrm{b}}$] Spectroscopic membership based on the radial velocity and/or H$\alpha$ emission \citep{Furesz08}.
\item[$^{\mathrm{c}}$] Astrometric membership based on proper motions \citep{Dias06}.
\item[$^*$] Unreliable parameters due to the lack of optical photometry and a likely strong reddening and/or NIR excess.
\end{list}
\end{table*} 

\addtocounter{table}{-1}

\begin{table*}
\vspace{-0.3cm}
\caption{\textit{Continued.}} 
\vspace{-0.2cm}
\begin{tabular}{ r l l l l l l l r c c c} 
\hline\hline 
  Id  &   2MASS		    & Other  & $T_{\rm eff}$ &  $L$           & $A_V$	  & $R$      & $M$	      &  Age  & \multicolumn{3}{c}{Membership} \\
      &                     & Id &  (K)          &  ($L_{\sun}$)  & (mag)	  & ($R_{\sun}$) & ($M_{\sun}$)   &  (Myr)  & Ph$^{\mathrm{a}}$ 
      & Sp$^{\mathrm{b}}$ & As$^{\mathrm{c}}$ \\ 
\hline 
\noalign{\smallskip}
  75  &  {\scriptsize 05350284-0551031}   &         	     	 &  4800     &   0.447 &   1.75  &   0.97 &	0.90 &    39.8  &  N &  ... &  ... \\  
  80  &  {\scriptsize 05353168-0542457}   &         	     	 &  3800     &   0.091 &   0.29  &   0.70 &	0.62 &    40.2  &  N &  ... &  ... \\ 
  86  &  {\scriptsize 05351895-0545117}   &         	     	 &  3900     &   0.032 &   0.03  &   0.39 &	...  &     ...  &  N &  ... &  ... \\ 
  89  &  {\scriptsize 05353692-0543172}   &         	     	 &  3400     &   0.042 &   0.09  &   0.59 &	0.25 &    11.3  &  P &  ... &  ... \\ 
  90  &  {\scriptsize 05352650-0548497}   &\object{Parenago 2049}&  3800     &   0.488 &   0.00  &   1.61 &	0.85 &     4.0  &  Y &  ... &  ... \\ 
  91  &  {\scriptsize 05352707-0548522}   &\object{Parenago 2050}&  3800     &   0.945 &   0.00  &   2.25 &	0.85 &     1.3  &  Y &  ... &  ... \\ 
  92  &  {\scriptsize 05351197-0541521}   &	            	 &  3800     &   0.041 &   0.25  &   0.47 &     ...  &     ...  &  N &  ... &  ... \\ 
{\it  93} & {\scriptsize 05351790-0542340}&\object{Parenago 1929}&{\it 4775} &   4.792 &   0.58  &   3.20 &     1.8  &     1.0  &  Y &  ... &  ... \\ 
  94  &  {\scriptsize 05352080-0544527}   &                 	 &  4200     &   0.025 &   0.52  &   0.30 &     ...  &     ...  &  N &  ... &  ... \\ 
\hline
\end{tabular}			         	                	                 
\begin{list}{}{}		         	                	                 
\item[$^{\mathrm{a}}$] Photometric membership (present work). 
\item[$^{\mathrm{b}}$] Spectroscopic membership based on the radial velocity and/or H$\alpha$ emission \citep{Furesz08}.
\item[$^{\mathrm{c}}$] Astrometric membership based on proper motions \citep{Dias06}.
\end{list}
\end{table*}

\section{Projected rotation velocity and spectral classification}	
\label{sec:vsini}

For the determination of the projected rotational velocity ($v\sin i$) and the spectral type, we analyzed the high-resolution 
spectra obtained in 2004 and 2005 by \citet{Furesz08} for some of the stars in our field with the Hectochelle multi-object 
spectrograph at the MMT telescope. These spectra, with a resolution of $R\simeq 34\,000$, are centered at H$\alpha$ and
span 190\,\AA. In addition, for nine stars we also used Hectochelle spectra obtained in 2006 and 2007 in the
spectral range $\lambda$5150--5300\,\AA, which includes the \ion{Mg}{i}\,b triplet.

We used ROTFIT, a code written by one of us \citep{Frasca03,Frasca06} in IDL to find simultaneously the spectral type and 
$v\sin i$ of the target by searching for the spectrum, 
among a library of standard star spectra, that best matches (minimum of the residuals) the target one.
Rotational broadening of the template spectrum is part of the fitting procedure, and a small amount 
of veiling is also allowed. 

As a standard star library, we used a sample of 101 stars whose atmospheric parameters ($T_{\rm eff}$, $\log g$, and [Fe/H])
are known \citep{Prugniel01,Cayrel01} and are rather well-distributed in effective temperature and gravity (see
Table~\ref{tab:references}\footnote{Available in electronic form only.}).
The standard star spectra were retrieved from the E{\sc lodie} Archive \citep{Moultaka04} and have a spectral resolution
$R\simeq 42\,000$ that is close to that of Hectochelle. However, the resolution of E{\sc lodie} spectra was degraded to
that of Hectochelle by convolving them with a Gaussian kernel of the proper width.

We excluded the H$\alpha$ and the two [\ion{N}{ii}] nebular lines at $\lambda$6548 and 
6583 \AA ~from the spectral region to be fitted.

For the stars with very low signal-to-noise ratio (S/N) at the continuum, we obtained unreliable values
for the $v\sin i$ and spectral type. These values were not considered and only the parameters derived 
from the SED were used for these objects.

The $v\sin i$ values found by ROTFIT for the spectra with a sufficient signal (S/N\,$\geq 30$) are 
nearly independent on the standard star giving the best fit, while the veiling has the effect 
of slightly reducing the $v\sin i$ whenever it exceeds $\approx$30\%. 
The $v\sin i$ and veiling values which minimize the residuals were adopted. 

\begin{figure}[ht]
  \begin{center}
    \includegraphics[width=8.8cm]{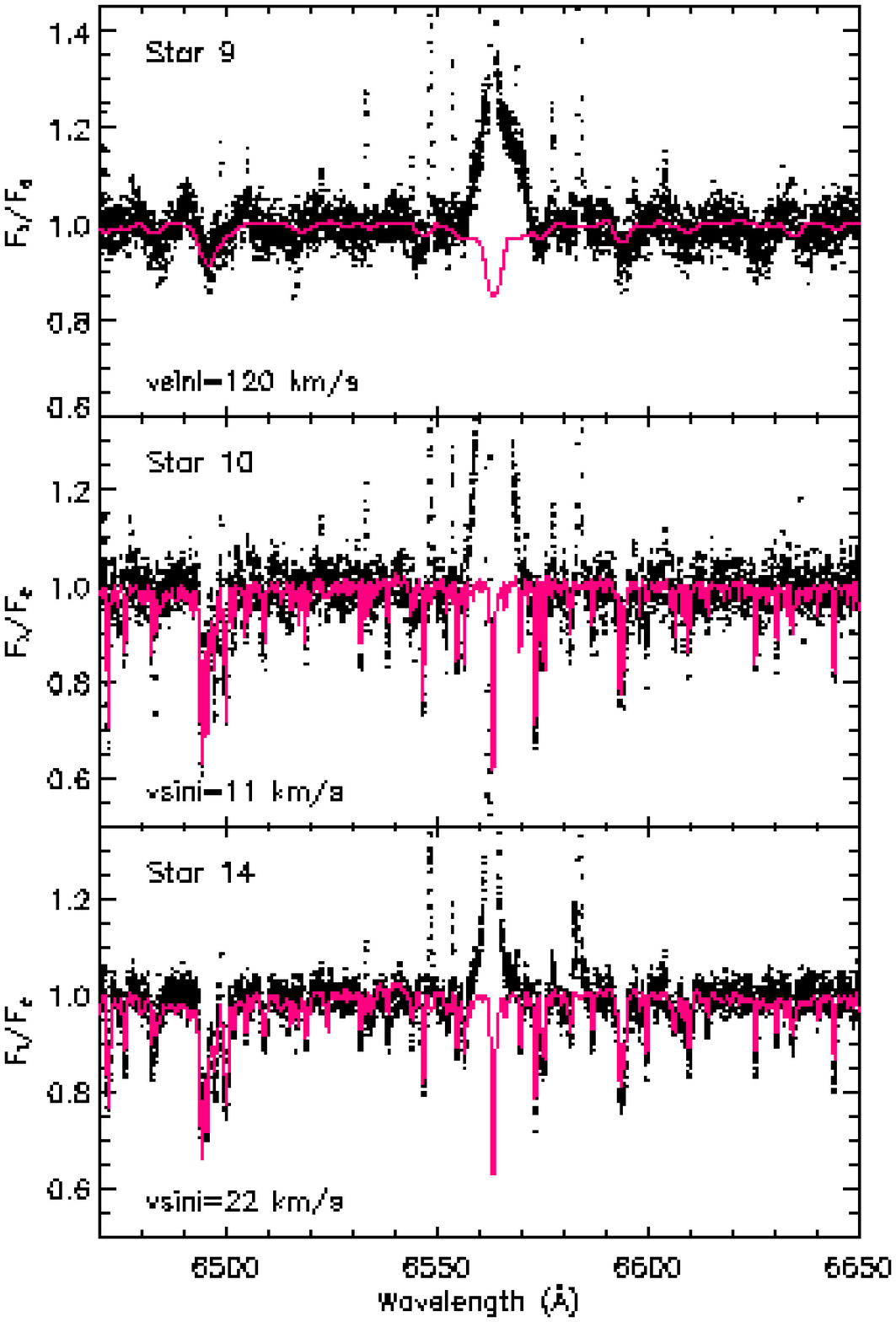}
  \end{center}
\caption{Example of observed Hectochelle spectra (dots) of our targets with the best template
superimposed (thin lines). The Id number and the $v\sin i$ value are also marked in each box.
}
\label{fig:spe}
\end{figure}

\begin{figure*}[ht]
  \begin{center}
    \includegraphics[width=5.9cm]{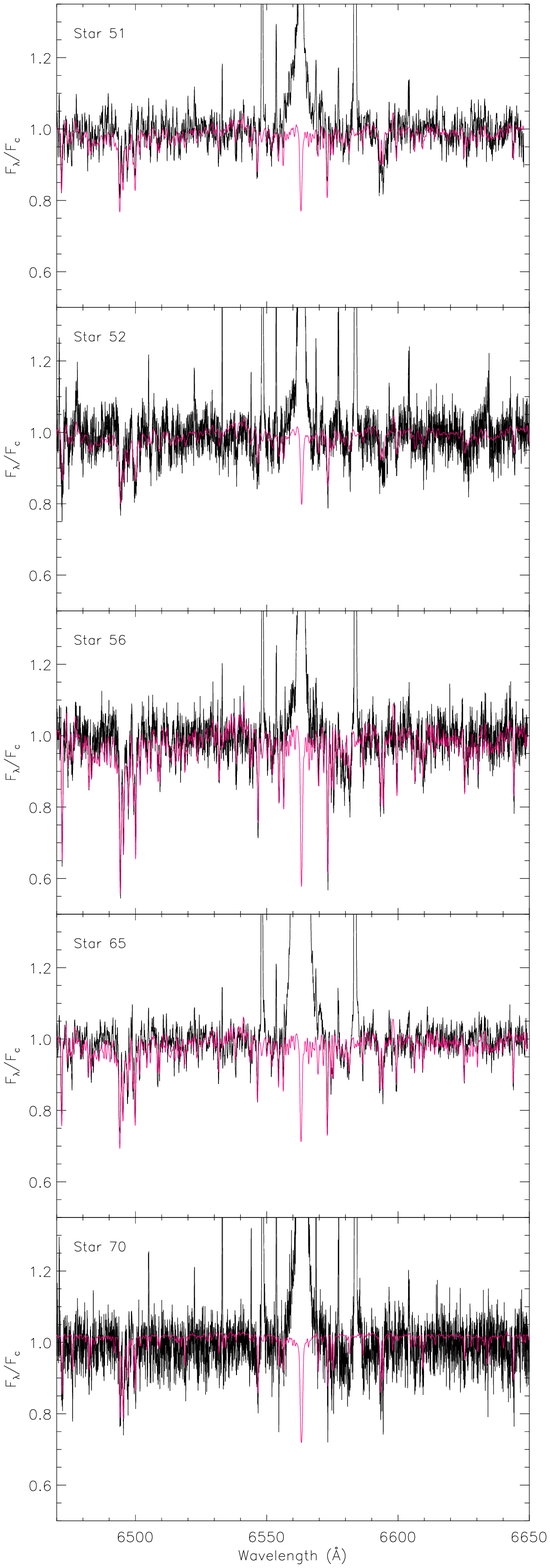}
    \includegraphics[width=5.9cm]{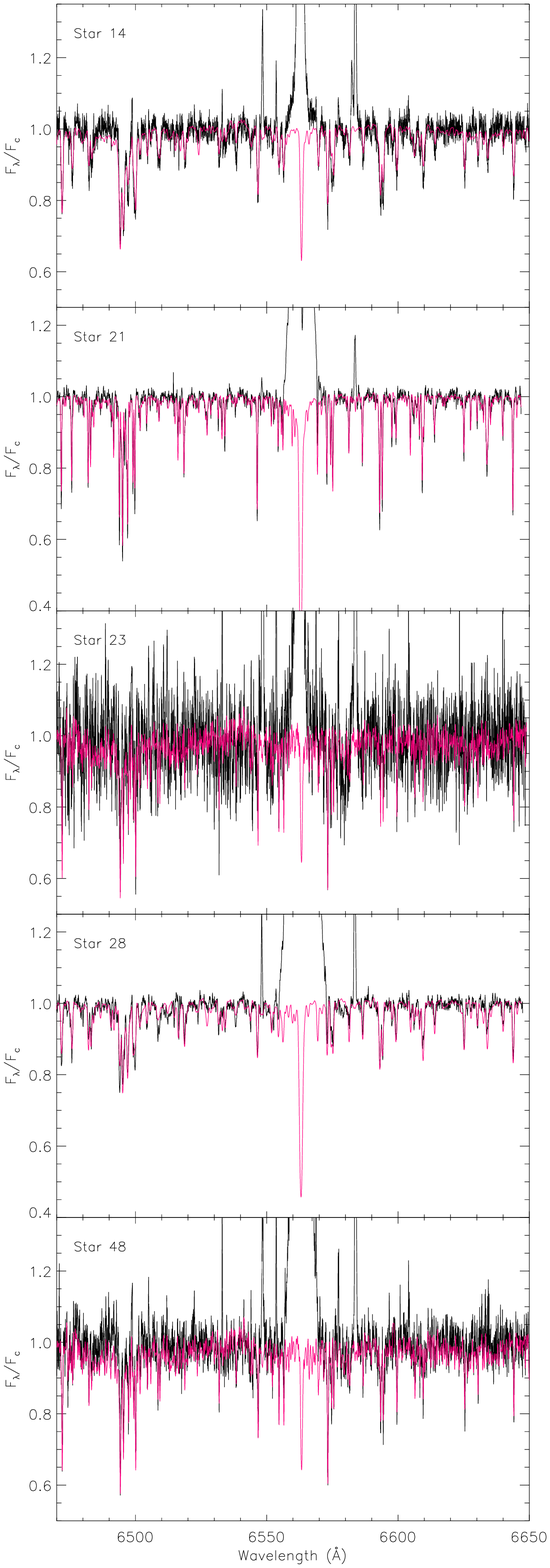}
    \includegraphics[width=5.9cm]{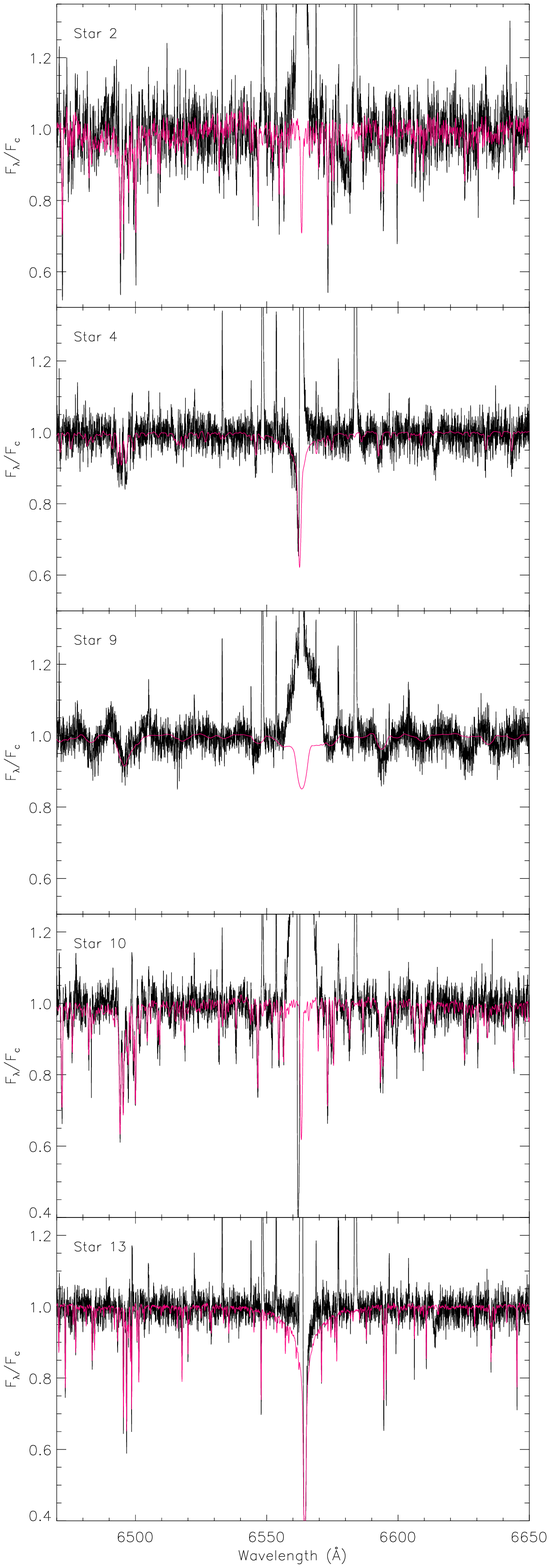}
  \end{center}
\caption{Hectochelle spectra in the H$\alpha$ region of our targets (black lines) with the best template
superimposed (thin red lines). The H$\alpha$ emission has been cut in order to display the photospheric
absorption lines. The Id number is marked in the top-left corner of each box.
}
\label{fig:spe_ha_all}
\end{figure*}

\begin{figure*}[ht]
  \begin{center}
    \includegraphics[width=5.5cm]{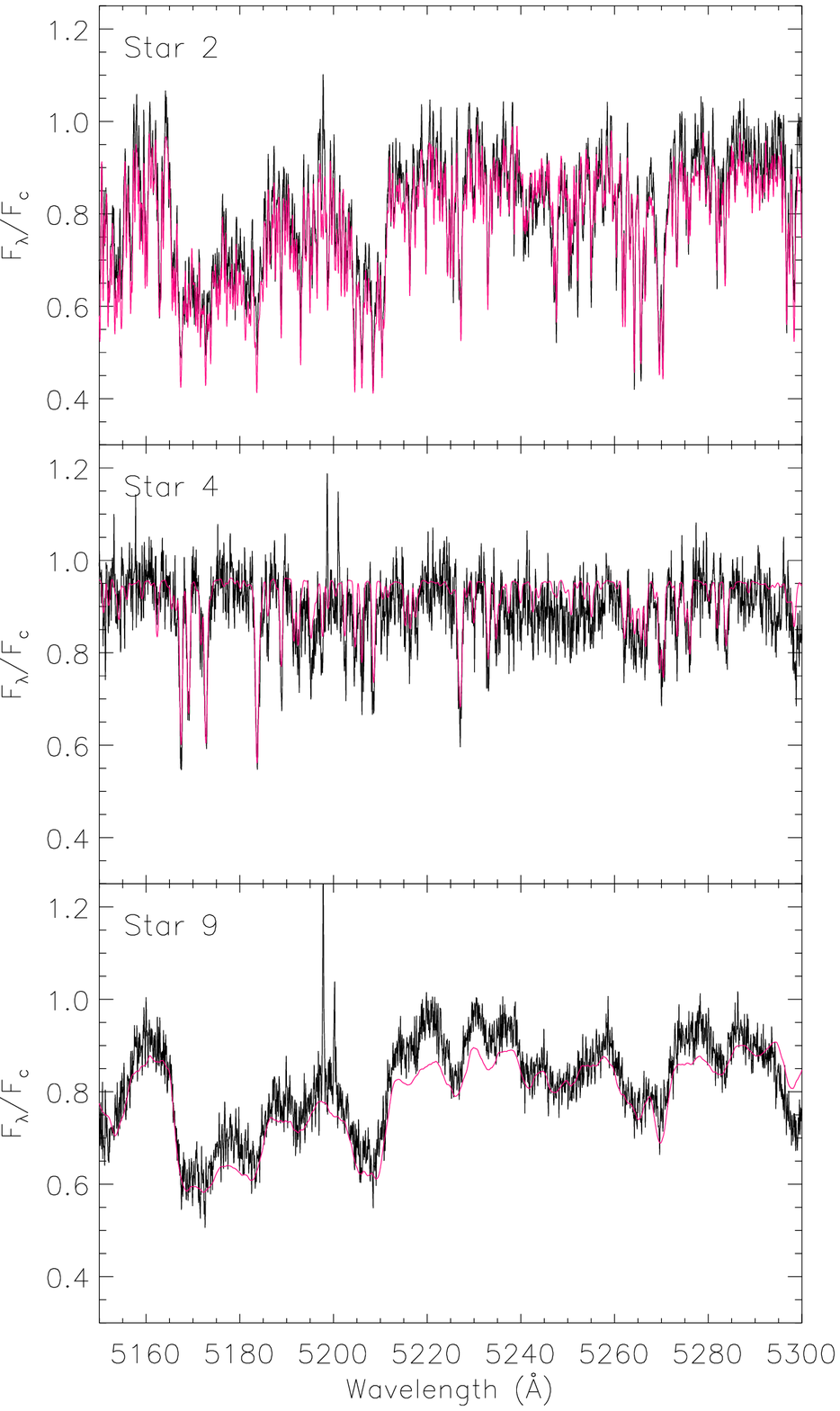}
    \includegraphics[width=5.5cm]{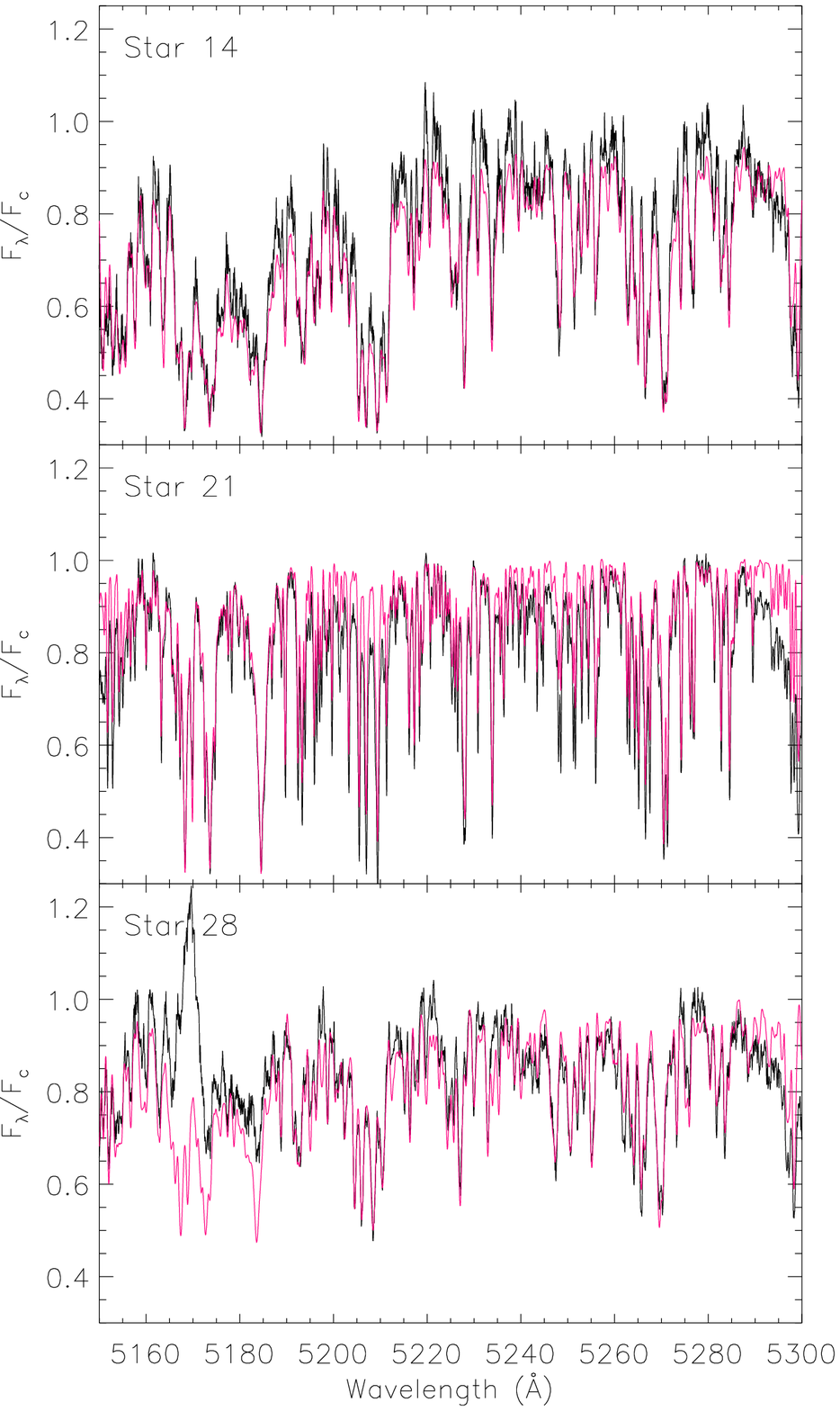}
    \includegraphics[width=5.5cm]{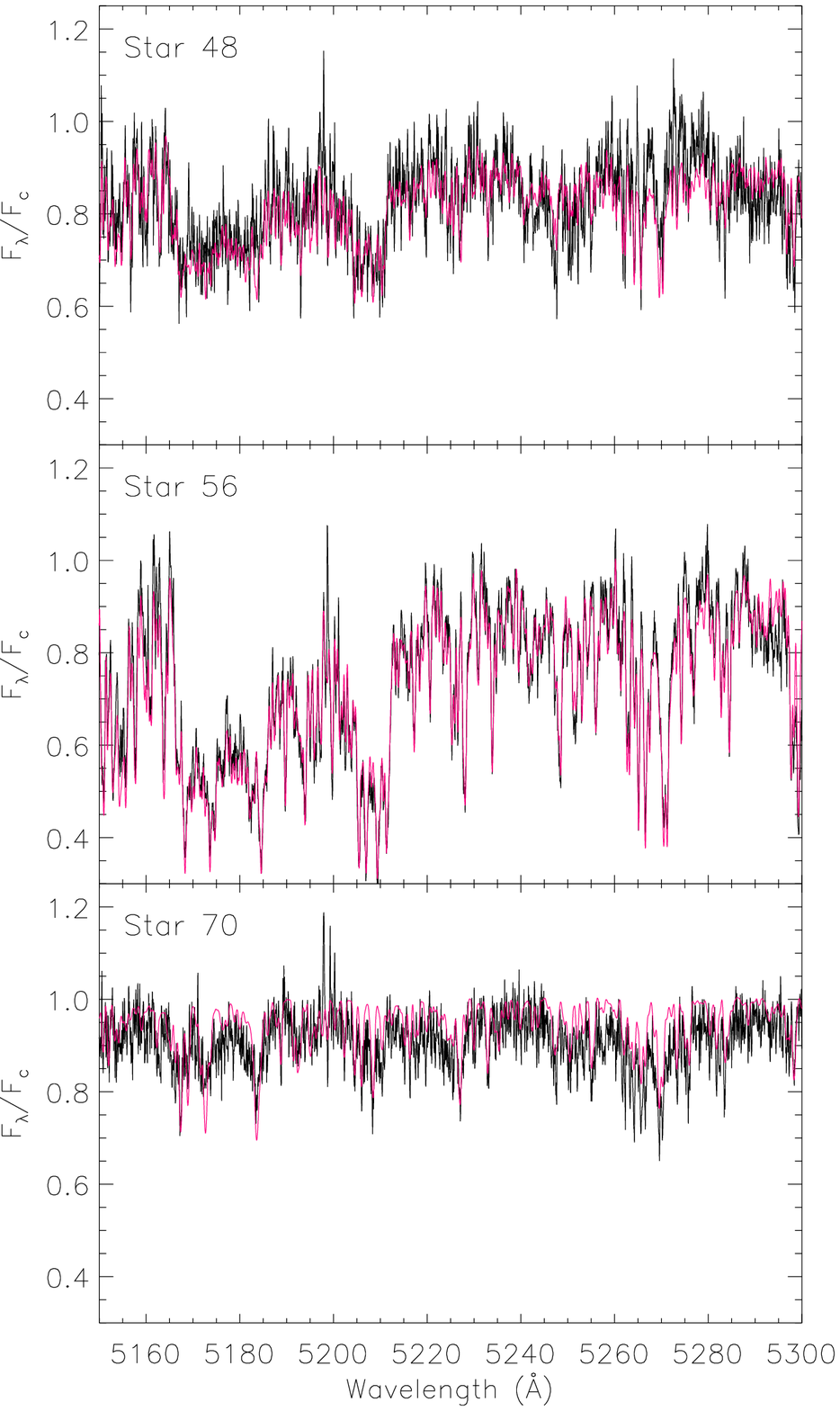}
  \end{center}
\caption{Hectochelle spectra in the \ion{Mg}{i}\,b region of our targets (black lines) with the best template
superimposed (thin red lines). The Id number is marked in the top-left corner of each box.
}
\label{fig:spe_mg_all}
\end{figure*}

The values of $v\sin i$ and $T_{\rm eff}$, reported in Table~\ref{tab:spec} for the 15 stars with sufficiently
exposed spectra, are the weighted averages of the five best templates. 
 The usual weights, $w_i=1/\sigma_i^2$, where $\sigma_i$ are the r.m.s. of each fit, were adopted. 
The standard errors on $v\sin i$ and $T_{\rm eff}$ are also given in Table~\ref{tab:spec}.
In the same table, the spectral types derived by us with ROTFIT and from previous works mainly
based on low-resolution spectra are also listed.
Some examples of the results of the fitting are shown in Fig.~\ref{fig:spe}.
The observed spectra with the fitted templates are displayed in Fig.~\ref{fig:spe_ha_all} and 
Fig.~\ref{fig:spe_mg_all}\footnote{Available in electronic form only.} 
for the H$\alpha$ and the \ion{Mg}{i}\,b spectral regions, respectively.

\begin{table*}
\caption{Physical parameters and spectral types for the stars with Hectochelle or literature spectra.} 
\label{tab:spec}
\begin{tabular}{ r l l l c l l l l l} 
\hline\hline 
\noalign{\smallskip}
  Id      &   2MASS		              &  $P$	&  $R$          &   $v\sin i$	&   $T_{\rm eff}$ &  Sp. Type	& $EW_{H\alpha}^{\rm g}$  & $EW_{Li}$ & $\sin i$ \\  
          &                                   & (days)  &  ($R_{\sun}$) & (km\,s$^{-1}$) &   (K)	   &		 & ~~(\AA)	  & ~~(\AA)   &	  \\ 
\noalign{\smallskip}
\hline 
\noalign{\smallskip}
   2      &  {\scriptsize 05352029-0546399}   & 7.751   & 1.24   &    7.0$\pm$1.0  & 3710$\pm$110 &  M2.5$^{\mathrm{a}}$  &  ~~7.5 &  & 0.864 \\
   4      &  {\scriptsize 05352007-0545526}   &         &        &   30$\pm$5  & 5720$\pm$100	   &  G2$^{\mathrm{a}}$   &  ~~0.0 &  &     \\ 
   5      &  {\scriptsize 05351981-0545409}   & 0.60    & 1.33   &  	    &	 & M3.5$^{\mathrm{b}}$ &   &  & \\ 
   9      &  {\scriptsize 05352281-0544428}   & 0.5627  & 1.98   &   126$\pm$6     & 3480$\pm$170  & M3$^{\mathrm{a}}$  & ~~3.6 & 0.72$^{\rm c}$ &  0.674  \\   
  10      &  {\scriptsize 05352617-0545084}   & 5.816   & 1.25   &   11.0$\pm$1.0  & 4110$\pm$170  & K8$^{\mathrm{a}}$  & ~~7.0 & 0.52$^{\rm d}$ & $\simeq 1$   \\    
  13      &  {\scriptsize 05351621-0547201}   &         &        &   4.0$\pm$2.0  & 5800$\pm$200   & G1$^{\mathrm{a}}$  &  ~~0.0 &  & \\ 
  14      &  {\scriptsize 05351192-0545379}   & 4.047   & 1.89   &   22.5$\pm$1.5 & 4250$\pm$170 &  K7$^{\mathrm{a}}$, K8$^{\mathrm{b}}$  &  ~~3.8 & 0.41$^{\rm d}$ &  0.952 \\  
  17      &  {\scriptsize 05350614-0545311}   &  8.435  & 1.26   &    &    & M0$^{\mathrm{b}}$   &     &   &   \\ 
  18      &  {\scriptsize 05353222-0544265}   & 12.461  & 2.16   &    &    & M3$^{\mathrm{b}}$   &     &   &   \\ 
  19      &  {\scriptsize 05352973-0548450}   &  0.605  & 2.29   &    &    & K8$^{\mathrm{b}}$   &     &   &   \\ 
  21      &  {\scriptsize 05351033-0546335}   &         & 4.72   &   12.0$\pm$2.0 & 4830$\pm$250  &  K0$^{\mathrm{a}}$    &  ~~3.9 &  & \\ 
  23      &  {\scriptsize 05351011-0548342}   &         & 0.84   &   1.0$\pm$0.5  & 3690$\pm$70   &  M2.5$^{\mathrm{a}}$  &  ~~9.0 &  & \\ 
  28      &  {\scriptsize 05351423-0543175}   &         & 3.28   &   20.0$\pm$1.5 & 4260$\pm$270  & K7$^{\mathrm{a}}$, K7e$^{\mathrm{e}}$  &  13.7  &   & \\ 
  29      &  {\scriptsize 05351236-0543184}   & 8.468   & 1.87   &    &    & K7e$^{\mathrm{d}}$  &   &  \\ 
  47      &  {\scriptsize 05353023-0551169}   & 10.02   & 1.21   &    &	  & K8$^{\mathrm{b}}$	&   &	&  \\  
  48      &  {\scriptsize 05353047-0549037}   &  7.721  & 1.04   &   7.5$\pm$1.5 &  3830$\pm$220  & M2$^{\mathrm{a}}$, M2$^{\mathrm{b}}$   &   ~~8.6 &  &  $\simeq 1$ \\ 
  51      &  {\scriptsize 05353316-0547074}   &  2.763  & 0.92   &   16.0$\pm$0.5 &  3690$\pm$100  & M3$^{\mathrm{a}}$, M3.5$^{\mathrm{b}}$   & ~~2.6 &   &  0.949 \\ 	   
  52      &  {\scriptsize 05352930-0545381}   &  1.696  & 1.03   &   25.0$\pm$2.0 &  3740$\pm$90   & M2.5$^{\mathrm{a}}$, M3.5$^{\mathrm{b}}$ &  ~~6.6 &  0.41$^{\rm c}$  &  0.813 \\ 
  56      &  {\scriptsize 05353092-0543053}   &  3.80   & 1.38   &   13.5$\pm$1.5 &  3920$\pm$260  & M1.5$^{\mathrm{a}}$, M2$^{\mathrm{b}}$   &  ~~4.9 &  0.50$^{\rm c}$  &  0.734 \\  
  60      &  {\scriptsize 05351643-0542396}   &  4.11   & 1.40   &    &    & K5$^{\mathrm{b}}$   &   &  0.45$^{\rm c}$   &  \\ 
  62      &  {\scriptsize 05351715-0541538}   &  2.05   & 1.18   &    &    & M5$^{\mathrm{b}}$   &   &  \\ 
  65      &  {\scriptsize 05345908-0544303}   &         & 2.07   &   13.0$\pm$1.5 & 3990$\pm$260  & M1.5$^{\mathrm{a}}$  &  ~~8.6 &  & \\ 
  70      &  {\scriptsize 05350054-0548591}   &         & 1.50   &   12.0$\pm$2.0 &  4600$\pm$500  & K5$^{\mathrm{a}}$  &   ~~8.2 &  & \\ 
  93      &  {\scriptsize 05351790-0542340}   & 5.70:   & 3.33   &   34$^{\rm f}$ &  4775$^{\rm f}$   &	 K3$^{\rm f}$   &         &  & $\simeq 1$ \\ 
\hline				                								        	 
\end{tabular}			         	                	                 
\begin{list}{}{}		         	                	                 
\item[$^{\mathrm{a}}$] Present work.
\item[$^{\mathrm{b}}$] From \citet{rebu00}.
\item[$^{\mathrm{c}}$] From \citet{stass99}.
\item[$^{\mathrm{d}}$] From \citet{Sicilia}.
\item[$^{\mathrm{e}}$] From \citet{Herbig88}.
\item[$^{\mathrm{f}}$] From \citet{wolf}.
\item[$^{\mathrm{g}}$] From \citet{Furesz08}.
\end{list}
\end{table*} 

 From the values of $v\sin i$, radius ($R$), and rotation period, $P$, reported in 
Table~\ref{tab:spec}, the inclination of the rotation axis can be estimated as 
\begin{equation}
\label{Eq:vsini}
\sin i = \frac{v\sin i \cdot P}{2\pi R} 
\end{equation}
{\noindent  This parameter is useful for the application of a spot model to the light curves, 
which is presented in Sect.~\ref{sec:spot}.}

\section{Spot modelling}	
\label{sec:spot}

The availability of simultaneous light curves from the optical to the NIR allows us to reconstruct 
roughly the starspot distribution and determine two basic spot parameters, i.e. temperature and dimension.  
Moreover, hot and cool spots produce a different effect on the $RIJH$ light curves, so that it is possible 
to discriminate between the two as the main cause for the observed variability \citep[see, e.g.][]{Bouvier89}.

In order to search for a unique solution for the multi-band light curves, we used MACULA, 
a spot model code developed by us in the IDL environment \citep{Frasca05}.
The model assumes circular dark (cool) or bright (hot) spots on the surface of a spherical limb-darkened star. 
The linear limb-darkening coefficients for $RIJHK$ bands are from \citet{claret00}, who calculated them for 
a grid of Phoenix NextGen models.

\citet{Frasca05} have shown that two circular spots are sufficient to reproduce the general shape of light
curves of spotted stars without introducing too many free parameters. 

The flux contrast ($F_{\rm sp}/F_{\rm ph}$) can be evaluated through the Planck spectral energy distribution, 
the ATLAS9 \citep{Kuru93} and PHOENIX NextGen \citep{hau99} atmosphere models. 
\citet{Frasca05} demonstrated that both ATLAS9 and NextGen provide values of 
the spot temperature ($T_{\rm sp}$) and area coverage ($A_{\rm rel}$) that are in close agreement, 
while the black-body assumption for the SED leads to underestimate the spot temperature. 
In the present work, we have preferred the NextGen flux ratios because they can go down to a temperature 
of 1700\,K, while the minimum temperature for ATLAS9 fluxes is 3500\,K, which is close to the photospheric 
temperature of the coolest stars in our sample.
For this purpose, we integrated the NextGen spectra, weighted with the transmission curves of the REM filters.
 
We applied our code only to the six stars with the highest-quality light curves in at least three photometric bands.
For three of them with Hectochelle spectra, namely \#9, \#10, and \#14, we have an estimate of 
the inclination of the rotation axis through Eq.~\ref{Eq:vsini}. For the remaining stars to which we applied
the spot modelling (\#11, \#17, and \#18) we adopted a value of $i=70\degr$, which is a reasonable choice given the
rather large amplitudes of their light curves. 
For each star, we started to fit the light curve in a reference band (usually $I$) with a 2-spot model  at a fixed temperature 
(normally $T_{\rm sp}/T_{\rm ph}\,=\,$ 0.80), and found the geometrical parameters of the spots (longitudes and latitudes)
as well as their area by minimizing the $\chi^2$.
Then, we fixed the spot positions and, for a few fixed values of the temperature ratio ($T_{\rm sp}/T_{\rm ph}$), 
we let the spot dimensions to vary, searching for the value of spot area which minimizes the $\chi^2$ of the fit for 
the given value of $T_{\rm sp}/T_{\rm ph}$. 
As already found by \citet{Frasca05}, the $\chi^2$ does not vary significantly for large intervals of $T_{\rm sp}/T_{\rm ph}$,
so that several models in the solution grid, for a given passband, are fitting the light curve with the same accuracy.
Only the use of at least two diagnostics permits to remove the degeneracy of $T_{\rm sp}$ and area in the space of
the parameters. 

The grids of solutions for different bands allow to define the ranges of values for the spot temperature and 
area as those for which all the curves are well fitted within the errors.

\begin{figure}  
\hspace{-.5cm}
   \includegraphics[width=9cm]{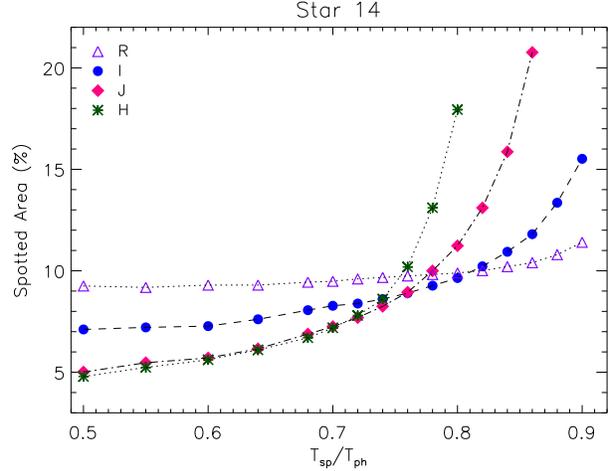}
\vspace{0cm}
\caption {Grids of solutions of $RIJH$ light curves for star \#14. Each pass-band is represented by a different symbol.}
\label{fig:Teff_aree_14}
\end{figure}

\begin{figure}
\hspace{-1.cm}
 \includegraphics[width=12cm]{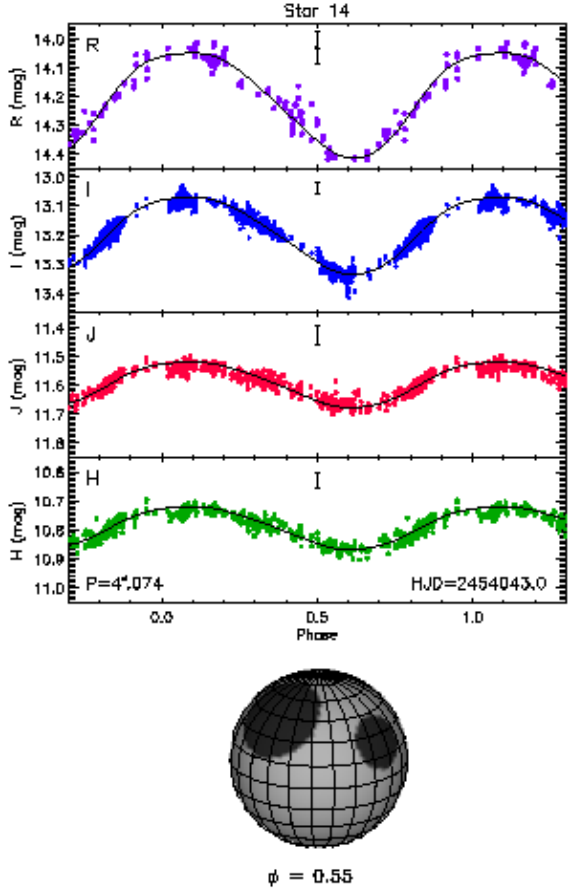}
\vspace{0.3cm}
\caption{Observed (dots) and synthetic (full lines) $R$, $I$, $J$, and $H$ light curves of 
star \#14. The mean error-bar is also displayed in the upper mid part of each box. 
A schematic map of the starspot distribution, as seen at the phase of maximum spot visibility, 
is also shown. The inclination of the rotation axis, as derived from Eq.~\ref{Eq:vsini}, is $i=70\degr$.} 
\label{fig:palle_14}
\end{figure}

The results of the solution grids for the $RIJH$ light curves of star \#14, for which we found $i=70\degr$ (see 
Table~\ref{tab:spot_parameters}), are displayed in Fig.~\ref{fig:Teff_aree_14}, where the percent spotted area 
(in units of star surface) versus the fractional spot temperature $T_{\rm sp}/T_{\rm ph}$ is plotted. 
The figure shows the different behaviour of the $RIJH$ solution grids, 
and the rather small region in the plane $T_{\rm sp}/T_{\rm ph}$--$A_{\rm rel}$ were the intersection of the four 
solution grids occurs. 
This allows us to assert that the fractional spot temperature is $T_{\rm sp}/T_{\rm ph}\simeq 0.76\pm 0.04$, i.e.
a spot temperature in between 3000 and 3400 K, and a spotted area in the range 7--9\%.
In Fig.~\ref{fig:palle_14}, the observed $RIJH$ light curves and the simultaneous solution provided by MACULA are 
displayed together with a map of the spotted photosphere, as seen at the rotational phase of maximum spot
visibility.

\begin{figure}  
\hspace{-.5cm}
   \includegraphics[width=9cm]{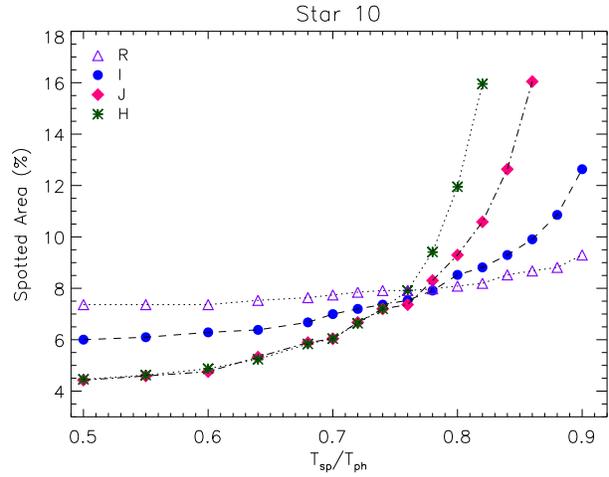}
\vspace{0cm}
\caption {Grids of solutions of $RIJH$ light curves for star \#10. Different symbols have been used for each pass-band.}
\label{fig:Teff_aree_10}
\end{figure}

Similar results are obtained for star \#10 ($i\simeq 90\degr$), for which the grids of solutions intersect each other for values 
of $T_{\rm sp}/T_{\rm ph}\simeq 0.76\pm 0.02$ (Fig.~\ref{fig:Teff_aree_10}). We also tried with hot spots, but
the solution grids in different bands do not intersect, ruling out accretion as the cause of the observed modulation.
This is a remarkable result because star \#10 displays the characteristics of a cTTS, such as the very strong 
and broad H$\alpha$ emission \citep{Furesz08} and the mid-IR excess (Fig.~\ref{Fig:SED}) which testifies the 
presence of a conspicuous accretion disc. It seems that cool photospheric spots, indicative 
of strong magnetic activity, are presumably the main responsible for the observed variability.

\begin{figure}
\hspace{-1.cm}
 \includegraphics[width=12cm]{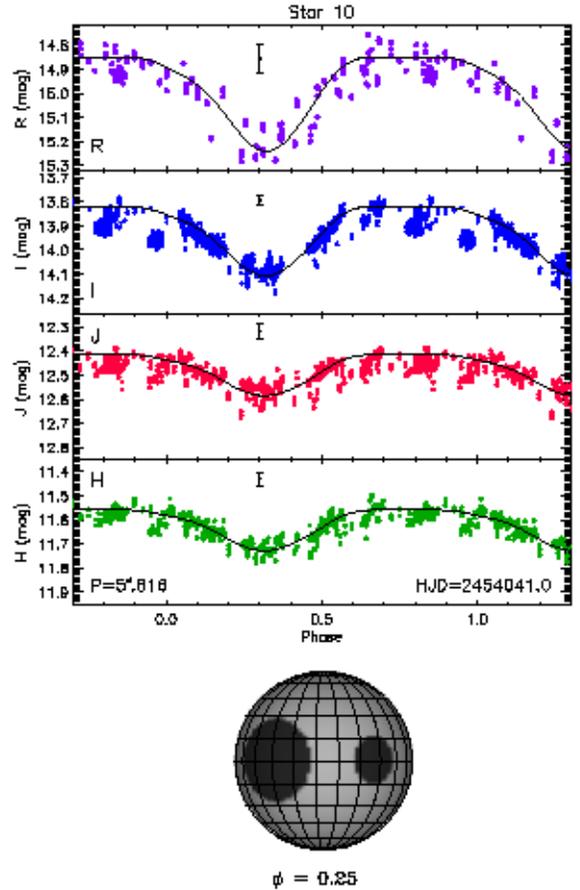}
\vspace{0.3cm}
\caption{Observed (dots) and synthetic (full lines) $R$, $I$, $J$, and $H$ light curves of 
star \#10. The mean error-bar is also displayed in the upper mid part of each box. 
A schematic map of the starspot distribution, as seen at the phase of maximum spot visibility, 
is also shown. The inclination of the rotation axis, as derived from Eq.~\ref{Eq:vsini}, is $i=90\degr$.} 
\label{fig:palle_10}
\end{figure}

However, superimposed to the rotational modulation, additional ``fluctuations'' larger than typical 
errors are clearly visible in the $RIJH$ light curves of star \#10 (Fig.~\ref{fig:palle_10}). 
These features appear to be correlated in the different bands. This kind of variability was already 
observed in some young stars in Orion by, e.g., \citet{Carpenter01} and could be ascribed to 
intrinsic variability due to accretion. 

\begin{figure}  
\hspace{-.5cm}
   \includegraphics[width=9cm]{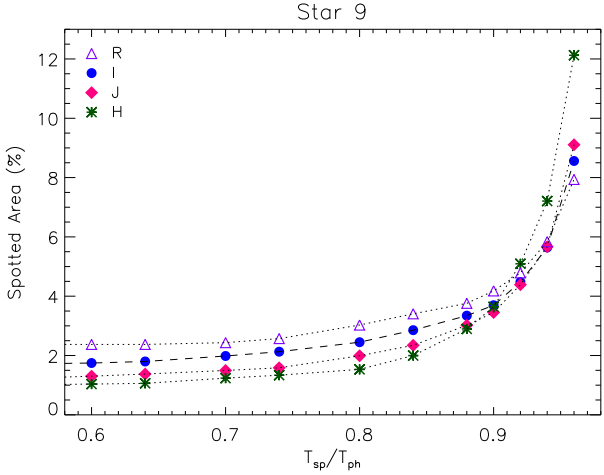}
\vspace{0cm}
\caption {Grids of solutions of $RIJH$ light curves for star \#9. Different symbols have been used for each pass-band.}
\label{fig:Teff_aree_9}
\end{figure}

The last star with known inclination ($i=45\degr$) analyzed with MACULA for the determination of spot 
parameters is \#9. This star displays fairly good light curves in $RIJH$ bands, although with rather low amplitudes, 
and is a very interesting case because of its very short rotation period.
As seen in Fig.~\ref{fig:Teff_aree_9}, the intersection of the solution grids occurs for larger values of
fractional spot temperature compared to the previous stars. We found $T_{\rm sp}/T_{\rm ph}\simeq 0.93\pm 0.03$,
which corresponds to a temperature difference $\Delta T \simeq 270\pm 100$\,K. 

Spot temperatures of a few hundreds Kelvin degrees cooler than the surrounding photosphere have been found in very active 
and fast-rotating cool stars \citep[see, e.g.,][]{olah97,Berdy05}.
We remark that we cannot assess if the spots are really uniform and warmer than in the other Orion stars 
analyzed by us or if there is a mixture of regions with different temperatures. In this case, warmer areas (analogous to
the sunspot penumbrae) could be dominating.
The lower temperature could be also the effect of the simultaneous presence
in the same area of cool spots and bright white-light faculae. There is no way, with the present photometric data,
to further investigate this point. A simultaneous light curve in the near-UV could be very helpful to detect bright
faculae which should have a much higher contrast at shorter wavelengths.

\begin{figure}
\hspace{-1.cm}
	\includegraphics[width=12cm]{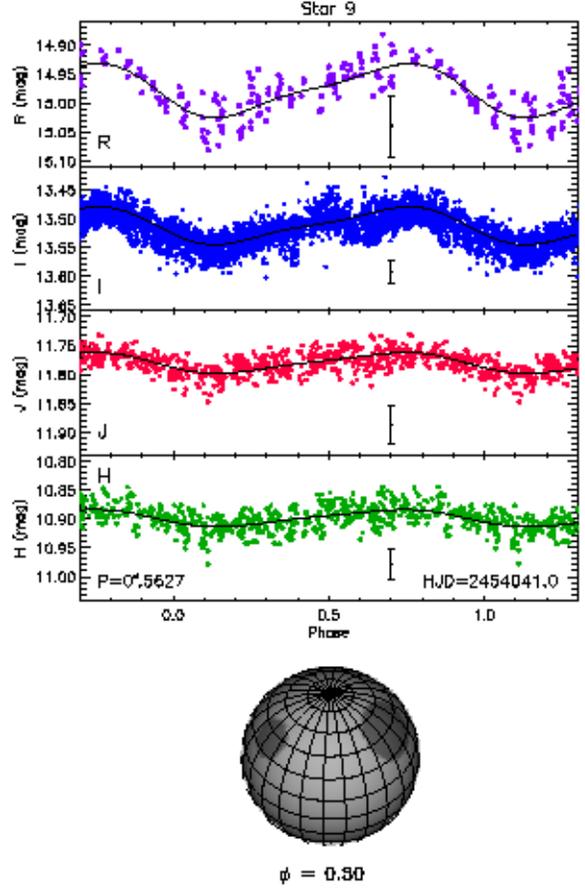}
\vspace{0.3cm}
\caption{Observed (dots) and synthetic (full lines) $R$, $I$, $J$, and $H$ light curves of 
star \#9. The mean error-bar is also displayed in the lower mid part of each box. 
A schematic map of the starspot distribution, as seen at the phase of maximum spot visibility, 
is also shown. The inclination of the rotation axis, as derived from Eq.~\ref{Eq:vsini}, is $i=45\degr$.} 
\label{fig:palle_9}
\end{figure}

In Fig.~\ref{fig:palle_9}, the observed $RIJH$ light curves and the simultaneous solution provided by MACULA are 
displayed together with a map of the spotted photosphere, as seen at the rotational phase of maximum spot
visibility.

\begin{figure}  
\hspace{-.5cm}
   \includegraphics[width=9cm]{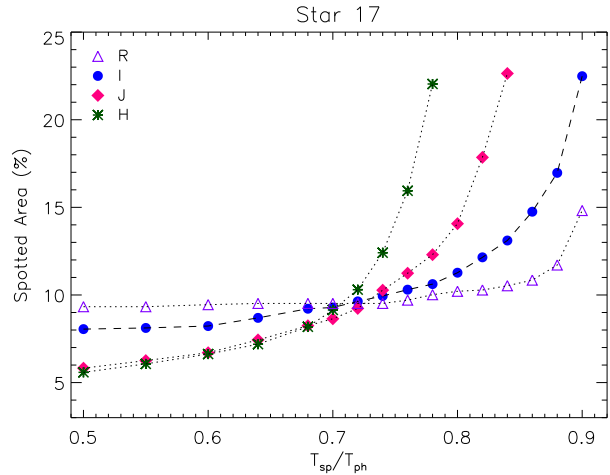}
\vspace{0cm}
\caption {Grids of solutions of $RIJH$ light curves for star \#17. Different symbols have been used for each pass-band.}
\label{fig:Teff_aree_17}
\end{figure}

\begin{figure}
\hspace{-1.cm}
 \includegraphics[width=12cm]{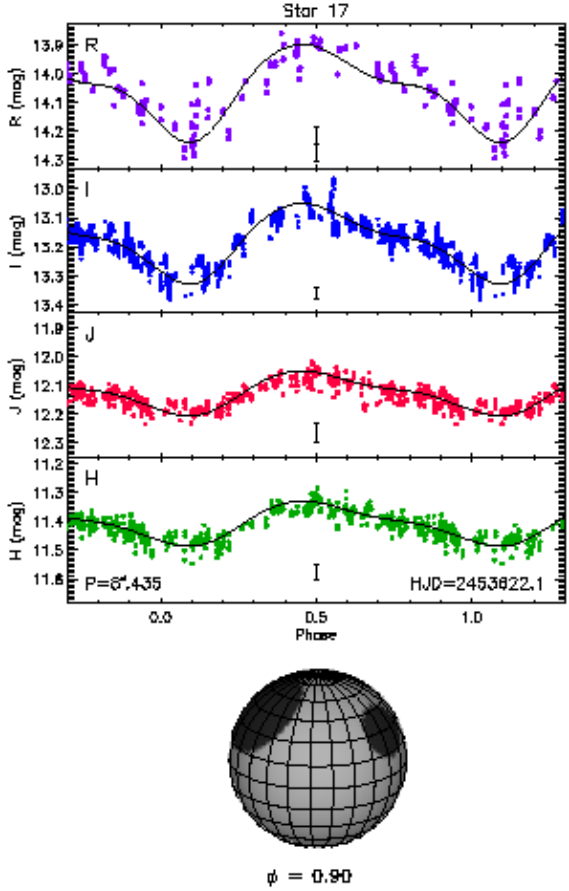}
\vspace{0.3cm}
\caption{Observed (dots) and synthetic (full lines) $R$, $I$, $J$, and $H$ light curves of 
star \#17. The mean error-bar is also displayed in the lower mid part of each box. 
A schematic map of the starspot distribution, as seen at the phase of maximum spot visibility, 
is also shown. An inclination of the rotation axis $i=70\degr$ was adopted.} 
\label{fig:palle_17}
\end{figure}

For the star \#17, we assumed an inclination $i=70\degr$ and obtained a spot temperature of 
$T_{\rm sp}/T_{\rm ph}\simeq 0.71\pm 0.03$ from the solution grids in the $RIJH$ bands displayed in 
Fig.~\ref{fig:Teff_aree_17}. This star shows spot properties similar to \#14 and \#10.
%
%
In Fig.~\ref{fig:palle_17}, the observed $RIJH$ light curves and the simultaneous solution provided by MACULA are 
displayed together with a map of the spotted photosphere, as seen at the rotational phase of maximum spot
visibility.

\begin{figure}  
\hspace{-.5cm}
   \includegraphics[width=9cm]{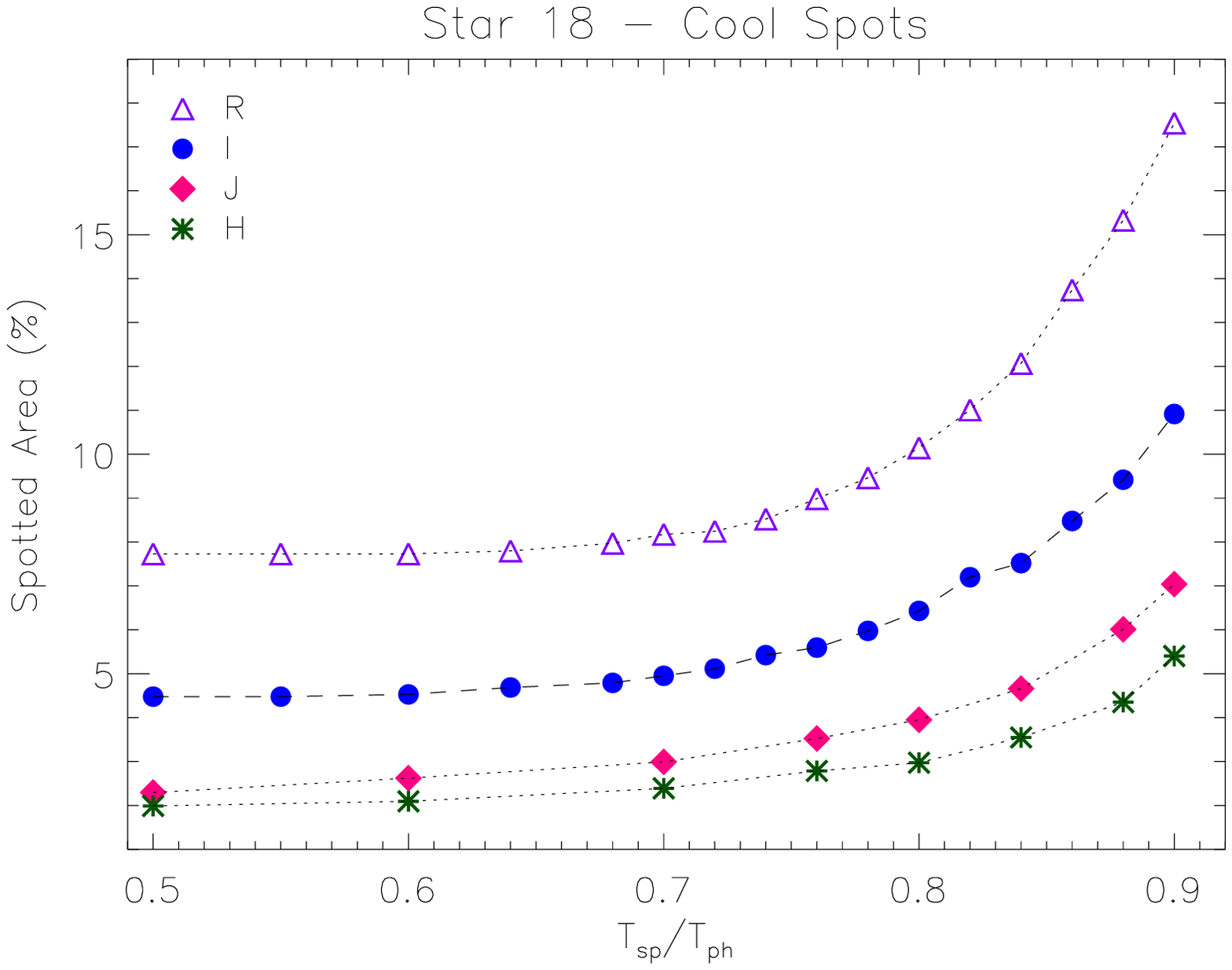}
   \includegraphics[width=9cm]{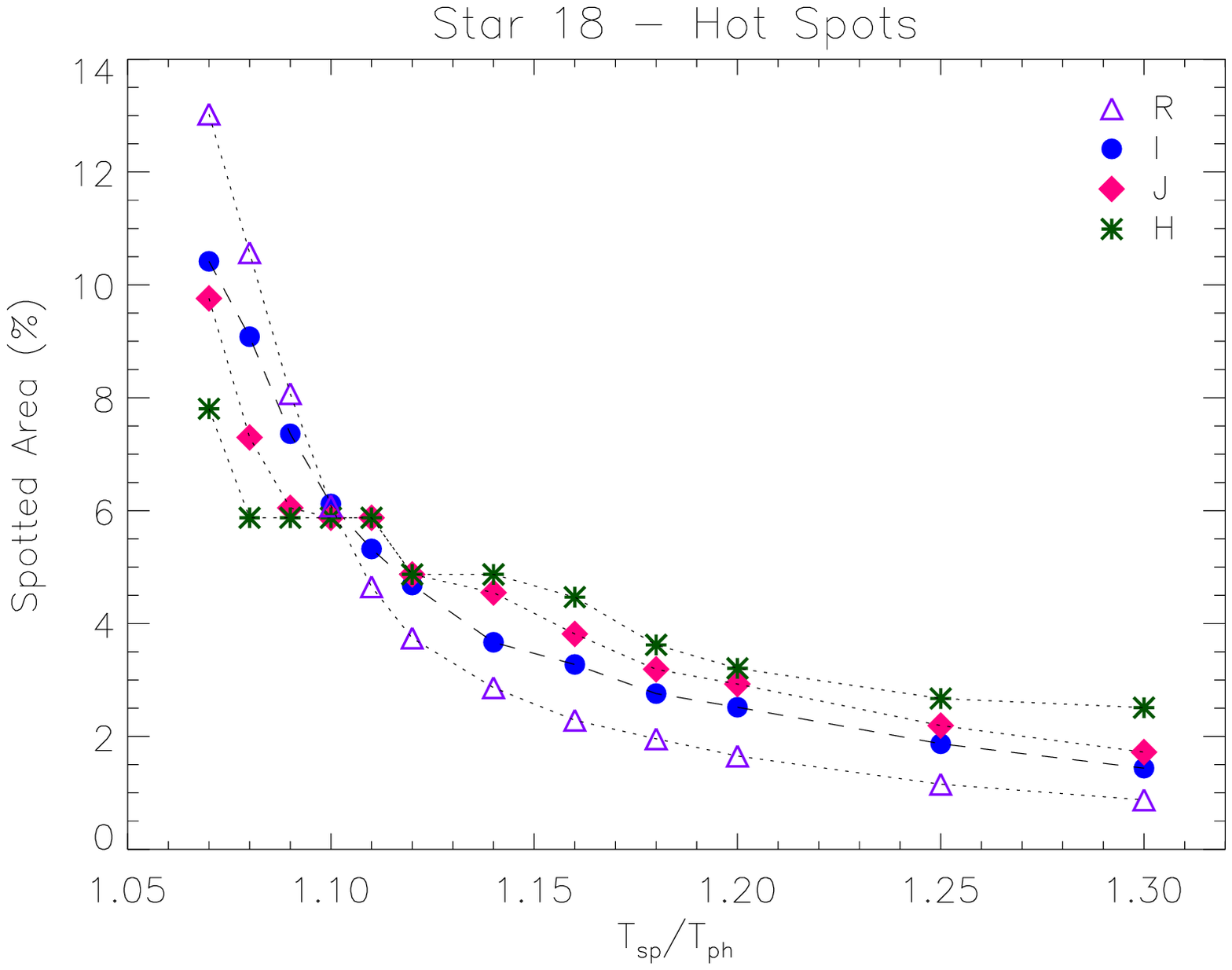}
\vspace{0cm}
\caption {{\it Top)} Grids of cool-spot solutions of $RIJH$ light curves for star \#18. 
Different symbols have been used for each pass-band. {\it Bottom)} Grids of hot-spot solutions. }
\label{fig:Teff_aree_18}
\end{figure}

We also tried simultaneous solutions of the $RIJH$ bands for star \#18. We were able to fit the 
individual light curves with a model with two cool spots but the grids of solutions for different bands do not 
show any intersection, as seen in the top panel of Fig.~\ref{fig:Teff_aree_18}. This is due to the 
strong decrease of the amplitude of the light curves that goes from 0$\fm$225 to 0$\fm$110 from $R$ to 
$I$ and becomes as low as 0$\fm$055 in $J$ and $H$ bands. There is no value of $T_{\rm sp}/T_{\rm ph}$
that allows such a rapid amplitude decrease.    
As shown by \citet{Bouvier89} for DF~Tau, a steep decrease of the modulation amplitude with the
increasing wavelength can be reproduced only  with hot spots.
Thus, we tried to model the light curves with hot spots and found intersection of all the solution grids for a value
of $T_{\rm sp}/T_{\rm ph}=1.10 \pm 0.02$ corresponding to spots 360$\pm 70$ K hotter than the surrounding 
photosphere. 
This does not imply that cool spots are not present on the photosphere of this star, but the main responsible
for the observed variations are hot spots.

We cannot state whether these features are tied to accretion because we have neither 
Hectochelle spectra nor mid-IR photometry, but the relatively long rotation period ($P_{\rm rot}\simeq 12.5$ days)
and the position on the HR diagram are consistent with a very young disc-locked object still in an active 
phase of mass accretion.

\begin{figure}
\hspace{-1.cm}
 \includegraphics[width=12cm]{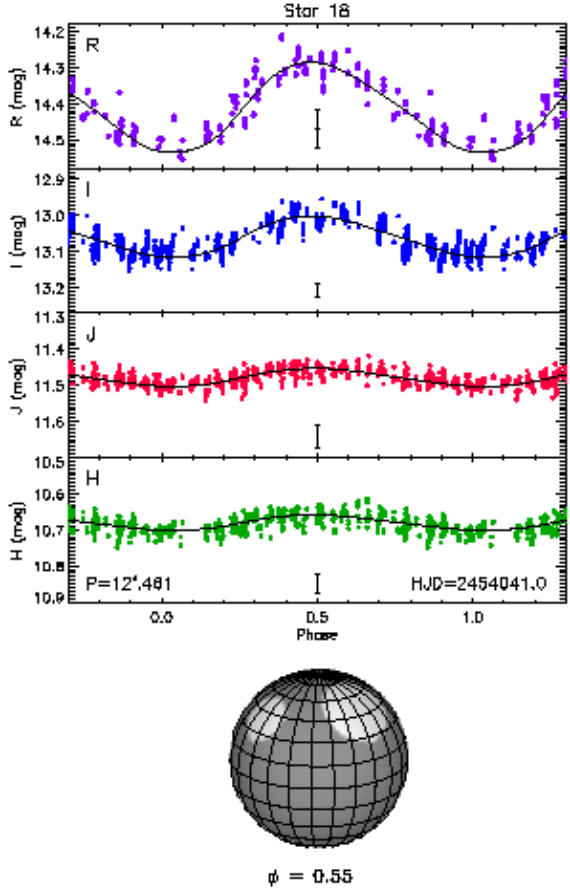}
\vspace{0.3cm}
\caption{Observed (dots) and synthetic (full lines) $R$, $I$, $J$, and $H$ light curves of 
star \#18. The mean error-bar is also displayed in the lower mid part of each box. 
A schematic map of the hot-spot distribution, as seen at the phase of maximum spot visibility, 
is also shown. An inclination of the rotation axis $i=70\degr$ was adopted.} 
\label{fig:palle_18}
\end{figure}

 The $RIJH$ light curves and the simultaneous solution provided by MACULA are 
displayed together with a map of the hot spots in Fig.~\ref{fig:palle_18}.

\begin{figure}  
\hspace{-.5cm}
   \includegraphics[width=9cm]{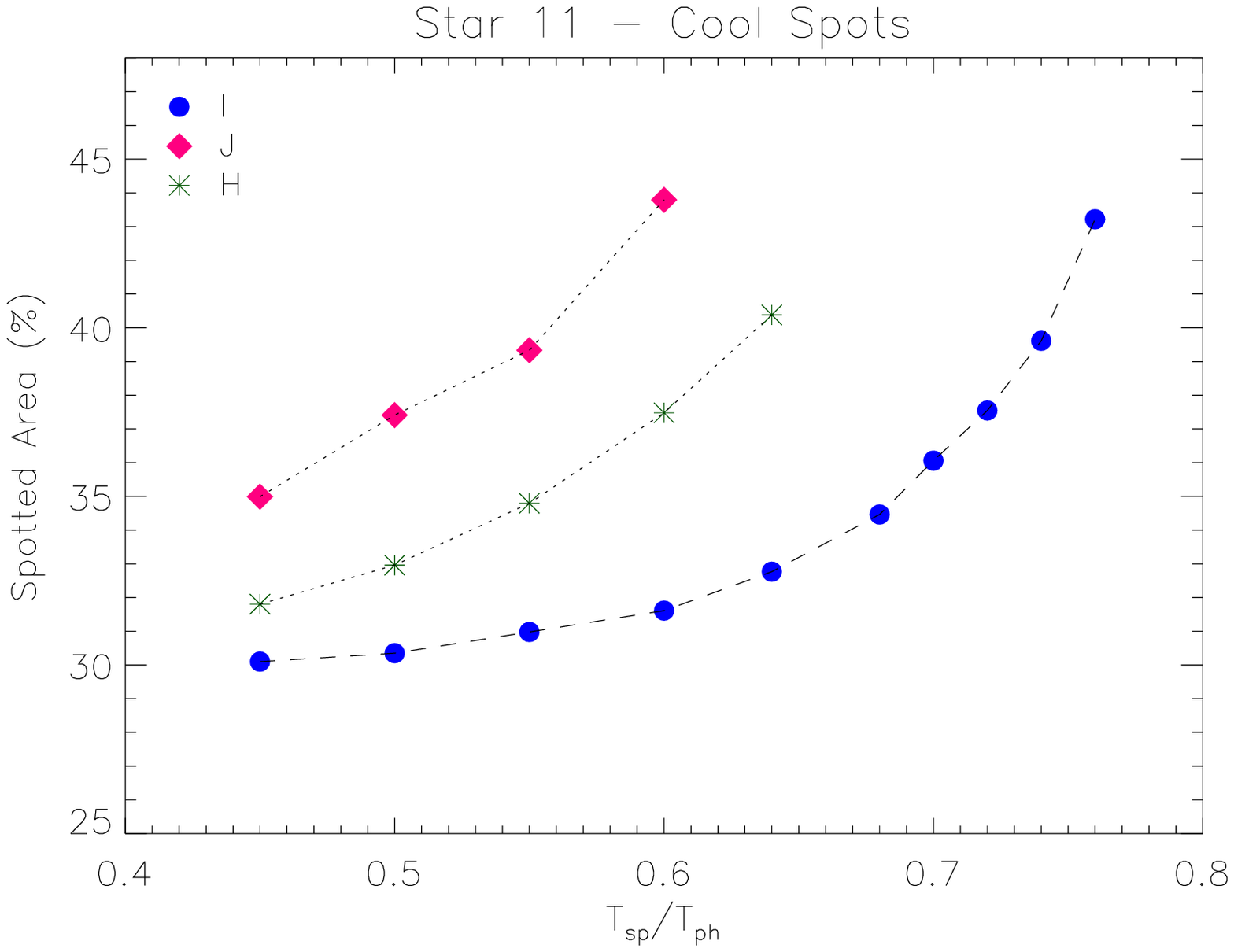}
   \includegraphics[width=9cm]{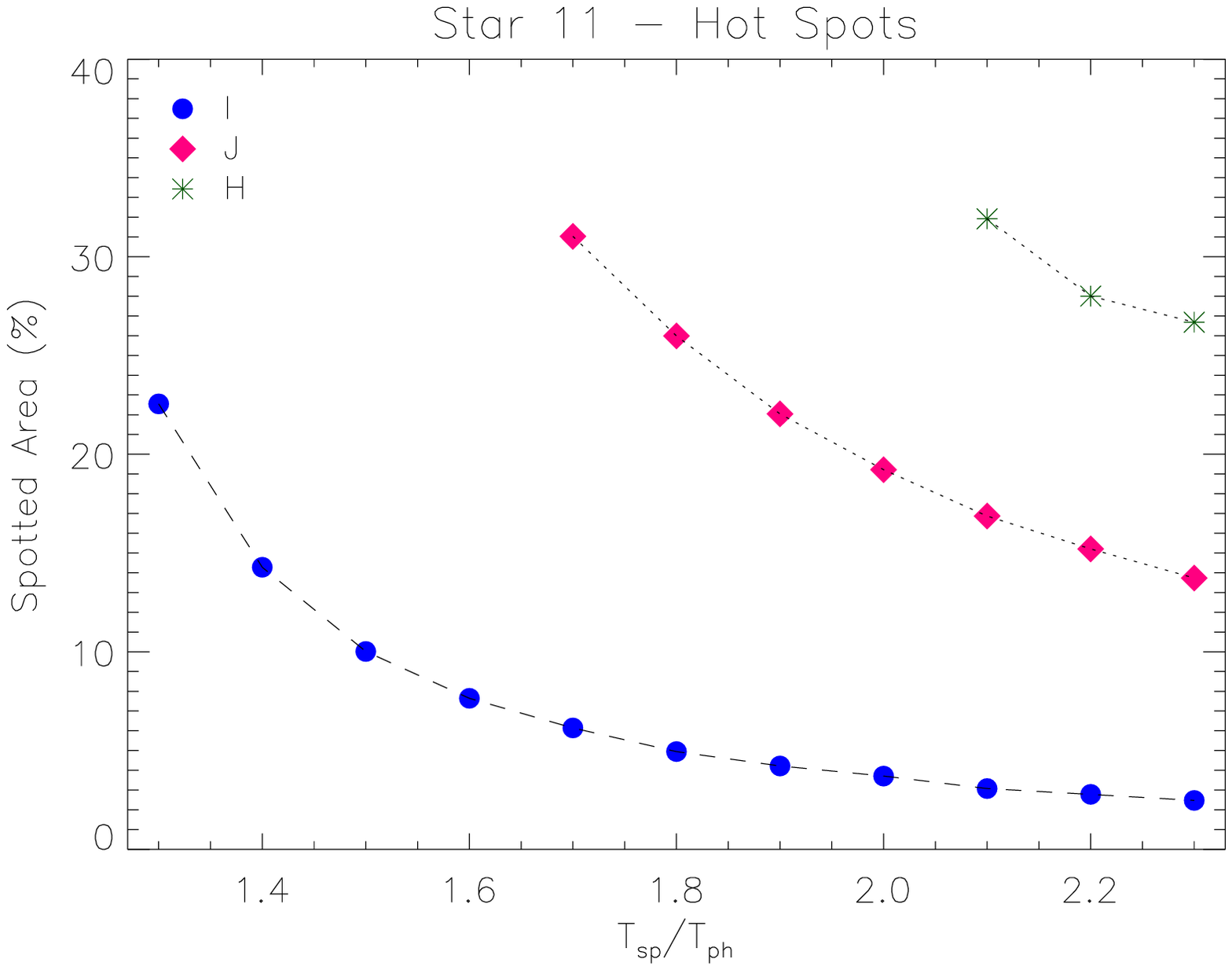}
\vspace{0cm}
\caption {{\it Top)} Grids of cool-spot solutions of $RIJH$ light curves for star \#11. 
Different symbols have been used for each pass-band. {\it Bottom)} Grids of hot-spot solutions. }
\label{fig:Teff_aree_11}
\end{figure}

The last object for which we attempted to model the rotational modulation with MACULA is star \#11, the one 
with the largest modulation amplitude in our sample. This object is very red and too faint in the $R$ band
for obtaining a suitable light curve. Only an indication of a modulation amplitude $\Delta R\simeq 0\fm 97$ can 
be derived. The $I$, $J$, and $H$ curves are instead fully usable.
We initially tried with cool spots and found solutions with very large active regions, covering a large fraction
of a stellar hemisphere, as expected for getting variation amplitudes of about 1 magnitude.
Anyway, as seen in the top panel of Fig.~\ref{fig:Teff_aree_11}, there is no intersection between the grids for 
different bands.
We also tried with hot spots without success (Fig.~\ref{fig:Teff_aree_11}, bottom panel). This result is not surprising 
if we consider that the amplitude of the light curves has a reverse trend, as a function of wavelength, compared 
to the rotational modulation produced by cool or hot spots for the other sources. For star \#11, the modulation amplitude 
slightly increases with the central wavelength of the band.
A possible explanation for this behaviour is that the main source of variation is not the stellar photosphere but the 
accretion disc from which most of the NIR flux originates (see Fig.~\ref{Fig:SED}). Inhomogeneities/condensations 
in the disc could give rise to wavelength-independent or increasing amplitudes, depending on the ratio of stellar to disc
luminosity at the wavelength of observation. Indeed, star \#11 is the object with the largest IR excess in our sample.

\section{Discussion}	\label{sec:discussion}

Having detected rotation periods for 29 stars in a field flanking the ONC and derived
spot parameters for a few of them, we now shortly discuss our results accounting for the 
evolutionary status of the investigated sources and the evidences of active accretion
and circumstellar discs for some of them. Moreover, the characteristics of the strong flare
of star \#19 (\object{2MASS\,J05352973-0548450}) are discussed  in some details.

As shown in Table~\ref{tab:param}, forty-three stars are likely members (either photometric and/or kinematic) 
of the Orion cloud, and their positions in the HR diagram appear consistent with those 
of PMS stars. 
Furthermore, our analysis of the SEDs reveals that five of these stars (namely, \#10, \#11,
\#59, \#61, and \#93) have significant IR excesses, signature of circumstellar discs. 
Star \#10 was reported by \citet{Sicilia} as a possible cTTS. 
This classification would be consistent with the IR excess of the object.
Stars \#59 and \#61 exhibit, beside to the NIR excess, erratic photometric variations, 
quite common in T\,Tauri stars.

\subsection{Rotation period distribution}	\label{sec:histograms}

\begin{figure}
	 \includegraphics[width=8.5cm]{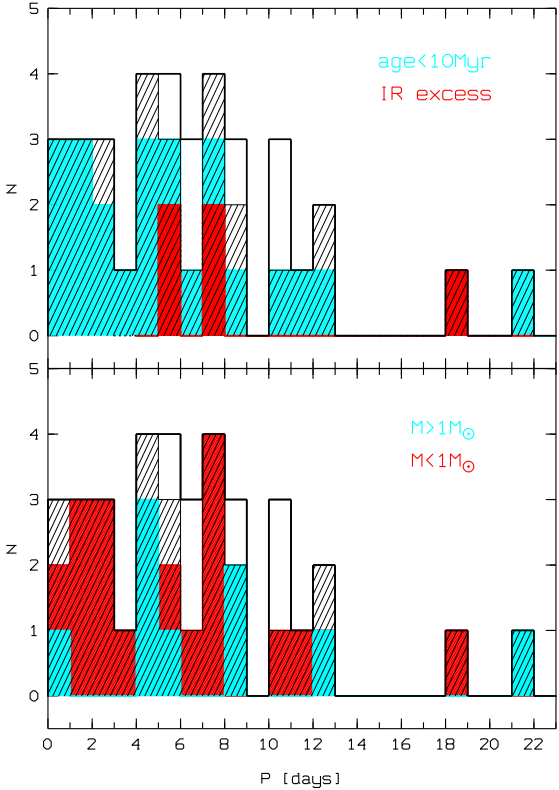}
\vspace{-0.3cm}
\caption{The two panels show the distributions of rotation periods for stars
in the REM field (thick line), and for the subsample classified as Orion
members (hatched area).
The upper panel highlights the stars younger than 10\,Myr (light-grey/cyan
shaded area) and the objects with NIR excess (dark-grey/red shaded area).
The lower panel highlights the members with masses respectively less than
(dark-grey/red) and higher than (light-grey/cyan) 1\,M$_{\odot}$. }
\label{fig:hist_Prot}
\end{figure}

 The distribution of rotation periods is shown in Fig.~\ref{fig:hist_Prot}.
From the upper panel, we note that the majority of the objects with detected period are classified 
as Orion members. The latter are dominated by stars younger than 10\,Myr, among which five objects 
with significant NIR excess are found. 
Notably, none of these stars with NIR excess has a rotation period shorter than 5\,days, 
in agreement with the hypothesis that the angular momentum of the stars is moderated 
by interaction with the disc.
The lower panel shows instead the two subsamples with masses less than and higher than 
 1\,$M_\odot$, respectively, but we do not see any obvious dependency upon mass.
The poor statistics prevent us to unambiguously distinguish the bimodal distribution with peaks 
near 2 and 8 days clearly displayed by ONC members \citep{Herbst02}.

\subsection{Spot properties}

 In most cases, the amplitudes of the modulation detected in different photometric bands 
decrease at longer wavelengths, with a wavelength dependency that is consistent with the presence 
of cool photospheric spots, except for star \#18, whose behaviour can only be reproduced 
with hot spots. 
Two stars (namely, \#11, and \#91) show instead either increasing or almost invariant amplitudes 
at longer wavelengths.
Star \#11, which exhibits the broadest modulation in our sample, shows 
light-curve amplitudes that increase from 0.85\,mag in $V$ to 1.25\,mag in $J$ and slightly 
decrease in $H$ and $K$. 
This star also shows a remarkable excess emission in the mid-IR, hence the cause of variability 
might reside in the inner circumstellar disc (e.g., possible disc inhomogeneities/instabilities). 
Indeed, the observed modulation in the $IJH$ bands could not be modelled neither with cool
nor with hot starspots. 
Star \#91, with relatively small photometric variations, has almost invariant light-curve amplitudes.
In contrast, star \#10, which displays the typical signatures of a cTTS, exhibits modulation 
amplitudes that decrease systematically from $R$ to $H$ and are well reproduced by cool spots with 
$T_{\rm sp}/T_{\rm ph}\simeq 0.76$. 

For five stars with fairly good light curves in the $RIJH$ bands we could find the spot parameters from
the simultaneous solution of the multi-band light curves (Sect.~\ref{sec:spot}).
These data are summarized in Table~\ref{tab:spot_parameters}. Apart from star \#18, we found cool spots.
It is interesting to note that the temperature of the cool spots is nearly the same ($ T_{\rm sp}\simeq 3100$\,K) for all the 
four stars regardless of the photospheric temperature. This average value is noticeably lower than sunspot umbrae ($T\approx 3800$\,K).
However, it is not easy to define the best parameter describing the blocking 
effect on energy convective transport produced by the spot magnetic fields. For a more meaningful comparison, the temperature ratio, 
$T_{\rm sp}/T_{\rm ph}$, or the temperature difference, $\Delta T=T_{\rm ph}-T_{\rm sp}$, should be adopted.  
As seen in Table~\ref{tab:spot_parameters}, for the three stars with fairly long rotation periods (\#10, \#14, and \#17) $\Delta T$ is
in the range 1000--1300\,K, i.e. smaller than sunspot umbrae but slightly larger than values of 400--900\,K found for the 
giant and sub-giants components of RS~CVn binaries adopting a similar approach \citep{Frasca05,Frasca08}. Conversely, the spotted area
(8--10\% of the star surface) is smaller, on average, than that found by \citet{Frasca05,Frasca08} for RS~CVn stars ($A_{\rm rel}$=11--18\%). 
Our results are in agreement with \citet{Bouvier89}, which, from multi-band photometry, find cool spots in T~Tau stars with
temperature mostly in the range 700--1200\,K and filling factors ranging from about 5\% to 10\%.
The difference between these
Orion population stars and the Sun is likely due to the clear tendency for spots to have a larger contrast with respect 
to the photosphere in hotter stars \citep[][and references therein]{Berdy05}.
However, the higher temperature contrast compared to RS~CVn stars is not due to this effect because our PMS stars have $T_{\rm eff}$ in the range 
4100--4400\,K, hence cooler than the aforementioned RS~CVn components ($T_{\rm eff}\sim$\,4600--4900\,K).
The different internal structure of the fully convective PMS stars, compared to sub-giant stars, could be responsible for such
behaviour.
The much lower $\Delta T$ found for star \#9 is in line with its lower effective temperature but could also be tied to its 
very high rotation rate.

\begin{table*}  
\caption{Spot parameters$^{*}$. }
\label{tab:spot_parameters}
\begin{tabular}{rccrrcrrclrl}
\hline\hline
\noalign{\medskip}
Id   &  $i$      & $R_1$     &  $Lon_1$  &  $Lat_1$  &  $R_2$    &  $Lon_2$  & $Lat_2$   &  $T_{\rm sp}/T_{\rm ph}$ & \multicolumn{1}{c}{$T_{\rm sp}$} & $\Delta T$  & \multicolumn{1}{c}{$A_{\rm rel}$}\\ 
     & ($\degr$) & ($\degr$) & ($\degr$) & ($\degr$) & ($\degr$) & ($\degr$) & ($\degr$) &                          & \multicolumn{1}{c}{(K)}          &     (K)     &     \multicolumn{1}{c}{(\%)}     \\ 
\noalign{\medskip}
\hline
\noalign{\medskip}
\#9  &  45   &  20.4    &     42    &    45     &    13.3   &    170    &    40	  &   0.92$\pm$0.03   & 3240$\pm$100 &  270    &    5.0$^{+3.5}_{-1.5}$	\\
\#10 &  90   &  27.6    &    126    &	  0     &    15.6   &	  55    &     0	  &   0.76$\pm$0.02   & 3120$\pm$~80 &  990    &    7.5$\pm$0.5	\\
\#14 &  70   &  29.3    &    240    &    45     &    18.2   &    150    &    25	  &   0.76$\pm$0.04   & 3230$\pm$170 & 1020    &    8.0$\pm$1       \\
\#17 &  70   &  31.6    &     33    &    45     &    18.3   &    265    &    30	  &   0.71$\pm$0.03   & 3120$\pm$130 & 1280    &    9.5$\pm$1.5   \\
\#18 &  70   &  23.6    &    155    &    50     &    16.0   &    252    &    40	  &   1.10$\pm$0.02   & 3960$\pm$~70 & $-$360  &    6.0$\pm$1       \\
\noalign{\medskip}
\hline
\end{tabular}
\begin{flushleft}
$^{*}$ $R_1$, $Lon_1$, and $Lat_1$  are radius, longitude, and latitude of the larger spot. $R_2$, $Lon_2$, and $Lat_2$ 
are the same for the smaller spot. The spot temperature is the same for both spots and is expressed in units of the photospheric temperature,
$T_{\rm sp}/T_{\rm ph}$, as well as in Kelvin degrees. The difference $\Delta T=T_{\rm ph}-T_{\rm sp}$ is also listed. The total spot area,
$A_{\rm rel}$, is expressed in percent units of the star surface.
\end{flushleft}
\end{table*}
\normalsize

\subsection{Flare events in 2MASS\,J05352973-0548450 and 2MASS~J05352550-0545448}
 Star\,\#\,19 (2MASS\,J05352973-0548450) is a known variable (\object{V498\,Ori}), already classified both as a flare star 
(\object{Parenago\,2078}), and an emission-line star (\object{Haro\,4-365}, PaCh\,274,\object{ Kiso\,A-0976 174}).

Here we report the serendipitous detection, on the 7th of December 2006 at 4:41\,UT, 
of a remarkable flare event from star\,\#\,19 in the $I$ band (see Fig.~\ref{fig:flare}). 
The peak amplitude ($\Delta\,I\simeq0.3$\,mag) and the duration of the flare are comparable 
to those observed by \citet{Scholz05} in a rapidly rotating ($P=0.61$ days), very low-mass 
(0.09\,$M_{\sun}$) member of the $\epsilon$\,Ori cluster. 
Though, we were able, thanks to the prolonged and dense temporal sampling of our $I$-band
observations, to follow the entire event and resolve both the rise and decay phases.

The luminosity in the $I$ band at the flare peak, $L_{I}^{\rm peak}$, can be evaluated from
the equation
\begin{equation}
\label{Eq:flare_erg}
L_{I}=4\pi d^2\cdot 10^{-0.4(I-A_I)} F_I W \mathrm{(erg\,s^{-1})}
\end{equation}
{\noindent where $d=414$\,pc is the distance to the star, $I$ is the apparent magnitude of the 
star, $A_I=0.58A_V$ is the extinction in the $I$ band, $F_I=1.196\times10^{-9}$ erg\,cm$^{-2}$\,s$^{-1}$\AA$^{-1}$ 
\citep{Lamla82} is the Earth flux (outside the atmosphere) in the $I$ band from a zero-magnitude 
star, and $W \approx 1000$\,\AA\  is the $I$ band-width.}
We find a flare peak luminosity $L_{I}^{\rm peak}\simeq 1.7\cdot10^{32}$ erg\,s$^{-1}$. The stellar luminosity before
the flare $L_{I_0}\simeq 5.3\cdot10^{32}$ erg\,s$^{-1}$, corresponding to a magnitude $I_0=11\fm91$, has been subtracted.

The energy released during the flare in the $I$ band, $E_I\simeq 2.7\cdot10^{35}$ erg, was computed by integrating 
the excess luminosity (over the pre-flare level) all along the duration of the flare.
The flare results to be a very powerful one.	

We estimated the coverage factor of the flaring area following the guidelines of \citet{Hawley95}. They showed that
the continuum emission from the near-UV to the R band of strong flares in dMe stars is well reproduced by a black-body 
spectrum with a temperature of $\sim$ 9000\,K. This is also in agreement with the findings of \citet{Machado74} for the
continuum emission in a solar flare that they find consistent with thermal emission from electrons at $T_e=8500\pm500$ K in the upper photosphere.
According to Eq.~3.1 of \citet{Hawley95}, one can estimate the temperature of the flaring region and the area coverage at the flare 
peak with observations in at least two bands. From $I$ and $J$ light curves of the flare of V498\,Ori we could derive both parameters.
Given the relatively large errors of the $J$ magnitudes ($\sim0.03$\,mag) compared to the flare intensification in the $J$ band
($\Delta J\sim0.09$\,mag), we
preferred to fix the temperature to $T_{\rm flare}=9000$\,K and deduced an area coverage $X$ of about 14\%. This is much larger than the 
typical values of $\sim 0.01$\% found by \citet{Hawley95} in dMe stars. An estimate of the total flare optical luminosity is then given by
$L_{\rm opt}= X\cdot 4\pi R^2\sigma T_{\rm flare}^4 \sim 1.6\cdot10^{34}$ erg\,s$^{-1}$ and the optical flare energy
turns out to be $\sim 6\cdot10^{36}$ erg. These values are about 3 orders of magnitude larger than those derived by \citet{Hawley95}
for an optical flare of the dMe star \object{AD~Leo}.

Other parameters characterizing the flare are the rise and decay time-scales. 
We fitted the rise and decay phase of the flare light curve with an exponential function 
of the form
\begin{equation}
I(t) = \Delta I e^{[(t-t_{\rm peak})/\tau)]} + I_0 
\end{equation}
\noindent{We found as rise and decay e-folding times the values $\tau_{r}\simeq 2.5$\,min and $\tau_{d}\simeq 24.9$\,min, respectively.}

\begin{figure}[tbh]
    \centering{\hspace{1cm} \includegraphics[width=7.5cm,height=9cm]{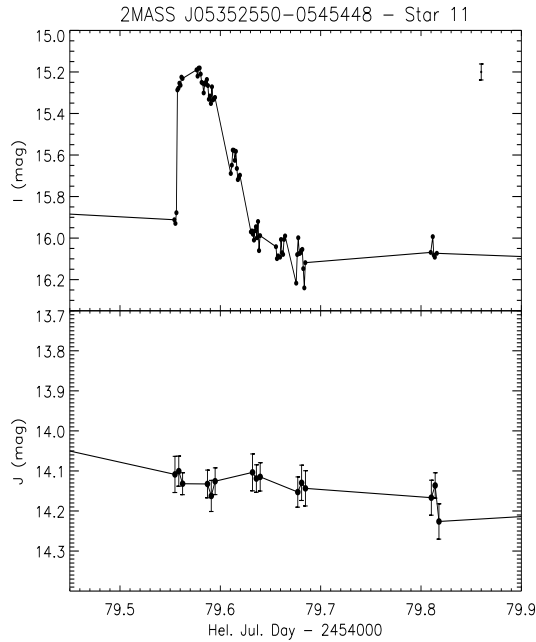} }
    \caption{The flare event observed on Star \#11 (2MASS~J05352550-0545448 = NR~Ori) on December 10th, 2006 at 3:30 UT in 
    the $I$ band (\textit{upper panel}).
    The simultaneous $J$ light curve (\textit{bottom panel}) does not show any clear variation.}
    \label{fig:flare_11}
    \end{figure}

We applied the same analysis to the flare observed on the 10th of December 2006 at 1:30\,UT 
on star\,\#\,11 (2MASS~J05352550-0545448 = \object{NR~Ori}) in the $I$ band (see Fig.~\ref{fig:flare_11})\footnote{Available in electronic 
form only.}. 
The peak amplitude, $\Delta\,I\simeq0.85$\,mag, was higher than in the previous event of star\,\#\,19 but the duration 
of the flare ($\simeq$\,110 minutes) was nearly the same. We found rise and decay e-folding times of $\tau_{r}\simeq 6$\,min and $\tau_{d}\simeq 59$\,min, 
respectively. No flux intensification was observed in the $J$ band.
The peak luminosty and the energy released in the $I$ band are $L_{I}^{\rm peak}\simeq 1.4\cdot10^{31}$ erg\,s$^{-1}$ and 
$E_I\simeq 6.4\cdot10^{34}$ erg, i.e. smaller than those derived for the flare on star\,\#\,19, notwithstanding the huge $I$-flux intensification.
This is due to contrast reasons, being star\,\#\,11 intrinsically much fainter than star\,\#\,19.
Assuming a flare temperature $T_{\rm flare}=9000$\,K, we found an area coverage factor $X\simeq$\,3\%. 
The total flare luminosty and the energy released at optical wavelengths, calculated as for star\,\#\,19, are
$L_{\rm opt}\sim 9.7\cdot10^{32}$ erg\,s$^{-1}$ and $E_{\rm opt}\sim 4.4\cdot10^{34}$ erg\,s$^{-1}$.

We can finally attempt a very rough estimate of the flare frequency among the stars 
in the REM field that we classify as members and likely members of the young Orion population, 
using the time interval covered by observations ($\approx50$\,hours) and the number of objects 
and found a value of the order of $8\times10^{-4} {\rm h}^{-1}$. 

\section{Conclusions}	
\label{sec:conclusions}

We presented the results of an intensive photometric monitoring of a 10$\arcmin\times 10\arcmin$ field 
flanking the Orion Nebula Cluster (ONC) conducted during three consecutive months. 
The main results of our work can be summarized as follows:
\begin{itemize}
\item We detected rotation periods for 29 stars, spanning from about 0.6 to 20 days, sixteen of which are new 
periodic variables. Thanks to the relatively long time-baseline we measured the periods with sufficient accuracy
($\Delta P/P<2\%$) also for the slowest rotators. 
We remark that none of the stars with NIR excess has a rotation period shorter than 5\,days, 
in agreement with the hypothesis that the angular momentum of the stars is moderated 
by interaction with the disc.

\item The analysis of the spectral energy distribution and, for some stars, the high-resolution spectra provided 
us with $T_{\rm eff}$ and luminosity, and that allowed us to construct the HR diagram of these stars. 
We could then assign a photometric membership to the objects by a comparison with PMS evolutionary tracks and 
derive masses and ages. The majority of the objects with detected period are classified as Orion members and result 
to be younger than 10\,Myr.

\item Our spot modelling code enabled us to derive the starspot properties for five of these star, based on the 
simultaneous analysis of the light curves in several optical and NIR bands. 
For one of these, the light curves could only be modelled with hot spots, which are likely related to magnetospheric
accretion. For three stars with $T_{\rm eff}$ in the range 4100--4400\,K and rotation periods between 4 and 8 days
we found cool spots with $\Delta T=T_{\rm ph}-T_{\rm sp}$ in the range 1000--1300\,K, i.e. larger, on average,  
than $\Delta T$ typically found for the sub-giant and giant components of RS~CVn systems. Conversely, the spotted area
(8--10\% of the star surface) is smaller. These differences could be due to the fully convective internal structure 
of PMS stars which gives rise to a different dynamo action.
For the cool ($T_{\rm eff}\simeq 3500$\,K) and ultra-fast ($P\simeq 0.56$\,days) star \#9, a smaller spot 
contrast ($\Delta T\simeq 270$\,K) is found. This could be due both to the lower effective temperature and the high rotation rate.

For the star with the highest modulation amplitudes (star \#11), which also displays a remarkable near- and mid-IR excess, the 
light curves were not consistent neither with hot nor with cool spots.
We suggest that the flux modulation could be produced by inhomogeneities of the circumstellar disk.

\item A very strong flare was detected on star \#19 (V498\,Ori) in the $I$ band. The temporal evolution of the event was fully 
resolved and we evaluated the rise and decay e-folding times as $\tau_{r}\simeq 2.5$\,min and $\tau_{d}\simeq 24.9$\, min, respectively.
We estimated an energy released in the $I$ band of nearly $3\cdot10^{35}$ erg and a 20 times higher energy released in the 
optical continuum. Another strong flare, which released an energy of about $6\cdot10^{34}$\,erg in the $I$ band, was observed on  
star\,\#\,11 (\object{NR~Ori}), which is cooler and less massive than star \#\,19 and displays the broadest modulation and a strong infrared excess.

\end{itemize}

\begin{acknowledgements}
We are grateful to the anonymous referee for very useful comments and suggestions that helped to improve the manuscript.
We acknowledge the REM team for technical support, and in particular Stefano Covino and Emilio Molinari, 
for their help in setting-up the observations. 
We thank Dr. Aurora Sicilia-Aguilar for providing us with 
Hectochelle spectra of some of our targets.
This research has made use of SIMBAD and VIZIER databases, operated at CDS, Strasbourg, France. 
We acknowledge financial support from INAF and Italian MIUR.

\end{acknowledgements}

\bibliographystyle{aa}

\Online

\onecolumn

\begin{landscape}
\begin{longtable}{ r l l l l l l l l l l l l l l l l l }
\caption[ ]{Standard $VRIJHK'$ magnitudes of all the sources.} \\ 
\scriptsize
\label{tab:VRIJHK} 
\begin{tabular}{ r l l l l l l l l l l l l l l l l l } 
\hline\hline
  Id  &   2MASS		    &   $V$     &  err	  &  $R$     &   err    &  $I$     &   err    &   $J$    &  err     &  $H$     &   err    &   $K$    &   err    & [3.6] & [4.5] & [5.8] & [8.0] \\ 
      &                     &  (mag)   &   (mag)  &  (mag)   &  (mag)   & (mag)    &  (mag)   &   (mag)  &  (mag)   &  (mag)   &  (mag)   &  (mag)   &  (mag)   & (mag) & (mag) & (mag) & (mag)	\\ 
\hline	     			      	   	 	    	           	  	     	   	   	   	      	   	       
   1  &  05352184-0546085   &  ...     &   ...    &  19.21   &   0.43   &  17.28   &   0.17   &   14.71  &   0.06   &  14.09   &   0.05   &  13.75   &   0.12	&  ...  &  ...  &  ...  &  ...  \\ 
   2  &  05352029-0546399   &  16.30   &   0.08   &  15.13   &   0.06   &  13.89   &   0.02   &   12.46  &   0.03   &  11.76   &   0.03   &  11.57   &   0.05	&  ...  &  ...  &  ...  &  ...  \\ 
   3  &  05352021-0546510   &  15.72   &   0.08   &  14.63   &   0.06   &  13.39   &   0.02   &   12.01  &   0.03   &  11.27   &   0.03   &  11.07   &   0.04	&  ...  &  ...  &  ...  &  ...  \\ 
   4  &  05352007-0545526   &  16.12   &   0.08   &  15.23   &   0.06   &  14.38   &   0.02   &   12.99  &   0.03   &  12.37   &   0.03   &  12.13   &   0.05	&  ...  &  ...  &  ...  &  ...  \\ 
   5  &  05351981-0545409   &  16.92   &   0.07   &  15.73   &   0.06   &  14.26   &   0.02   &   12.70  &   0.03   &  12.03   &   0.03   &  11.80   &   0.05	&  ...  &  ...  &  ...  &  ...  \\ 
   6  &  05351903-0546163   &  16.26   &   0.08   &  15.18   &   0.06   &  13.81   &   0.02   &   12.36  &   0.03   &  11.64   &   0.03   &  11.43   &   0.05	&  ...  &  ...  &  ...  &  ...  \\ 
   7  &  05351924-0547012   &  18.60   &   0.11   &  17.03   &   0.09   &  15.26   &   0.03   &   13.51  &   0.04   &  12.92   &   0.03   &  12.65   &   0.06	&  ...  &  ...  &  ...  &  ...  \\ 
   8  &  05351950-0547457   &  16.72   &   0.08   &  15.57   &   0.06   &  14.27   &   0.02   &   12.93  &   0.03   &  12.22   &   0.03   &  12.03   &   0.05	&  ...  &  ...  &  ...  &  ...  \\ 
   9  &  05352281-0544428   &  16.31   &   0.08   &  14.97   &   0.06   &  13.52   &   0.02   &   11.78  &   0.03   &  10.90   &   0.03   &  10.62   &   0.04	& 10.45 & 10.49 & 10.50 & 10.43 \\ 
  10  &  05352617-0545084   &  16.05   &   0.08   &  14.99   &   0.06   &  13.97   &   0.02   &   12.49  &   0.03   &  11.63   &   0.03   &  11.27   &   0.05	& 10.55 & 10.23 &  9.72 &  8.96 \\ 
  11  &  05352550-0545448   &  16.80   &   0.07   &  16.33   &   0.08   &  15.46   &   0.04   &   13.45  &   0.04   &  12.10   &   0.03   &  11.10   &   0.05	&  9.93 &  9.56 &  9.22 &  8.39 \\ 
  12  &  05351695-0545558   &  17.17   &   0.07   &  16.01   &   0.06   &  14.21   &   0.02   &   12.28  &   0.03   &  11.59   &   0.03   &  11.28   &   0.05	&  ...  &  ...  &  ...  &  ...  \\ 
  13  &  05351621-0547201   &  15.89   &   0.08   &  14.97   &   0.06   &  14.07   &   0.02   &   12.68  &   0.03   &  11.99   &   0.03   &  11.77   &   0.05	&  ...  &  ...  &  ...  &  ...  \\ 
  14  &  05351192-0545379   &  15.23   &   0.08   &  14.21   &   0.06   &  13.19   &   0.02   &   11.60  &   0.03   &  10.80   &   0.03   &  10.55   &   0.04	& 10.27 & 10.37 & 10.29 & 10.50 \\ 
  15  &  05351290-0545376   &  18.43   &   0.16   &  17.10   &   0.09   &  15.40   &   0.04   &   13.31  &   0.04   &  12.53   &   0.03   &  12.17   &   0.05	&  ...  &  ...  &  ...  &  ...  \\ 
  16  &  05351797-0549375   &  15.26   &   0.08   &  14.35   &   0.06   &  13.13   &   0.02   &   11.94  &   0.03   &  11.23   &   0.03   &  11.04   &   0.04	&  ...  &  ...  &  ...  &  ...  \\ 
  17  &  05350614-0545311   &  15.02   &   0.09   &  14.06   &   0.06   &  13.18   &   0.02   &   12.14  &   0.03   &  11.42   &   0.03   &  11.25   &   0.05	& 11.09 & 11.18 & 11.25 & 10.97 \\ 
  18  &  05353222-0544265   &  15.62   &   0.08   &  14.40   &   0.06   &  13.06   &   0.02   &   11.48  &   0.03   &  10.68   &   0.03   &  10.42   &   0.04	&  ...  &  ...  &  ...  &  ...  \\ 
  19  &  05352973-0548450   &  13.59   &   0.09   &  12.73   &   0.06   &  11.88   &   0.02   &   10.75  &   0.03   &  10.07   &   0.03   &   9.90   &   0.05	&  ...  &  ...  &  ...  &  ...  \\ 
  20  &  05350738-0548010   &  17.50   &   0.26   &  16.86   &   0.08   &  14.81   &   0.03   &   12.45  &   0.03   &  11.41   &   0.03   &  10.99   &   0.05	&  ...  &  ...  &  ...  &  ...  \\ 
  21  &  05351033-0546335   &  12.88   &   0.09   &  11.91   &   0.06   &  11.06   &   0.02   &    9.56  &   0.04   &	9.20   &   0.03   &   8.47   &   0.05	&  ...  &  ...  &  ...  &  ...  \\ 
  22  &  05351205-0547296   &  ...     &   ...    &  18.48   &   0.23   &  16.41   &   0.08   &   14.08  &   0.04   &  13.31   &   0.04   &  12.91   &   0.07	&  ...  &  ...  &  ...  &  ...  \\ 
  23  &  05351011-0548342   &  17.07   &   0.07   &  16.02   &   0.07   &  14.56   &   0.03   &   13.27  &   0.03   &  12.60   &   0.03   &  12.38   &   0.05	&  ...  &  ...  &  ...  &  ...  \\ 
  24  &  05351400-0549362   &  ...     &   ...    &   ...    &   ...    &  ...     &  ...     &    8.66  &   0.06   &   8.64   &   0.03   &   8.73   &   0.05   &  ...  &  ...  &  ...  &  ...  \\ 
  25  &  05351493-0549348   &  13.97   &   0.08   &  13.80   &   0.05   &  13.34   &   0.03   &   12.03  &   0.03   &  11.38   &   0.03   &  11.19   &   0.05   &  ...  &  ...  &  ...  &  ...  \\ 
  26  &  05352616-0548166   &  ...     &   ...    &   ...    &    ...   &  ...     &  ...     &    9.39  &   0.03   &   9.27   &   0.03   &   9.27   &   0.05   &  ...  &  ...  &  ...  &  ...  \\ 
  27  &  05352637-0544340   &  17.65   &   0.08   &  16.98   &   0.09   &  15.45   &   0.04   &   13.88  &   0.04   &  13.20   &   0.03   &  12.85   &   0.07	&  ...  &  ...  &  ...  &  ...  \\ 
  28  &  05351423-0543175   &  13.50   &   0.09   &  12.71   &   0.06   &  11.84   &   0.02   &   10.41  &   0.03   &	9.56   &   0.03   &   9.03   &   0.05	&  ...  &  ...  &  ...  &  ...  \\ 
  29  &  05351236-0543184   &  13.85   &   0.09   &  12.98   &   0.06   &  12.23   &   0.02   &   11.19  &   0.03   &  10.46   &   0.03   &  10.25   &   0.05	&  ...  &  ...  &  ...  &  ...  \\ 
  30  &  05351146-0544183   & ...      &   ...    &  17.65   &   0.12   &  15.82   &   0.05   &   13.80  &   0.04   &  13.09   &   0.03   &  12.76   &   0.06   &  ...  &  ...  &  ...  &  ...  \\ 
  31  &  05350575-0544001   &  16.63   &   0.08   &  15.48   &   0.06   &  14.09   &   0.02   &   12.83  &   0.03   &  12.12   &   0.03   &  11.87   &   0.05   &  ...  &  ...  &  ...  &  ...  \\ 
  32  &  05350567-0543046   & ...      &   ...    &  17.59   &   0.13   &  15.71   &   0.05   &   14.27  &   0.05   &  13.62   &   0.05   &  13.51   &   0.11   &  ...  &  ...  &  ...  &  ...  \\ 
  33  &  05350486-0544267   & ...      &   ...    &  18.97   &   0.41   &  17.85   &   0.32   &   15.20  &   0.08   &  13.57   &   0.05   &  12.69   &   0.06   &  ...  &  ...  &  ...  &  ...  \\ 
  34  &  05350278-0544429   & ...      &   ...    &   ...    &   ...    &  ...     &  ...     &   13.81  &   0.04   &  13.15   &   0.04   &  12.87   &   0.07   &  ...  &  ...  &  ...  &  ...  \\ 
  35  &  05350303-0545333&18.63$^{\rm b}$&   0.05 &   ...   & ... &15.46$^{\rm b}$ &   0.02   &   12.83  &   0.03   &  11.70   &   0.03   &  11.24   &   0.04   &  ...  &  ...  &  ...  &  ...  \\ 
  36  &  05350421-0545469   & ...      &   ...    &   ...    &   ...    &  ...     &  ...     &   14.12  &   0.05   &  13.03   &   0.04   &  12.62   &   0.06   &  ...  &  ...  &  ...  &  ...  \\ 
  37  &  05350215-0547200   & ...      &   ...    &   ...    &   ...    &  ...     &  ...     &   14.54  &   0.05   &  13.49   &   0.04   &  13.08   &   0.08   &  ...  &  ...  &  ...  &  ...  \\ 
  38  &  05350176-0548140   & ...      &   ...    &   ...    &   ...    &  ...     &  ...     &   14.65  &   0.06   &  13.42   &   0.04   &  12.70   &   0.06   &  ...  &  ...  &  ...  &  ...  \\ 
  39  &  05350406-0548540   &  17.12   &   0.07   &  16.01   &   0.07   &  14.50   &   0.03   &   13.40  &   0.04   &  12.70   &   0.03   &  12.53   &   0.06   &  ...  &  ...  &  ...  &  ...  \\ 
  40  &  05350627-0549021   & ...      &   ...    &   ... & ... & 18.48$^{\rm a}$  &   0.51   &   14.33  &   0.05   &  12.59   &   0.03   &  11.70   &   0.05   &  ...  &  ...  &  ...  &  ...  \\ 
  41  &  05350825-0550003   &  16.31   &   0.08   &  15.35   &   0.06   &  13.99   &   0.02   &   12.90  &   0.03   &  12.19   &   0.03   &  12.03   &   0.05   &  ...  &  ...  &  ...  &  ...  \\ 
  42  &  05351088-0549485   & ...      &   ...    &  17.45   &   0.12   &  15.56   &   0.05   &   14.10  &   0.04   &  13.49   &   0.04   &  13.33   &   0.09   &  ...  &  ...  &  ...  &  ...  \\ 
  43  &  05351602-0550448   & ...      &   ...    &   ...    &   ... 	&  ...     &  ...     &   13.31  &   0.04   &  12.44   &   0.03   &  12.05   &   0.05   &  ...  &  ...  &  ...  &  ...  \\ 
  44  &  05351473-0551040   & ...      &   ...    &   ...    &   ... 	&  ...     &  ...     &   12.59  &   0.03   &  11.36   &   0.03   &  10.78   &   0.04   &  ...  &  ...  &  ...  &  ...  \\ 
  45  &  05352193-0551071   & ...      &   ...    &   ...    &   ... 	&  ...     &  ...     &    9.72  &   0.03   &   9.32   &   0.03   &   9.26   &   0.04   &  ...  &  ...  &  ...  &  ...  \\ 
  46  &  05352538-0551087   & ...      &   ...    &   ...    &   ... 	&  ...     &  ...     &    8.83  &   0.04   &   8.52   &   0.03   &   8.36   &   0.04   &  ...  &  ...  &  ...  &  ...  \\ 
  47  &  05353023-0551169& 14.76$^{\rm c}$ & 0.05 &  ... &  ... & 13.31$^{\rm d}$  &   0.02   &   12.10  &   0.03   &  11.39   &   0.03   &  11.22   &   0.05   & 11.13 & 11.03 & 11.11 & 11.01 \\
  48  &  05353047-0549037   &  16.50   &   0.08   &  15.46   &   0.06   &  14.22   &   0.03   &   12.89  &   0.04   &  12.06   &   0.03   &  11.76   &   0.05   &  ...  &  ...  &  ...  &  ...  \\ 
  49  &  05353294-0549326   & ...      &   ...    &  18.01   &   0.17   &  16.21   &   0.08   &   14.56  &   0.05   &  13.77   &   0.04   &  13.49   &   0.10	&  ...  &  ...  &  ...  &  ...  \\ 
  50  &  05353385-0548211   & ...      &   ...    &  17.65   &   0.13   &  16.22   &   0.08   &   14.24  &   0.05   &  13.29   &   0.04   &  12.98   &   0.07	&  ...  &  ...  &  ...  &  ...  \\ 
  51  &  05353316-0547074   &  16.72   &   0.08   &  15.69   &   0.06   &  14.35   &   0.02   &   13.08  &   0.03   &  12.37   &   0.03   &  12.17   &   0.05	& 11.90 & 11.72 & 11.58 & 10.94 \\	     
  52  &  05352930-0545381   &  16.91   &   0.07   &  15.80   &   0.06   &  14.37   &   0.02   &   12.91  &   0.03   &  12.23   &   0.03   &  12.02   &   0.05	&  ...  &  ...  &  ...  &  ...  \\ 
  53  &  05353372-0546162   &  16.98   &   0.07   &  16.03   &   0.07   &  14.63   &   0.03   &   13.33  &   0.03   &  12.63   &   0.03   &  12.42   &   0.06	&  ...  &  ...  &  ...  &  ...  \\ 
  54  &  05353266-0545284   &  ...     &   ...    &  19.34   &   0.59   &  16.98   &   0.14   &   14.33  &   0.05   &  13.50   &   0.04   &  12.98   &   0.09   &  ...  &  ...  &  ...  &  ...  \\ 
  55  &  05353229-0544060   &  17.21   &   0.07   &  16.01   &   0.07   &  14.60   &   0.03   &   13.24  &   0.03   &  12.54   &   0.03   &  12.34   &   0.05   &  ...  &  ...  &  ...  &  ...  \\ 
\hline  	     
\medskip
\end{tabular}	     
\end{longtable}

\addtocounter{table}{-1}

\begin{longtable}{ r l l l l l l l l l l l l l l l l l }
\caption[ ]{\textit{Continued.}} \\
\scriptsize
\label{tab:VRIJHKbis} 
\begin{tabular}{ r l l l l l l l l l l l l l l l l l } 
\hline\hline
  Id  &   2MASS		    &   $V$     &  err	  &  $R$     &   err    &  $I$     &   err    &   $J$    &  err     &  $H$     &   err    &   $K$    &   err    & [3.6] & [4.5] & [5.8] & [8.0] \\ 
      &                     &  (mag)   &   (mag)  &  (mag)   &  (mag)   & (mag)    &  (mag)   &   (mag)  &  (mag)   &  (mag)   &  (mag)   &  (mag)   &  (mag)   & (mag) & (mag) & (mag) & (mag)	\\ 
\hline	     			      	   	 	    	           	  	     	   	   	   	      	   	       
  56  &  05353092-0543053   &  15.27   &   0.08   &  14.23   &   0.06   &  13.18   &   0.02   &  12.06   &   0.03   &  11.32   &   0.03   &  11.14   &   0.05	&  ...  &  ...  &  ...  &  ...  \\ 
  57  &  05352765-0542551   &  15.89   &   0.08   &  14.78   &   0.06   &  13.72   &   0.02   &  11.98   &   0.03   &  10.97   &   0.03   &  10.56   &   0.05	&  ...  &  ...  &  ...  &  ...  \\ 
  58  &  05352634-0543364   & ...      &   ...    &  18.17   &   0.20   &  16.23   &   0.07   &  14.20   &   0.04   &  13.56   &   0.04   &  13.29   &   0.09   &  ...  &  ...  &  ...  &  ...  \\ 
  59  &  05352125-0542123   &  15.95   &  0.08    &  14.99   &   0.06	&  13.84   &   0.02   &  12.29   &   0.04   &  11.45   &   0.03   &  10.94   &   0.05	& 10.11 &  9.84 &  9.45 &  8.46 \\ 
  60  &  05351643-0542396   &  13.58   &  0.09    &  12.84   &   0.06	&  12.32   &   0.02   &  11.51   &   0.03   &  10.87   &   0.03   &  10.75   &   0.05	& 10.70 & 10.89 & 10.75 & 10.48 \\ 
  61  &  05351734-0542145   &  16.49   &  0.08    &  15.41   &   0.06	&  14.08   &   0.02   &  12.46   &   0.04   &  11.56   &   0.03   &  11.00   &   0.05	&  9.85 &  9.12 &  8.67 &  7.55 \\ 
  62  &  05351715-0541538   &  17.83   &  0.06    &  16.44   &   0.08	&  14.73   &   0.03   &  13.16   &   0.04   &  12.53   &   0.03   &  12.26   &   0.06	& 11.68 & 11.57 & 11.18 & 10.48 \\ 
  63  &  05351123-0541361   & ...      &   ...    &  17.90   &   0.14   &  15.95   &   0.07   &  13.58   &   0.04   &  12.41   &   0.03   &  11.94   &   0.05   & 11.56 & 11.64 & 11.30 & 10.96 \\ 
  64  &  05350133-0541135   & ...      &   ...    &  ...     &   ...    &  ...     &   ...    &  12.91   &   0.04   &  11.27   &   0.03   &  10.36   &   0.04   & ...	& ...	& ...	& ...	\\ 
  65  &  05345908-0544303   & ...      &   ...    &  ...   & ... & 12.90$^{\rm c}$ &   0.03   &  11.31   &   0.03   &  10.47   &   0.03   &  10.08   &   0.04   & ...	& ...	& ...	& ...	\\ 
  66  &  05345923-0544553   & 18.26$^{\rm b}$ & 0.05 & ... & ... & 15.29$^{\rm b}$ &   0.02   &  12.64   &   0.03   &  11.16   &   0.03   &  10.27   &   0.04   &  8.48 &  7.99 &  7.50 &  6.89 \\
  67  &  05345827-0545056   & ...      &   ...    &  ...     &   ...    &  ...     &   ...    &  13.41   &   0.04   &  11.97   &   0.03   &  11.11   &   0.04   & ...	& ...	& ...	& ...	\\ 
  68  &  05345923-0545588   & ...      &   ...    &  ...     &   ...    &  ...     &   ...    &  11.57   &   0.03   &  10.21   &   0.03   &   9.48   &   0.04   & ...	& ...	& ...	& ...	\\ 
  69  &  05345898-0547596   & 19.08$^{\rm b}$ & 0.07 & ... & ... & 15.31$^{\rm b}$ &   0.01   &  12.15   &   0.03   &  10.79   &   0.03   &  10.18   &   0.04   & ...	& ...	& ...	& ...	\\ 
  70  &  05350054-0548591   & ...      &   ...    &  ...  &  ... & 15.01$^{\rm c}$ &   0.05   &  12.72   &   0.03   &  11.45   &   0.03   &  10.93   &   0.04   & ...	& ...	& ...	& ...	\\ 
  71  &  05345821-0549346   & ...      &   ...    &  ...     &   ... 	&  ...     &   ...    &  13.02   &   0.03   &  12.37   &   0.03   &  12.12   &   0.05   & ...	& ...	& ...	& ...	\\ 
  72  &  05350183-0549326   & ...      &   ...    &  ...     &   ... 	&  ...     &   ...    &  14.14   &   0.04   &  13.56   &   0.05   &  13.30   &   0.09   & ...	& ...	& ...	& ...	\\ 
  73  &  05350313-0550015   & ...      &   ...    &  ...     &   ... 	&  ...     &   ...    &  14.04   &   0.04   &  13.40   &   0.04   &  13.16   &   0.08   & ...	& ...	& ...	& ...	\\ 
  74  &  05350015-0550274   & ...      &   ...    &  ...     &   ... 	&  ...     &   ...    &  13.54   &   0.04   &  12.14   &   0.03   &  11.49   &   0.05   & ...	& ...	& ...	& ...	\\ 
  75  &  05350284-0551031   & 16.34$^{\rm b}$ & 0.03 & ... & ... & 14.10$^{\rm b}$ &   0.01   &  12.86   &   0.03   &  12.13   &   0.03   &  11.94   &   0.05   & ...	& ...	& ...	& ...	\\
  76  &  05350676-0551014   & ...      &   ...    &  ...     &   ... 	&  ...     &   ...    &  12.72   &   0.03   &  11.33   &   0.03   &  10.76   &   0.04   & ...	& ...	& ...	& ...	\\ 
  77  &  05351132-0550552   & ...      &   ...    &  ...     &   ... 	&  ...     &   ...    &  14.26   &   0.05   &  13.42   &   0.04   &  12.89   &   0.07   & ...	& ...	& ...	& ...	\\ 
  78  &  05351258-0551288   & ...      &   ...    &  ...     &   ... 	&  ...     &   ...    &  12.70   &   0.03   &  12.05   &   0.03   &  11.88   &   0.05   & ...	& ...	& ...	& ...	\\ 
  79  &  05352388-0551279   & ...      &   ...    &  ...     &   ... 	&  ...     &   ...    &  13.82   &   0.04   &  13.16   &   0.04   &  12.92   &   0.08   & ...	& ...	& ...	& ...	\\ 
  80  &  05353168-0542457   &  17.70   &  0.06    &  16.34   &   0.07   &  14.93   &   0.03   &  13.78   &   0.04   &  13.14   &   0.04   &  12.92   &   0.08   & ...	& ...	& ...	& ...	\\ 
  81  &  05345963-0547040   & ...      &   ...    &  ...     &   ... 	&  ...     &   ...    &  14.63   &   0.06   &  14.06   &   0.06   &  13.87   &   0.14   & ...	& ...	& ...	& ...	\\ 
  82  &  05345886-0547067   & ...      &   ...    &  ...     &   ... 	&  ...     &   ...    &  14.35   &   0.05   &  13.84   &   0.06   &  13.50   &   0.11   & ...	& ...	& ...	& ...	\\ 
  83  &  05345767-0548191   & ...      &   ...    &  ...     &   ... 	&  ...     &   ...    &  14.32   &   0.05   &  13.73   &   0.06   &  13.71   &   0.13   & ...	& ...	& ...	& ...	\\ 
  84  &  05345798-0548426   & ...      &   ...    &  ...     &   ... 	&  ...     &   ...    &  14.58   &   0.06   &  14.06   &   0.07   &  13.48   &   0.10   & ...	& ...	& ...	& ...	\\ 
  85  &  05345607-0549352   & ...      &   ...    &  ...     &   ... 	&  ...     &   ...    &  14.30   &   0.05   &  13.22   &   0.04   &  12.76   &   0.07   & ...	& ...	& ...	& ...	\\ 
  86  &  05351895-0545117   &  18.29   &  0.10    &  17.28   &   0.10   &  16.31   &   0.07   &  14.84   &   0.06   &  14.09   &   0.05   &  14.04   &   0.16   & ...	& ...	& ...	& ...	\\ 
  87  &  05351486-0547429   & ...      &   ...    &  ...     &   ... 	&  ...     &   ...    &  15.37   &   0.08   &  14.33   &   0.06   &  13.94   &   0.14   & ...	& ...	& ...	& ...	\\
  88  &  05353743-0543160   & ...      &   ...    &  ...     &   ... 	&  ...     &   ...    &  15.03   &   0.07   &  13.75   &   0.05   &  13.24   &   0.09   & ...	& ...	& ...	& ...	\\ 
  89  &  05353692-0543172   & ...      &   ...    &  17.85   &   0.15   &  16.09   &   0.07   &  14.48   &   0.05   &  13.82   &   0.05   &  13.52   &   0.11   & ...	& ...	& ...	& ...	\\ 
  90  &  05352650-0548497   &  13.15   &  0.09    &  12.65   &   0.06	&  12.13   &   0.02   &  11.24   &   0.03   &  10.62   &   0.03   &  10.50   &   0.04	& ...	& ...	& ...	& ...	\\ 
  91  &  05352707-0548522   &  12.36   &  0.09    &  11.78   &   0.06	&  11.30   &   0.02   &  10.49   &   0.03   &	9.90   &   0.03   &   9.70   &   0.05	& ...	& ...	& ...	& ...	\\ 
  92  &  05351197-0541521   &  18.33   &  0.11    &  17.13   &   0.10	&  16.01   &   0.06   &  14.60   &   0.06   &  13.83   &   0.06   &  13.61   &   0.13	& ...	& ...	& ...	& ...	\\ 
  93  &  05351790-0542340   &  12.32   &  0.09    &  11.75   &   0.06	&  11.18   &   0.02   &   9.89   &   0.03   &	9.19   &   0.03   &   8.67   &   0.05	&  7.98 &  7.65 &  7.33 &  6.53 \\ 
  94  &  05352080-0544527   &  18.50   &  0.25    &  17.89   &   0.15	&  16.87   &   0.12   &  15.35   &   0.08   &  14.61   &   0.08   &  14.40   &   0.22	& ...	& ...	& ...	& ...	\\ 
  95  &  05350499-0549297   & ...      &   ...    &  ...     &   ... 	&  ...     &   ...    &  15.05   &   0.07   &  14.49   &   0.08   &  14.43   &   0.24   & ...	& ...	& ...	& ...	\\ 
  96  &  05345751-0550324   & ...      &   ...    &  ...     &   ... 	&  ...     &   ...    &  14.80   &   0.08   &  14.28   &   0.15   &  13.51   &   0.16   & ...	& ...	& ...	& ...	\\ 
  97  &  05345782-0546008   & ...      &   ...    &  ...     &   ... 	&  ...     &   ...    &  15.02   &   0.09   &  14.35   &   0.10   &  14.09   &   0.20   & ...	& ...	& ...	& ...	\\
\hline  	     
\medskip
\end{tabular}	     
~\\  
$^{\rm a}$ Not detected in either of the ROSS bands. $I_{\rm N} = 18.48$ is the photographic $I$ magnitude 
($\lambda_{\rm eq}\simeq 0.8 \mu$m) from the GSC2.3 \citep{lask08}. \\
$^{\rm b}$ $V$ and $I$ from \citet{rebu00}.~~~~~~~~~~~~~~~~~~~~~~~~~~~~~~~~~~~~~~~~~~~~~~~~~~~~~~~~~~~~~~~~~~~~~~~~~~~~~~~~~~~~~~~~~~~~~~~~~~~~~~~~~~~~~~~~~~~~~~~~~~~~~~~~~~~~~~~~~~~~ \\
$^{\rm c}$ $I$ from \citet{DENIS05} catalogue. ~~~~~~~~~~~~~~~~~~~~~~~~~~~~~~~~~~~~~~~~~~~~~~~~~~~~~~~~~~~~~~~~~~~~~~~~~~~~~~~~~~~~~~~~~~~~~~~~~~~~~~~~~~~~~~~~~~~~~~~~~~~~~~~~~~~~~~~~~~~~~~~\\
$^{\rm d}$ $V-I$ from  \citet{rebu01}. ~~~~~~~~~~~~~~~~~~~~~~~~~~~~~~~~~~~~~~~~~~~~~~~~~~~~~~~~~~~~~~~~~~~~~~~~~~~~~~~~~~~~~~~~~~~~~~~~~~~~~~~~~~~~~~~~~~~~~~~~~~~~~~~~~~~~~~~~~~~~~~~~~~~~~~~\\
\end{longtable}
\end{landscape} 

\twocolumn

\begin{table*}
\caption{Reference star with Elodie spectra used for the spectral classification.} 
\label{tab:references}
\begin{tabular}{ l l c r r c c l l c r r} 
\hline\hline 
\noalign{\smallskip}
  Name  &  Sp. Type   & $T_{\rm eff}$ & $\log g$ &  [Fe/H]  &   &    &	Name	&    Sp. Type	& $T_{\rm eff}$ & $\log g$ &  [Fe/H] \\
        &             & (K)	      &  	 &          &   &    &          &		& (K)		&	   &	     \\ 
\noalign{\smallskip}
\hline 
\noalign{\smallskip}
\multicolumn{5}{c}{Main Sequence}                           &   &    &  	\multicolumn{5}{c}{Sub-giants}\\
 HD~128167  &     F2V	   &	6730  &  4.31  &  $-0.41$   &   &    &  HD~216385  &	 F7IV	  &    6290  &  3.97  &  $-0.25$  \\   
 HD~116568  &     F3V	   &	6824  &  4.39  &  $ 0.00$   &   &    &  HD~51530   &	 F8Vbw    &    6000  &  3.99  &  $-0.38$  \\   
 HD~61421   &     F5IV-V   &	6500  &  4.04  &  $ 0.00$   &   &    &  HD~204613  &	 G0IIIw   &    5650  &  3.80  &  $-0.35$  \\   
 HD~173667  &     F6V	   &	6310  &  3.99  &  $ 0.03$   &   &    &  HD~196755  &	 G5IV	  &    5510  &  3.60  &  $-0.09$  \\   
 HD~107213  &     F8Vs	   &	6260  &  4.01  &  $ 0.19$   &   &    &  HD~161797  &	 G5IV	  &    5590  &  3.94  &  $ 0.24$  \\   
 HD~181096  &     F6IV	   &	6250  &  3.90  &  $-0.29$   &   &    &  HD~191026  &	 K0IV	  &    5200  &  3.49  &  $-0.10$  \\   
 HD~693     &     F7V	   &	6200  &  4.16  &  $-0.37$   &   &    &  HD~188512  &	 G8IV	  &    5100  &  3.50  &  $-0.27$  \\   
 HD~219623  &     F7V	   &	6140  &  4.23  &  $ 0.01$   &   &    &  HD~198149  &	 K0IV	  &    5000  &  3.41  &  $-0.32$  \\   
 HD~187691  &     F8V	   &	6090  &  4.07  &  $ 0.07$   &   &    &  HD~127243  &	 G3IV	  &    5400  &  3.50  &  $-0.59$  \\   
 HD~19994   &     F8V	   &	6090  &  3.97  &  $ 0.14$   &   &    &  HD~168723  &	 K0III-IV &    4890  &  3.21  &  $-0.42$  \\   
 HD~114710  &     G0V	   &	6010  &  4.30  &  $-0.03$   &   &    &  HD~222404  &	 K1IV	  &    4800  &  3.00  &  $ 0.04$  \\   
 HD~109358  &     G0V	   &	5885  &  4.42  &  $-0.09$   &   &    & \multicolumn{5}{c}{Giants}\\				       
 HD~186408  &     G2V	   &	5800  &  4.26  &  $ 0.06$   &   &    &  HD~27022   &	G5III	 &    5275  &  2.60  &  $ 0.29$  \\    
 HD~217014  &     G2V	   &	5800  &  4.33  &  $ 0.20$   &   &    &  HD~216131  &	G8III	 &    4950  &  2.50  &  $-0.03$  \\    
 HD~86728   &     G1V	   &	5680  &  4.28  &  $ 0.11$   &   &    &  HD~3546    &	G8III	 &    4930  &  3.16  &  $-0.64$  \\    
 HD~076151  &     G3V	   &	5690  &  4.37  &  $ 0.02$   &   &    &  HD~27348   &	G8III	 &    4875  &  2.55  &  $ 0.04$  \\    
 HD~157214  &     G2V	   &	5740  &  4.24  &  $-0.34$   &   &    &  HD~35369   &	G8III	 &    4850  &  2.24  &  $-0.19$  \\    
 HD~126053  &     G1V	   &	5690  &  4.45  &  $-0.35$   &   &    &  HD~19476   &	K0III	 &    4940  &  3.08  &  $ 0.04$  \\    
 HD~013403  &     G3V	   &	5653  &  4.00  &  $-0.31$   &   &    &  HD~135722  &	G8III	 &    4847  &  2.56  &  $-0.44$  \\    
 HD~106116  &     G4V	   &	5512  &  4.50  &  $-0.10$   &   &    &  HD~180711  &	G9III	 &    4820  &  2.98  &  $-0.27$  \\    
 HD~117176  &     G5V	   &	5480  &  3.83  &  $-0.11$   &   &    &  HD~23183   &	G8III	 &    4782  &  2.15  &  $-0.45$  \\    
 HD~174719  &     G6V	   &	5518  &  4.30  &  $-0.20$   &   &    &  HD~13530   &	G8III	 &    4920  &  3.16  &  $-0.50$  \\    
 HD~169822  &     G7V	   &	5500  &  4.40  &  $ 0.00$   &   &    &  HD~40460   &	K1III	 &    4741  &  2.00  &  $-0.50$  \\    
 HD~068638  &     G8V	   &	5430  &  4.40  &  $-0.24$   &   &    &  HD~41597   &	G8III	 &    4700  &  2.30  &  $-0.54$  \\    
 HD~13783   &     G8V	   &	5338  &  4.35  &  $-0.55$   &   &    &  HD~37160   &	G8III-IV &    4800  &  2.46  &  $-0.53$  \\    
 HD~185144  &     K0V	   &	5260  &  4.55  &  $-0.24$   &   &    &  HD~39003   &	K0III	 &    4550  &  1.90  &  $-0.04$  \\    
 HD~117635  &     G9V	   &	5200  &  4.10  &  $-0.48$   &   &    &  HD~212943  &	K0III	 &    4586  &  2.81  &  $-0.34$  \\    
 HD~145675  &     K0V	   &	5330  &  4.51  &  $ 0.45$   &   &    &  HD~177463  &	K1III	 &    4560  &  2.65  &  $-0.25$  \\    
 HD~10476   &     K1V	   &	5150  &  4.44  &  $-0.17$   &   &    &  HD~96833   &	K1III	 &    4520  &  2.36  &  $-0.16$  \\    
 HD~165341  &     K0V	   &	5260  &  5.00  &  $-0.25$   &   &    &  HD~85503   &	K0III	 &    4540  &  2.20  &  $ 0.29$  \\    
 HD~23439   &     K1V	   &	5050  &  4.5   &  $-1.10$   &   &    &  HD~48433   &	K1III	 &    4460  &  1.88  &  $-0.25$  \\    
 HD~132142  &     K1V	   &	5108  &  4.50  &  $-0.55$   &   &    &  HD~9927    &	K3III	 &    4417  &  2.06  &  $-0.01$  \\    
 HD~17382   &     K1V	   &	5065  &  4.50  &  $-0.13$   &   &    &  HD~102224  &	K0III	 &    4350  &  1.15  &  $-0.43$  \\    
 HD~190404  &     K1V	   &	5051  &  4.45  &  $-0.17$   &   &    &  HD~124897  &	K2IIIp   &    4300  &  1.50  &  $-0.63$  \\    
 HD~4628    &     K2V	   &	4960  &  4.60  &  $-0.29$   &   &    &  HD~30834   &	K3III	 &    4115  &  1.73  &  $-0.21$  \\    
 HD~166620  &     K2V	   &	4944  &  4.47  &  $-0.23$   &   &    &  HD~10380   &	K3III	 &    4057  &  1.43  &  $-0.25$  \\    
 HD~82106   &     K3V	   &	4827  &  4.10  &  $-0.11$   &   &    &  HD~161074  &	K4III	 &    3980  &  1.73  &  $-0.27$  \\    
 HD~219134  &     K3V	   &	4710  &  4.50  &  $ 0.20$   &   &    &  HD~29139   &	K5III	 &    3850  &  0.55  &  $-0.10$  \\    
 HD~32147   &     K3V	   &	4600  &  4.55  &  $ 0.28$   &   &    &  HD~46784   &	M0III	 &    3616  &  1.45  &  $ 0.07$  \\    
 HD~131977  &     K4V	   &	4600  &  4.70  &  $ 0.04$   &   &    &  HD~168720  &	M1III	 &    3790  &  1.83  &  $ 0.04$  \\    
 HD~190007  &     K4V	   &	4563  &  4.36  &  $ 0.00$   &   &    &  HD~169931  &	M7II	 &    3106  & $-0.47$&  $-0.21$  \\    
 HD~29697   &     K3V	   &	4454  &  4.50  &  $-0.04$   &   &    &  	   &		 &	    &	     &  	 \\ 
 HD~101206  &     K5V	   &	4601  &  4.00  &  $-0.58$   &   &    &  	   &		 &	    &	     &  	 \\ 
 HD~201091  &     K5V	   &	4500  &  4.56  &  $-0.43$   &   &    &  	   &		 &	    &	     &  	 \\ 
 HD~28343   &     K7V	   &	4284  &  4.60  &  $ 0.01$   &   &    &  	   &		 &	    &	     &  	 \\ 
 HD~157881  &     K7V	   &	4200  &  4.70  &  $-0.20$   &   &    &  	   &		 &	    &	     &  	 \\ 
 HD~201092  &     K7V	   &	4200  &  4.40  &  $-0.63$   &   &    &  	   &		 &	    &	     &  	 \\ 
 HD~88230   &     K8V	   &	4000  &  4.50  &  $ 0.28$   &   &    &  	   &		 &	    &	     &  	 \\  
 HD~79210   &     M0V	   &	3868  &  4.61  &  $-0.38$   &   &    &  	   &		 &	    &	     &  	 \\ 
 BD+36\,2219&     M1V	   &	3748  &  4.76  &  $-0.45$   &   &    &  	   &		 &	    &	     &  	 \\ 
 HD~119850  &     M1.5V    &    3623  &  4.80  &  $-0.10$   &   &    &  	   &		 &	    &	     &  	 \\  
 HD~36395   &     M1.5V    &    3626  &  4.80  &  $ 0.60$   &   &    &  	   &		 &	    &	     &  	 \\  
 BD+44\,2051&     M2V	   &	3650  &  4.85  &  $-0.43$   &   &    &  	   &		 &	    &	     &  	 \\  
 GJ~2066    &     M2V	   & ~3650$^{\rm a}$  &  ---~  &  ---~~ &  &  & 	   &		 &	    &	     &  	 \\  
 HD~95735   &     M2V	   &   3620   &  4.90  &  $-0.20$   &   &    &	           &		 &	    &	     &  	 \\  
 HD~1326A   &     M2V	   & ~3520$^{\rm a}$  &  ---~  &  ---~~ &  &  & 	   &		 &	    &	     &  	 \\  
 GJ~408     &     M2.5V    & ~3460$^{\rm a}$  &  ---~  &  ---~~ &  &  & 	   &		 &	    &	     &  	 \\  
 HD~173739  &     M3V	   & ~3400$^{\rm a}$  &  ---~  &  ---~~ &  &  & 	   &		 &	    &	     &  	 \\  
 G~103-68   &     M3V	   & ~3400$^{\rm a}$  &  ---~  &  ---~~ &  &  & 	   &		 &	    &	     &  	 \\  
 GJ~896A    &     M3.5V    &    3630  &  4.90  &  $ 0.10$   &   &    &  	   &		 &	    &	     &  	 \\  
 HD~173740  &     M3.5V    &	3395  &  4.93  &  $-0.54$   &   &    &  	   &		 &	    &	     &  	 \\  
\hline				                								        	 
\end{tabular}			         	                	                 
\begin{list}{}{}		         	                	                 
\item[$^{\mathrm{a}}$] Effective temperature from spectral-type according to \citet{DeJa87}.
\end{list}
\end{table*} 

\end{document}